\documentclass[11pt]{article}

\newcommand{\marcnbox}
{\stln \lower1.4ex \hbox{
\begin{picture} (6.6, 3.1)
\multiput(.3, .3) (1, 0) {2} {\fr}
\put(2.3, .3) {\framebox(3,1){$\cdots$}}
\put(5.3, .3){\fr}
\put(.3, 1.4)
{$\overbrace{~~~~~~~~~~~~~~~~~~}^{n}$}
\end{picture}}}

\newcommand{\marcnplusonebox}
{\stln \lower1.4ex \hbox{
\begin{picture} (6.6, 3.1)
\multiput(.3, .3) (1, 0) {2} {\fr}
\put(2.3, .3) {\framebox(3,1){$\cdots$}}
\put(5.3, .3){\fr}
\put(.3, 1.4)
{$\overbrace{~~~~~~~~~~~~~~~~~~}^{n+1}$}
\end{picture}}}

\newcommand{\marcnplustwobox}
{\stln \lower1.4ex \hbox{
\begin{picture} (6.6, 3.1)
\multiput(.3, .3) (1, 0) {2} {\fr}
\put(2.3, .3) {\framebox(3,1){$\cdots$}}
\put(5.3, .3){\fr}
\put(.3, 1.4)
{$\overbrace{~~~~~~~~~~~~~~~~~~}^{n+2}$}
\end{picture}}}

\newcommand{\marcnplusthreebox}
{\stln \lower1.4ex \hbox{
\begin{picture} (6.6, 3.1)
\multiput(.3, .3) (1, 0) {2} {\fr}
\put(2.3, .3) {\framebox(3,1){$\cdots$}}
\put(5.3, .3){\fr}
\put(.3, 1.4)
{$\overbrace{~~~~~~~~~~~~~~~~~~}^{n+3}$}
\end{picture}}}

\newcommand{\marcnplusfourbox}
{\stln \lower1.4ex \hbox{
\begin{picture} (6.6, 3.1)
\multiput(.3, .3) (1, 0) {2} {\fr}
\put(2.3, .3) {\framebox(3,1){$\cdots$}}
\put(5.3, .3){\fr}
\put(.3, 1.4)
{$\overbrace{~~~~~~~~~~~~~~~~~~}^{n+4}$}
\end{picture}}}

\newcommand{\oneoneonebox}
{\stln \lower3.8ex \hbox{
\begin{picture}(1.6,3.6)
\multiput(.3,.3)(0,1){3}{\fr}
\end{picture}}}

\newcommand{\gencolbox}
{\stln \lower8.6ex \hbox{
\begin{picture}(1.6,7.6)
\multiput(.3, 4.3)(0,1){3}{\fr}
\put(.3,1.3){\framebox(1,3){$\vdots$}}
\put(.3,.3){\fr}
\end{picture}}}

\newcommand{\soneoneoneonebox}
{\stln \lower5ex\hbox{
\begin{picture}(1.6,4.6)
\multiput(.3,.3)(0,1){4}{\sfr}
\end{picture}}}

\newcommand{\sthreethreebox}
{\stln \lower2.6ex\hbox{
\begin{picture}(3.6,2.6)
\multiput(.3,1.3)(1,0){3}{\sfr}
\multiput(.3,.3)(1,0){3}{\sfr}
\end{picture}}}

\newcommand{\stwotwooneonebox}
{\stln \lower5ex\hbox{
\begin{picture}(2.6,4.6)
\multiput(.3,.3)(0,1){4}{\sfr}
\put(1.3,2.3){\sfr}
\put(1.3,3.3){\sfr}
\end{picture}}}

\newcommand{\soneoneoneoneoneonebox}
{\stln \lower7.4ex\hbox{
\begin{picture}(1.6,6.6)
\multiput(.3,.3)(0,1){6}{\sfr}
\end{picture}}}

\newcommand{\soneoneonebox}
{\stln \lower3.8ex\hbox{
\begin{picture}(1.6,3.6)
\multiput(.3,.3)(0,1){3}{\sfr}
\end{picture}}}

\newcommand{\sgencolbox}
{\stln \lower8.6ex\hbox{
\begin{picture}(1.6,7.6)
\multiput(.3,4.3)(0,1){3}{\sfr}
\put(.3,1.3){\framebox(1,3){$\vdots$}}
\put(.3,.3){\sfr}
\end{picture}}}

\newcommand{\stln}{\setlength{\unitlength}{2.2ex}}
\newcommand{\fr}{\framebox(1,1){}}
\newcommand{\sfr}{\framebox(1,1){\begin{picture}(1,1)
  \put(0,0){\line(1,1){1}}\end{picture}}}

\newcommand{\marcgenrowbox}
{\stln \lower1.4ex \hbox{
\begin{picture} (6.6, 3.1)
\multiput(.3, .3) (1, 0) {2} {\fr}
\put(2.3, .3) {\framebox(3,1){$\cdots$}}
\put(5.3, .3){\fr}
\put(.3, 1.4) {$\overbrace{~~~~~~~~~~~~~~~~~~}^{n}$}
\end{picture}}}

\newcommand{\onebox}
{\stln \lower1.4ex\hbox{
\begin{picture}(1.6,1.6)
\put(.3,.3){\fr}
\end{picture}}}

\newcommand{\twobox}
{\stln \lower1.4ex\hbox{
\begin{picture}(2.6,1.6)
\put(.3,.3){\fr}
\put(1.3,.3){\fr}
\end{picture}}}

\newcommand{\threebox}
{\stln \lower1.4ex\hbox{
\begin{picture}(3.6,1.6)
\multiput(.3,.3)(1,0){3}{\fr}
\end{picture}}}

\newcommand{\fourbox}
{\stln \lower1.4ex\hbox{
\begin{picture}(4.6,1.6)
\multiput(.3,.3)(1,0){4}{\fr}
\end{picture}}}

\newcommand{\fivebox}
{\stln \lower1.4ex\hbox{
\begin{picture}(5.6,1.6)
\multiput(.3,.3)(1,0){5}{\fr}
\end{picture}}}

\newcommand{\sixbox}
{\stln \lower1.4ex\hbox{
\begin{picture}(6.6,1.6)
\multiput(.3,.3)(1,0){6}{\fr}
\end{picture}}}

\newcommand{\genrowbox}
{\stln \lower1.4ex\hbox{
\begin{picture}(7.6,1.6)
\multiput(.3,.3)(1,0){3}{\fr}
\put(3.3,.3){\framebox(3,1){$\cdots$}}
\put(6.3,.3){\fr}
\end{picture}}}

\newcommand{\oneonebox}
{\stln \lower2.6ex\hbox{
\begin{picture}(1.6,2.6)
\put(.3,.3){\fr}
\put(.3,1.3){\fr}
\end{picture}}}

\newcommand{\twoonebox}
{\stln \lower2.6ex\hbox{
\begin{picture}(2.6,2.6)
\put(.3,1.3){\fr}
\put(1.3,1.3){\fr}
\put(0.3,0.3){\fr}
\end{picture}}}

\newcommand{\threeonebox}
{\stln \lower2.6ex\hbox{
\begin{picture}(3.6,2.6)
\multiput(.3,1.3)(1,0){3}{\fr}
\put(.3,.3){\fr}
\end{picture}}}

\newcommand{\fouronebox}
{\stln \lower2.6ex\hbox{
\begin{picture}(4.6,2.6)
\multiput(.3,1.3)(1,0){4}{\fr}
\put(.3,.3){\fr}
\end{picture}}}

\newcommand{\fiveonebox}
{\stln \lower2.6ex\hbox{
\begin{picture}(5.6,2.6)
\multiput(.3,1.3)(1,0){5}{\fr}
\put(.3,.3){\fr}
\end{picture}}}

\newcommand{\sixonebox}
{\stln \lower2.6ex\hbox{
\begin{picture}(6.6,2.6)
\multiput(.3,1.3)(1,0){6}{\fr}
\put(.3,.3){\fr}
\end{picture}}}

\newcommand{\twotwobox}
{\stln \lower2.6ex\hbox{
\begin{picture}(2.6,2.6)
\put(.3,.3){\fr}
\put(.3,1.3){\fr}
\put(1.3,.3){\fr}
\put(1.3,1.3){\fr}
\end{picture}}}

\newcommand{\threetwobox}
{\stln \lower2.6ex\hbox{
\begin{picture}(3.6,2.6)
\multiput(.3,1.3)(1,0){3}{\fr}
\put(.3,.3){\fr}
\put(1.3,.3){\fr}
\end{picture}}}

\newcommand{\fourtwobox}
{\stln \lower2.6ex\hbox{
\begin{picture}(4.6,2.6)
\multiput(.3,1.3)(1,0){4}{\fr}
\put(.3,.3){\fr}
\put(1.3,.3){\fr}
\end{picture}}}

\newcommand{\fivetwobox}
{\stln \lower2.6ex\hbox{
\begin{picture}(5.6,2.6)
\multiput(.3,1.3)(1,0){5}{\fr}
\put(.3,.3){\fr}
\put(1.3,.3){\fr}
\end{picture}}}

\newcommand{\sixtwobox}
{\stln \lower2.6ex\hbox{
\begin{picture}(6.6,2.6)
\multiput(.3,1.3)(1,0){6}{\fr}
\put(.3,.3){\fr}
\put(1.3,.3){\fr}
\end{picture}}}

\newcommand{\threethreebox}
{\stln \lower2.6ex\hbox{
\begin{picture}(3.6,2.6)
\multiput(.3,1.3)(1,0){3}{\fr}
\multiput(.3,.3)(1,0){3}{\fr}
\end{picture}}}

\newcommand{\fourthreebox}
{\stln \lower2.6ex\hbox{
\begin{picture}(4.6,2.6)
\multiput(.3,1.3)(1,0){4}{\fr}
\multiput(.3,.3)(1,0){3}{\fr}
\end{picture}}}

\newcommand{\fivethreebox}
{\stln \lower2.6ex\hbox{
\begin{picture}(5.6,2.6)
\multiput(.3,1.3)(1,0){5}{\fr}
\multiput(.3,.3)(1,0){3}{\fr}
\end{picture}}}

\newcommand{\sixthreebox}
{\stln \lower2.6ex\hbox{
\begin{picture}(6.6,2.6)
\multiput(.3,1.3)(1,0){6}{\fr}
\multiput(.3,.3)(1,0){3}{\fr}
\end{picture}}}

\newcommand{\fourfourbox}
{\stln \lower2.6ex\hbox{
\begin{picture}(4.6,2.6)
\multiput(.3,1.3)(1,0){4}{\fr}
\multiput(.3,.3)(1,0){4}{\fr}
\end{picture}}}

\newcommand{\fivefourbox}
{\stln \lower2.6ex\hbox{
\begin{picture}(5.6,2.6)
\multiput(.3,1.3)(1,0){5}{\fr}
\multiput(.3,.3)(1,0){4}{\fr}
\end{picture}}}

\newcommand{\sixfourbox}
{\stln \lower2.6ex\hbox{
\begin{picture}(6.6,2.6)
\multiput(.3,1.3)(1,0){6}{\fr}
\multiput(.3,.3)(1,0){4}{\fr}
\end{picture}}}

\newcommand{\fivefivebox}
{\stln \lower2.6ex\hbox{
\begin{picture}(5.6,2.6)
\multiput(.3,1.3)(1,0){5}{\fr}
\multiput(.3,.3)(1,0){5}{\fr}
\end{picture}}}

\newcommand{\sixfivebox}
{\stln \lower2.6ex\hbox{
\begin{picture}(6.6,2.6)
\multiput(.3,1.3)(1,0){6}{\fr}
\multiput(.3,.3)(1,0){5}{\fr}
\end{picture}}}

\newcommand{\sixsixbox}
{\stln \lower2.6ex\hbox{
\begin{picture}(6.6,2.6)
\multiput(.3,1.3)(1,0){6}{\fr}
\multiput(.3,.3)(1,0){6}{\fr}
\end{picture}}}

\newcommand{\sonebox}
{\stln \lower1.4ex\hbox{
\begin{picture}(1.6,1.6)
\put(.3,.3){\sfr}
\end{picture}}}

\newcommand{\stwobox}
{\stln \lower1.4ex\hbox{
\begin{picture}(2.6,1.6)
\put(.3,.3){\sfr}
\put(1.3,.3){\sfr}
\end{picture}}}

\newcommand{\sthreebox}
{\stln \lower1.4ex\hbox{
\begin{picture}(3.6,1.6)
\multiput(.3,.3)(1,0){3}{\sfr}
\end{picture}}}

\newcommand{\sfourbox}
{\stln \lower1.4ex\hbox{
\begin{picture}(4.6,1.6)
\multiput(.3,.3)(1,0){4}{\sfr}
\end{picture}}}

\newcommand{\sfivebox}
{\stln \lower1.4ex\hbox{
\begin{picture}(5.6,1.6)
\multiput(.3,.3)(1,0){5}{\sfr}
\end{picture}}}

\newcommand{\ssixbox}
{\stln \lower1.4ex\hbox{
\begin{picture}(6.6,1.6)
\multiput(.3,.3)(1,0){6}{\sfr}
\end{picture}}}

\newcommand{\sgenrowbox}
{\stln \lower1.4ex\hbox{
\begin{picture}(7.6,1.6)
\multiput(.3,.3)(1,0){3}{\sfr}
\put(3.3,.3){\framebox(3,1){$\cdots$}}
\put(6.3,.3){\sfr}
\end{picture}}}

\newcommand{\soneonebox}
{\stln \lower2.6ex\hbox{
\begin{picture}(1.6,2.6)
\put(.3,.3){\sfr}
\put(.3,1.3){\sfr}
\end{picture}}}

\newcommand{\stwoonebox}
{\stln \lower2.6ex\hbox{
\begin{picture}(2.6,2.6)
\put(.3,1.3){\sfr}
\put(1.3,1.3){\sfr}
\put(0.3,0.3){\sfr}
\end{picture}}}

\newcommand{\sthreeonebox}
{\stln \lower2.6ex\hbox{
\begin{picture}(3.6,2.6)
\multiput(.3,1.3)(1,0){3}{\sfr}
\put(.3,.3){\sfr}
\end{picture}}}

\newcommand{\sfouronebox}
{\stln \lower2.6ex\hbox{
\begin{picture}(4.6,2.6)
\multiput(.3,1.3)(1,0){4}{\sfr}
\put(.3,.3){\sfr}
\end{picture}}}

\newcommand{\sfiveonebox}
{\stln \lower2.6ex\hbox{
\begin{picture}(5.6,2.6)
\multiput(.3,1.3)(1,0){5}{\sfr}
\put(.3,.3){\sfr}
\end{picture}}}

\newcommand{\ssixonebox}
{\stln \lower2.6ex\hbox{
\begin{picture}(6.6,2.6)
\multiput(.3,1.3)(1,0){6}{\sfr}
\put(.3,.3){\sfr}
\end{picture}}}

\newcommand{\stwotwobox}
{\stln \lower2.6ex\hbox{
\begin{picture}(2.6,2.6)
\put(.3,.3){\sfr}
\put(.3,1.3){\sfr}
\put(1.3,.3){\sfr}
\put(1.3,1.3){\sfr}
\end{picture}}}

\newcommand{\sthreetwobox}
{\stln \lower2.6ex\hbox{
\begin{picture}(3.6,2.6)
\multiput(.3,1.3)(1,0){3}{\sfr}
\put(.3,.3){\sfr}
\put(1.3,.3){\sfr}
\end{picture}}}

\newcommand{\sfourtwobox}
{\stln \lower2.6ex\hbox{
\begin{picture}(4.6,2.6)
\multiput(.3,1.3)(1,0){4}{\sfr}
\put(.3,.3){\sfr}
\put(1.3,.3){\sfr}
\end{picture}}}

\newcommand{\sfivetwobox}
{\stln \lower2.6ex\hbox{
\begin{picture}(5.6,2.6)
\multiput(.3,1.3)(1,0){5}{\sfr}
\put(.3,.3){\sfr}
\put(1.3,.3){\sfr}
\end{picture}}}

\newcommand{\ssixtwobox}
{\stln \lower2.6ex\hbox{
\begin{picture}(6.6,2.6)
\multiput(.3,1.3)(1,0){6}{\sfr}
\put(.3,.3){\sfr}
\put(1.3,.3){\sfr}
\end{picture}}}

\newcommand{\sfourthreebox}
{\stln \lower2.6ex\hbox{
\begin{picture}(4.6,2.6)
\multiput(.3,1.3)(1,0){4}{\sfr}
\multiput(.3,.3)(1,0){3}{\sfr}
\end{picture}}}

\newcommand{\sfivethreebox}
{\stln \lower2.6ex\hbox{
\begin{picture}(5.6,2.6)
\multiput(.3,1.3)(1,0){5}{\sfr}
\multiput(.3,.3)(1,0){3}{\sfr}
\end{picture}}}

\newcommand{\ssixthreebox}
{\stln \lower2.6ex\hbox{
\begin{picture}(6.6,2.6)
\multiput(.3,1.3)(1,0){6}{\sfr}
\multiput(.3,.3)(1,0){3}{\sfr}
\end{picture}}}

\newcommand{\sfourfourbox}
{\stln \lower2.6ex\hbox{
\begin{picture}(4.6,2.6)
\multiput(.3,1.3)(1,0){4}{\sfr}
\multiput(.3,.3)(1,0){4}{\sfr}
\end{picture}}}

\newcommand{\sfivefourbox}
{\stln \lower2.6ex\hbox{
\begin{picture}(5.6,2.6)
\multiput(.3,1.3)(1,0){5}{\sfr}
\multiput(.3,.3)(1,0){4}{\sfr}
\end{picture}}}

\newcommand{\ssixfourbox}
{\stln \lower2.6ex\hbox{
\begin{picture}(6.6,2.6)
\multiput(.3,1.3)(1,0){6}{\sfr}
\multiput(.3,.3)(1,0){4}{\sfr}
\end{picture}}}

\newcommand{\sfivefivebox}
{\stln \lower2.6ex\hbox{
\begin{picture}(5.6,2.6)
\multiput(.3,1.3)(1,0){5}{\sfr}
\multiput(.3,.3)(1,0){5}{\sfr}
\end{picture}}}

\newcommand{\ssixfivebox}
{\stln \lower2.6ex\hbox{
\begin{picture}(6.6,2.6)
\multiput(.3,1.3)(1,0){6}{\sfr}
\multiput(.3,.3)(1,0){5}{\sfr}
\end{picture}}}

\newcommand{\ssixsixbox}
{\stln \lower2.6ex\hbox{
\begin{picture}(6.6,2.6)
\multiput(.3,1.3)(1,0){6}{\sfr}
\multiput(.3,.3)(1,0){6}{\sfr}
\end{picture}}}

\newcommand{\sgenrowonebox}
{\stln \lower2.6ex\hbox{
\begin{picture}(7.6,2.6)
\multiput(.3,1.3)(1,0){3}{\sfr}
\put(3.3,1.3){\framebox(3,1){$\cdots$}}
\put(6.3,1.3){\sfr}
\put(.3,.3){\sfr}
\end{picture}}}

\newcommand{\sgenrowtwobox}
{\stln \lower2.6ex\hbox{
\begin{picture}(7.6,2.6)
\multiput(.3,1.3)(1,0){3}{\sfr}
\put(3.3,1.3){\framebox(3,1){$\cdots$}}
\put(6.3,1.3){\sfr}
\put(.3,.3){\sfr}
\put(1.3,.3){\sfr}
\end{picture}}}

\newcommand{\marcshapirowbox}
{\stln \lower2.6ex\hbox{
\begin{picture}(12.6,2.6)
\multiput(.3,1.3)(1,0){3}{\sfr}
\put(3.3,1.3){\framebox(3,1){$\cdots$}}
\put(6.3,1.3){\sfr}
\put(7.3,1.3){\sfr}
\put(8.3,1.3){\framebox(3,1){$\cdots$}}
\put(11.3,1.3){\sfr}
\put(7.3,1){$\underbrace{~~~~~~~~~~~~~~~}_{k}$}
\multiput(.3,.3)(1,0){3}{\sfr}
\put(3.3,.3){\framebox(3,1){$\cdots$}}
\put(6.3,.3){\sfr}
\put(.3,0){$\underbrace{~~~~~~~~~~~~~~~~~~~~~}_{n}$}
\end{picture}}}

\newcommand{\sgentworowonerowbox}
{\stln \lower2.6ex\hbox{
\begin{picture}(11.6,2.6)
\multiput(.3,1.3)(1,0){3}{\sfr}
\put(3.3,1.3){\framebox(3,1){$\cdots$}}
\put(6.3,1.3){\sfr}
\put(7.3,1.3){\framebox(3,1){$\cdots$}}
\put(10.3,1.3){\sfr}
\multiput(.3,.3)(1,0){3}{\sfr}
\put(3.3,.3){\framebox(3,1){$\cdots$}}
\put(6.3,.3){\sfr}
\end{picture}}}

\newcommand{\sgenrowrowbox}
{\stln \lower2.6ex\hbox{
\begin{picture}(7.6,2.6)
\multiput(.3,1.3)(1,0){3}{\sfr}
\put(3.3,1.3){\framebox(3,1){$\cdots$}}
\put(6.3,1.3){\sfr}
\multiput(.3,.3)(1,0){3}{\sfr}
\put(3.3,.3){\framebox(3,1){$\cdots$}}
\put(6.3,.3){\sfr}
\end{picture}}}

\newcommand{\marctwojbox}
{\stln \lower1.4ex \hbox{
\begin{picture} (6.6, 3.1)
\multiput(.3, .3) (1, 0) {2} {\fr}
\put(2.3, .3) {\framebox(3,1){$\cdots$}}
\put(5.3, .3){\fr}
\put(.3, 1.4) {$\overbrace{~~~~~~~~~~~~~~~~~~}^{2j}$}
\end{picture}}}

\newcommand{\marctwojplusonebox}
{\stln \lower1.4ex \hbox{
\begin{picture} (6.6, 3.1)
\multiput(.3, .3) (1, 0) {2} {\fr}
\put(2.3, .3) {\framebox(3,1){$\cdots$}}
\put(5.3, .3){\fr}
\put(.3, 1.4) {$\overbrace{~~~~~~~~~~~~~~~~~~}^{2j+1}$}
\end{picture}}}

\newcommand{\marctwojplustwobox}
{\stln \lower1.4ex \hbox{
\begin{picture} (6.6, 3.1)
\multiput(.3, .3) (1, 0) {2} {\fr}
\put(2.3, .3) {\framebox(3,1){$\cdots$}}
\put(5.3, .3){\fr}
\put(.3, 1.4) {$\overbrace{~~~~~~~~~~~~~~~~~~}^{2j+2}$}
\end{picture}}}

\newcommand{\marctwojminusonebox}
{\stln \lower1.4ex \hbox{
\begin{picture} (6.6, 3.1)
\multiput(.3, .3) (1, 0) {2} {\fr}
\put(2.3, .3){\framebox(3,1){$\cdots$}}
\put(5.3, .3){\fr}
\put(.3, 1.4) {$\overbrace{~~~~~~~~~~~~~~~~~~}^{2j-1}$}
\end{picture}}}

\newcommand{\marctwojminustwobox}
{\stln \lower1.4ex \hbox{
\begin{picture} (6.6, 3.1)
\multiput(.3, .3) (1, 0) {2} {\fr}
\put(2.3, .3) {\framebox(3,1){$\cdots$}}
\put(5.3, .3){\fr}
\put(.3, 1.4) {$\overbrace{~~~~~~~~~~~~~~~~~~}^{2j-2}$}
\end{picture}}}

\newcommand{\marctwojonebox}
{\stln \lower2.6ex \hbox{
\begin{picture}(6.6, 4.1)
\multiput(.3, 1.3)(1,0){2}{\fr}
\put(2.3, 1.3) {\framebox(3, 1){$\cdots$}}
\put(5.3, 1.3) {\fr}
\put(.3,.3) {\fr}
\put(.3, 2.4){$\overbrace{~~~~~~~~~~~~~~~~~~}^{2j}$}
\end{picture}}}

\newcommand{\marctwojplustwoonebox}
{\stln \lower2.6ex \hbox{
\begin{picture}(6.6, 4.1)
\multiput(.3, 1.3)(1,0){2}{\fr}
\put(2.3, 1.3) {\framebox(3, 1){$\cdots$}}
\put(5.3, 1.3) {\fr}
\put(.3,.3) {\fr}
\put(.3, 2.4){$\overbrace{~~~~~~~~~~~~~~~~~~}^{2j+2}$}
\end{picture}}}

\newcommand{\marctwojminusoneonebox}
{\stln \lower2.6ex \hbox{
\begin{picture}(6.6, 4.1)
\multiput(.3, 1.3)(1,0){2}{\fr}
\put(2.3, 1.3) {\framebox(3, 1){$\cdots$}}
\put(5.3, 1.3) {\fr}
\put(.3,.3) {\fr}
\put(.3, 2.4){$\overbrace{~~~~~~~~~~~~~~~~~~}^{2j-1}$}
\end{picture}}}

\newcommand{\marctwojplusoneonebox}
{\stln \lower2.6ex \hbox{
\begin{picture}(6.6, 4.1)
\multiput(.3, 1.3)(1,0){2}{\fr}
\put(2.3, 1.3) {\framebox(3, 1){$\cdots$}}
\put(5.3, 1.3) {\fr}
\put(.3, .3) {\fr}
\put(.3, 2.4){$\overbrace{~~~~~~~~~~~~~~~~~~}^{2j+1}$}
\end{picture}}}

\newcommand{\marctwojtwobox}
{\stln \lower2.6ex \hbox{
\begin{picture}(7.6, 4.1)
\multiput(.3, 1.3)(1,0){3}{\fr}
\put(3.3, 1.3){\framebox(3, 1){$\cdots$}}
\put(6.3, 1.3){\fr}
\multiput(.3,.3)(1,0){2}{\fr}
\put(.3, 2.4) {$\overbrace{~~~~~~~~~~~~~~~~~~~~~}^{2j}$}
\end{picture}}}

\newcommand{\marctwojplusonetwobox}
{\stln \lower2.6ex \hbox{
\begin{picture}(7.6, 4.1)
\multiput(.3, 1.3)(1,0){3}{\fr}
\put(3.3, 1.3){\framebox(3, 1){$\cdots$}}
\put(6.3, 1.3){\fr}
\multiput(.3,.3)(1,0){2}{\fr}
\put(.3, 2.4) {$\overbrace{~~~~~~~~~~~~~~~~~~~~~}^{2j+1}$}
\end{picture}}}

\newcommand{\marctwojminusonetwobox}
{\stln \lower2.6ex \hbox{
\begin{picture}(7.6, 4.1)
\multiput(.3, 1.3)(1,0){3}{\fr}
\put(3.3, 1.3){\framebox(3, 1){$\cdots$}}
\put(6.3, 1.3){\fr}
\multiput(.3,.3)(1,0){2}{\fr}
\put(.3, 2.4) {$\overbrace{~~~~~~~~~~~~~~~~~~~~~}^{2j-1}$}
\end{picture}}}

\newcommand{\marctwojminustwoonebox}
{\stln \lower2.6ex \hbox{
\begin{picture}(6.6, 4.1)
\multiput(.3, 1.3)(1,0){2}{\fr}
\put(2.3, 1.3) {\framebox(3, 1){$\cdots$}}
\put(5.3, 1.3) {\fr}
\put(.3, .3) {\fr}
\put(.3, 2.4){$\overbrace{~~~~~~~~~~~~~~~~~~}^{2j-2}$}
\end{picture}}}

\newcommand{\marctwojplustwotwobox}
{\stln \lower2.6ex \hbox{
\begin{picture}(7.6, 4.1)
\multiput(.3, 1.3)(1,0){3}{\fr}
\put(3.3, 1.3){\framebox(3, 1){$\cdots$}}
\put(6.3, 1.3){\fr}
\multiput(.3, .3)(1,0){2}{\fr}
\put(.3, 2.4) {$\overbrace{~~~~~~~~~~~~~~~~~~~~~}^{2j+2}$}
\end{picture}}}

\newcommand{\marctwojminustwotwobox}
{\stln \lower2.6ex \hbox{
\begin{picture}(7.6, 4.1)
\multiput(.3, 1.3)(1,0){3}{\fr}
\put(3.3, 1.3){\framebox(3, 1){$\cdots$}}
\put(6.3, 1.3){\fr}
\multiput(.3,.3)(1,0){2}{\fr}
\put(.3, 2.4) {$\overbrace{~~~~~~~~~~~~~~~~~~~~~}^{2j-2}$}
\end{picture}}}

\newcommand{\marctwojtwojbox}
{\stln \lower2.6ex \hbox{
\begin{picture}(6.6, 4.1)
\multiput(.3, 1.3)(1,0) {2} {\fr}
\put(2.3,.3){\framebox(3,2){$\cdots$}}
\put(5.3, 1.3){\fr}
\multiput(.3,.3)(1,0){2}{\fr}
\put(5.3,.3){\fr}
\put(.3, 2.4){$\overbrace{~~~~~~~~~~~~~~~~~~}^{2j}$}
\end{picture}}}

\newcommand{\marctwojplusonetwojbox}
{\stln \lower2.6ex \hbox{
\begin{picture}(7.6, 4.1)
\multiput(.3, 1.3)(1,0) {2} {\fr}
\put(2.3,.3) {\framebox(3, 2){$\cdots$}}
\multiput(5.3, 1.3)(1,0){2} {\fr}
\multiput(.3, .3)(1,0){2} {\fr}
\put(5.3, .3) {\fr}
\put(.3, 2.4){$\overbrace{~~~~~~~~~~~~~~~~~~~~~}^{2j+1}$}
\end{picture}}}

\newcommand{\marctwojtwojminusonebox}
{\stln \lower2.6ex \hbox{
\begin{picture}(7.6, 4.1)
\multiput(.3, 1.3)(1,0) {2} {\fr}
\put(2.3, .3) {\framebox(3, 2){$\cdots$}}
\multiput(5.3, 1.3)(1,0){2} {\fr}
\multiput(.3, .3)(1,0){2} {\fr}
\put(5.3,.3) {\fr}
\put(.3,2.4){$\overbrace{~~~~~~~~~~~~~~~~~~~~~}^{2j}$}
\end{picture}}}

\newcommand{\marctwojplusonetwojplusonebox}
{\stln \lower2.6ex \hbox{
\begin{picture}(6.6, 4.1)
\multiput(.3, 1.3)(1,0) {2} {\fr}
\put(2.3, .3){\framebox(3,2){$\cdots$}}
\put(5.3, 1.3){\fr}
\multiput(.3, .3)(1,0){2}{\fr}
\put(5.3, .3){\fr}
\put(.3, 2.4){$\overbrace{~~~~~~~~~~~~~~~~~~}^{2j+1}$}
\end{picture}}}

\newcommand{\marctwojminusonetwojminusonebox}
{\stln \lower2.6ex \hbox{
\begin{picture}(6.6, 4.1)
\multiput(.3, 1.3)(1,0) {2} {\fr}
\put(2.3, .3){\framebox(3,2){$\cdots$}}
\put(5.3, 1.3){\fr}
\multiput(.3, .3)(1,0){2}{\fr}
\put(5.3, .3){\fr}
\put(.3, 2.4){$\overbrace{~~~~~~~~~~~~~~~~~~}^{2j-1}$}
\end{picture}}}

\newcommand{\marctwojplusonetwojminusonebox}
{\stln \lower2.6ex \hbox{
\begin{picture}(8.6, 4.1)
\multiput(.3, 1.3)(1,0){2}{\fr}
\put(2.3,.3){\framebox(3,2) {$\cdots$}}
\multiput(5.3, 1.3)(1,0){3}{\fr}
\multiput(.3,.3) (1,0) {2}{\fr}
\put(5.3,.3) {\fr}
\put(.3, 2.4) {$\overbrace{~~~~~~~~~~~~~~~~~~~~~~~~}^{2j+1}$}
\end{picture}}}

\newcommand{\marctwojplustwotwojbox}
{\stln \lower2.6ex \hbox{
\begin{picture}(8.6, 4.1)
\multiput(.3, 1.3)(1,0){2}{\fr}
\put(2.3, .3){\framebox(3,2){$\cdots$}}
\multiput(5.3,1.3)(1,0){3}{\fr}
\multiput(.3,.3)(1,0){2}{\fr}
\put(5.3,.3){\fr}
\put(.3, 2.4){$\overbrace{~~~~~~~~~~~~~~~~~~~~~~~~}^{2j+2}$}
\end{picture}}}

\newcommand{\marctwojtwojminustwobox}
{\stln \lower2.6ex \hbox{
\begin{picture}(8.6, 4.1)
\multiput(.3, 1.3)(1,0){2}{\fr}
\put(2.3, .3){\framebox(3,2){$\cdots$}}
\multiput(5.3,1.3)(1,0){3}{\fr}
\multiput(.3,.3)(1,0){2}{\fr}
\put(5.3,.3){\fr}
\put(.3, 2.4){$\overbrace{~~~~~~~~~~~~~~~~~~~~~~~~}^{2j}$}
\end{picture}}}

\newcommand{\marctwojplustwotwojminusonebox}
{\stln \lower2.6ex \hbox{
\begin{picture}(9.6,4.1)
\multiput(.3,1.3)(1,0){2}{\fr}
\put(2.3,.3){\framebox(3,2){$\cdots$}}
\multiput(5.3,1.3)(1,0){4}{\fr}
\multiput(.3,.3)(1,0){2}{\fr}
\put(5.3,.3){\fr}
\put(.3,2.4){$\overbrace{~~~~~~~~~~~~~~~~~~~~~~~~~~~}^{2j+2}$}
\end{picture}}}

\newcommand{\marctwojminusonetwojminustwobox}
{\stln \lower2.6ex \hbox{
\begin{picture}(7.6, 4.1)
\multiput(.3, 1.3)(1,0) {2} {\fr}
\put(2.3,.3) {\framebox(3, 2){$\cdots$}}
\multiput(5.3, 1.3)(1,0){2} {\fr}
\multiput(.3, .3)(1,0){2} {\fr}
\put(5.3, .3) {\fr}
\put(.3, 2.4){$\overbrace{~~~~~~~~~~~~~~~~~~~~~}^{2j-1}$}
\end{picture}}}

\newcommand{\marctwojplusonetwojminustwobox}
{\stln \lower2.6ex \hbox{
\begin{picture}(9.6,4.1)
\multiput(.3,1.3)(1,0){2}{\fr}
\put(2.3,.3){\framebox(3,2){$\cdots$}}
\multiput(5.3,1.3)(1,0){4}{\fr}
\multiput(.3,.3)(1,0){2}{\fr}
\put(5.3,.3){\fr}
\put(.3,2.4){$\overbrace{~~~~~~~~~~~~~~~~~~~~~~~~~~~}^{2j+1}$}
\end{picture}}}

\newcommand{\marctwojplustwotwojplusonebox}
{\stln \lower2.6ex \hbox{
\begin{picture}(7.6, 4.1)
\multiput(.3, 1.3)(1,0) {2} {\fr}
\put(2.3,.3) {\framebox(3, 2){$\cdots$}}
\multiput(5.3, 1.3)(1,0){2} {\fr}
\multiput(.3, .3)(1,0){2} {\fr}
\put(5.3, .3) {\fr}
\put(.3, 2.4){$\overbrace{~~~~~~~~~~~~~~~~~~~~~}^{2j+2}$}
\end{picture}}}

\newcommand{\marctwojminustwotwojminustwobox}
{\stln \lower2.6ex \hbox{
\begin{picture}(6.6, 4.1)
\multiput(.3, 1.3)(1,0) {2} {\fr}
\put(2.3, .3){\framebox(3,2){$\cdots$}}
\put(5.3, 1.3){\fr}
\multiput(.3, .3)(1,0){2}{\fr}
\put(5.3, .3){\fr}
\put(.3, 2.4){$\overbrace{~~~~~~~~~~~~~~~~~~}^{2j-2}$}
\end{picture}}}

\newcommand{\marctwojplustwotwojminustwobox}
{\stln \lower2.6ex \hbox{
\begin{picture}(10.6,4.1)
\multiput(.3,1.3)(1,0){2}{\fr}
\put(2.3,.3){\framebox(3,2){$\cdots$}}
\multiput(5.3,1.3)(1,0){5}{\fr}
\multiput(.3,.3)(1,0){2}{\fr}
\put(5.3,.3){\fr}
\put(.3,2.4){$\overbrace{~~~~~~~~~~~~~~~~~~~~~~~~~~~~~~}^{2j+2}$}
\end{picture}}}

\newcommand{\marctwojplustwotwojplustwobox}
{\stln \lower2.6ex \hbox{
\begin{picture}(6.6, 4.1)
\multiput(.3, 1.3)(1,0) {2} {\fr}
\put(2.3, .3){\framebox(3,2){$\cdots$}}
\put(5.3, 1.3){\fr}
\multiput(.3, .3)(1,0){2}{\fr}
\put(5.3, .3){\fr}
\put(.3, 2.4){$\overbrace{~~~~~~~~~~~~~~~~~~}^{2j+2}$}
\end{picture}}}

\newcommand{\marctwojplusthreetwojbox}
{\stln \lower2.6ex \hbox{
\begin{picture}(9.6,4.1)
\multiput(.3,1.3)(1,0){2}{\fr}
\put(2.3,.3){\framebox(3,2){$\cdots$}}
\multiput(5.3,1.3)(1,0){4}{\fr}
\multiput(.3,.3)(1,0){2}{\fr}
\put(5.3,.3){\fr}
\put(.3,2.4){$\overbrace{~~~~~~~~~~~~~~~~~~~~~~~~~~~}^{2j+3}$}
\end{picture}}}

\newcommand{\marctwojplusthreetwojplusonebox}
{\stln \lower2.6ex \hbox{
\begin{picture}(8.6, 4.1)
\multiput(.3, 1.3)(1,0){2}{\fr}
\put(2.3, .3){\framebox(3,2){$\cdots$}}
\multiput(5.3,1.3)(1,0){3}{\fr}
\multiput(.3,.3)(1,0){2}{\fr}
\put(5.3,.3){\fr}
\put(.3, 2.4){$\overbrace{~~~~~~~~~~~~~~~~~~~~~~~~}^{2j+3}$}
\end{picture}}}

\newcommand{\marctwojplusthreetwojminusonebox}
{\stln \lower2.6ex \hbox{
\begin{picture}(10.6,4.1)
\multiput(.3,1.3)(1,0){2}{\fr}
\put(2.3,.3){\framebox(3,2){$\cdots$}}
\multiput(5.3,1.3)(1,0){5}{\fr}
\multiput(.3,.3)(1,0){2}{\fr}
\put(5.3,.3){\fr}
\put(.3,2.4){$\overbrace{~~~~~~~~~~~~~~~~~~~~~~~~~~~~~~}^{2j+3}$}
\end{picture}}}

\newcommand{\marctwojplusthreetwojplustwobox}
{\stln \lower2.6ex \hbox{
\begin{picture}(7.6, 4.1)
\multiput(.3, 1.3)(1,0) {2} {\fr}
\put(2.3,.3) {\framebox(3, 2){$\cdots$}}
\multiput(5.3, 1.3)(1,0){2} {\fr}
\multiput(.3, .3)(1,0){2} {\fr}
\put(5.3, .3) {\fr}
\put(.3, 2.4){$\overbrace{~~~~~~~~~~~~~~~~~~~~~}^{2j+3}$}
\end{picture}}}

\newcommand{\marctwojplusthreetwojminustwobox}
{\stln \lower2.6ex \hbox{
\begin{picture}(11.6,4.1)
\multiput(.3, 1.3)(1,0){2}{\fr}
\put(2.3,.3){\framebox(3,2){$\cdots$}}
\multiput(5.3,1.3)(1,0){6}{\fr}
\multiput(.3,.3)(1,0){2}{\fr}
\put(5.3,.3){\fr}
\put(.3,2.4){$\overbrace{~~~~~~~~~~~~~~~~~~~~~~~~~~~~~~~~~}^{2j+3}$}
\end{picture}}}

\newcommand{\marctwojplusfourtwojbox}
{\stln \lower2.6ex \hbox{
\begin{picture}(10.6,4.1)
\multiput(.3,1.3)(1,0){2}{\fr}
\put(2.3,.3){\framebox(3,2){$\cdots$}}
\multiput(5.3,1.3)(1,0){5}{\fr}
\multiput(.3,.3)(1,0){2}{\fr}
\put(5.3,.3){\fr}
\put(.3,2.4){$\overbrace{~~~~~~~~~~~~~~~~~~~~~~~~~~~~~~}^{2j+4}$}
\end{picture}}}

\newcommand{\marctwojplusfourtwojplusonebox}
{\stln \lower2.6ex \hbox{
\begin{picture}(9.6,4.1)
\multiput(.3,1.3)(1,0){2}{\fr}
\put(2.3,.3){\framebox(3,2){$\cdots$}}
\multiput(5.3,1.3)(1,0){4}{\fr}
\multiput(.3,.3)(1,0){2}{\fr}
\put(5.3,.3){\fr}
\put(.3,2.4){$\overbrace{~~~~~~~~~~~~~~~~~~~~~~~~~~~}^{2j+4}$}
\end{picture}}}

\newcommand{\marctwojplusfourtwojminusonebox}
{\stln \lower2.6ex \hbox{
\begin{picture}(11.6,4.1)
\multiput(.3, 1.3)(1,0){2}{\fr}
\put(2.3,.3){\framebox(3,2){$\cdots$}}
\multiput(5.3,1.3)(1,0){6}{\fr}
\multiput(.3,.3)(1,0){2}{\fr}
\put(5.3,.3){\fr}
\put(.3,2.4){$\overbrace{~~~~~~~~~~~~~~~~~~~~~~~~~~~~~~~~~}^{2j+4}$}
\end{picture}}}

\newcommand{\marctwojplusfourtwojplustwobox}
{\stln \lower2.6ex \hbox{
\begin{picture}(8.6, 4.1)
\multiput(.3, 1.3)(1,0){2}{\fr}
\put(2.3, .3){\framebox(3,2){$\cdots$}}
\multiput(5.3,1.3)(1,0){3}{\fr}
\multiput(.3,.3)(1,0){2}{\fr}
\put(5.3,.3){\fr}
\put(.3, 2.4){$\overbrace{~~~~~~~~~~~~~~~~~~~~~~~~}^{2j+4}$}
\end{picture}}}

\newcommand{\marctwojplusfourtwojminustwobox}
{\stln \lower2.6ex \hbox{
\begin{picture}(12.6, 4.1)
\multiput(.3,1.3)(1,0){2}{\fr}
\put(2.3,.3){\framebox(3,2){$\cdots$}}
\multiput(5.3,1.3)(1,0){7}{\fr}
\multiput(.3,.3)(1,0){2}{\fr}
\put(5.3,.3){\fr}
\put(.3, 2.4){$\overbrace{~~~~~~~~~~~~~~~~~~~~~~~~~~~~~~~~~~~~}^{2j+4}$}
\end{picture}}}

\newcommand{\marctwojplusfivetwojminusonebox}
{\stln \lower2.6ex \hbox{
\begin{picture}(12.6,4.1)
\multiput(.3, 1.3)(1,0){2}{\fr}
\put(2.3,.3){\framebox(3,2){$\cdots$}}
\multiput(5.3,1.3)(1,0){7}{\fr}
\multiput(.3,.3)(1,0){2}{\fr}
\put(5.3,.3){\fr}
\put(.3,2.4){$\overbrace{~~~~~~~~~~~~~~~~~~~~~~~~~~~~~~~~~~~~}^{2j+5}$}
\end{picture}}}

\newcommand{\marctwojplusfivetwojplustwobox}
{\stln \lower2.6ex \hbox{
\begin{picture}(9.6,4.1)
\multiput(.3,1.3)(1,0){2}{\fr}
\put(2.3,.3){\framebox(3,2){$\cdots$}}
\multiput(5.3,1.3)(1,0){4}{\fr}
\multiput(.3,.3)(1,0){2}{\fr}
\put(5.3,.3){\fr}
\put(.3,2.4){$\overbrace{~~~~~~~~~~~~~~~~~~~~~~~~~~~}^{2j+5}$}
\end{picture}}}

\newcommand{\marctwojplusfivetwojplusonebox}
{\stln \lower2.6ex \hbox{
\begin{picture}(10.6, 4.1)
\multiput(.3,1.3)(1,0){2}{\fr}
\put(2.3,.3){\framebox(3,2){$\cdots$}}
\multiput(5.3,1.3)(1,0){5}{\fr}
\multiput(.3,.3)(1,0){2}{\fr}
\put(5.3,.3){\fr}
\put(.3, 2.4){$\overbrace{~~~~~~~~~~~~~~~~~~~~~~~~~~~~~~}^{2j+5}$}
\end{picture}}}

\newcommand{\marctwojplusfivetwojbox}
{\stln \lower2.6ex \hbox{
\begin{picture}(11.6, 4.1)
\multiput(.3,1.3)(1,0){2}{\fr}
\put(2.3,.3){\framebox(3,2){$\cdots$}}
\multiput(5.3,1.3)(1,0){6}{\fr}
\multiput(.3,.3)(1,0){2}{\fr}
\put(5.3,.3){\fr}
\put(.3, 2.4){$\overbrace{~~~~~~~~~~~~~~~~~~~~~~~~~~~~~~~~}^{2j+5}$}
\end{picture}}}

\newcommand{\marctwojplusfivetwojplusthreebox}
{\stln \lower2.6ex \hbox{
\begin{picture}(9.6,4.1)
\multiput(.3,1.3)(1,0){2}{\fr}
\put(2.3,.3){\framebox(3,2){$\cdots$}}
\multiput(5.3,1.3)(1,0){4}{\fr}
\multiput(.3,.3)(1,0){2}{\fr}
\put(5.3,.3){\fr}
\put(.3,2.4){$\overbrace{~~~~~~~~~~~~~~~~~~~~~~~~~~~}^{2j+5}$}
\end{picture}}}

\newcommand{\marctwojplusfivetwojminustwobox}
{\stln \lower2.6ex \hbox{
\begin{picture}(13.6,4.1)
\multiput(.3,1.3)(1,0){2}{\fr}
\put(2.3,.3){\framebox(3,2){$\cdots$}}
\multiput(5.3,1.3)(1,0){8}{\fr}
\multiput(.3,.3)(1,0){2}{\fr}
\put(5.3,.3){\fr}
\put(.3,2.4){$\overbrace{~~~~~~~~~~~~~~~~~~~~~~~~~~~~~~~~~~~~~~~}^{2j+5}$}
\end{picture}}}

\newcommand{\smarctwojbox}
{\stln \lower1.4ex \hbox{
\begin{picture}(6.6,3.1)
\multiput(.3,.3)(1,0){2}{\sfr}
\put(2.3,.3){\framebox(3,1){$\cdots$}}
\put(5.3,.3){\sfr}
\put(.3,1.4){$\overbrace{~~~~~~~~~~~~~~~~~~}^{2j}$}
\end{picture}}}

\newcommand{\smarctwojtwojbox}
{\stln \lower2.6ex \hbox{
\begin{picture}(6.6,4.1)
\multiput(.3,1.3)(1,0){2}{\sfr}
\put(2.3,.3){\framebox(3,2){$\cdots$}}
\put(5.3,1.3){\sfr}
\multiput(.3,.3)(1,0){2}{\sfr}
\put(5.3,.3){\sfr}
\put(.3,2.4){$\overbrace{~~~~~~~~~~~~~~~~~~}^{2j}$}
\end{picture}}}

\newcommand{\smarctwojplusonetwojbox}
{\stln \lower2.6ex \hbox{
\begin{picture}(7.6,4.1)
\multiput(.3,1.3)(1,0){2}{\sfr}
\put(2.3,.3){\framebox(3,2){$\cdots$}}
\multiput(5.3,1.3)(1,0){2}{\sfr}
\multiput(.3,.3)(1,0){2}{\sfr}
\put(5.3,.3){\sfr}
\put(.3,2.4){$\overbrace{~~~~~~~~~~~~~~~~~~~~~}^{2j+1}$}
\end{picture}}}

\newcommand{\smarctwojplustwotwojbox}
{\stln \lower2.6ex \hbox{
\begin{picture}(8.6,4.1)
\multiput(.3,1.3)(1,0){2}{\sfr}
\put(2.3,.3){\framebox(3,2){$\cdots$}}
\multiput(5.3,1.3)(1,0){3}{\sfr}
\multiput(.3,.3)(1,0){2}{\sfr}
\put(5.3,.3){\sfr}
\put(.3,2.4){$\overbrace{~~~~~~~~~~~~~~~~~~~~~~~~}^{2j+2}$}
\end{picture}}}

\newcommand{\smarctwojplusthreetwojbox}
{\stln \lower2.6ex \hbox{
\begin{picture}(9.6,4.1)
\multiput(.3,1.3)(1,0){2}{\sfr}
\put(2.3,.3){\framebox(3,2){$\cdots$}}
\multiput(5.3,1.3)(1,0){4}{\sfr}
\multiput(.3,.3)(1,0){2}{\sfr}
\put(5.3,.3){\sfr}
\put(.3,2.4){$\overbrace{~~~~~~~~~~~~~~~~~~~~~~~~~~~}^{2j+3}$}
\end{picture}}}

\newcommand{\smarctwojplusntwojbox}
{\stln \lower2.6ex \hbox{
\begin{picture}(12.6,4.1)
\multiput(.3,1.3)(1,0){2}{\sfr}
\put(2.3,.3){\framebox(3,2){$\cdots$}}
\multiput(5.3,1.3)(1,0){3}{\sfr}
\put(8.3,1.3){\framebox(3,1){$\cdots$}}
\put(11.3,1.3){\sfr}
\multiput(.3,.3)(1,0){2}{\sfr}
\put(5.3,.3){\sfr}
\put(.3,2.4){$\overbrace{~~~~~~~~~~~~~~~~~~}^{2j}$}
\put(6.3,2.4){$\overbrace{~~~~~~~~~~~~~~~~~~}^{n}$}
\end{picture}}}

\newcommand{\marctwojplusntwojbox}
{\stln \lower2.6ex \hbox{
\begin{picture}(12.6,4.1)
\multiput(.3,1.3)(1,0){2}{\fr}
\put(2.3,.3){\framebox(3,2){$\cdots$}}
\multiput(5.3,1.3)(1,0){3}{\fr}
\put(8.3,1.3){\framebox(3,1){$\cdots$}}
\put(11.3,1.3){\fr}
\multiput(.3,.3)(1,0){2}{\fr}
\put(5.3,.3){\fr}
\put(.3,2.4){$\overbrace{~~~~~~~~~~~~~~~~~~}^{2j}$}
\put(6.3,2.4){$\overbrace{~~~~~~~~~~~~~~~~~~}^{n}$}
\end{picture}}}

\newcommand{\marctwojplusnplusonetwojbox}
{\stln \lower2.6ex \hbox{
\begin{picture}(12.6,4.1)
\multiput(.3,1.3)(1,0){2}{\fr}
\put(2.3,.3){\framebox(3,2){$\cdots$}}
\multiput(5.3,1.3)(1,0){3}{\fr}
\put(8.3,1.3){\framebox(3,1){$\cdots$}}
\put(11.3,1.3){\fr}
\multiput(.3,.3)(1,0){2}{\fr}
\put(5.3,.3){\fr}
\put(.3,2.4){$\overbrace{~~~~~~~~~~~~~~~~~~}^{2j}$}
\put(6.3,2.4){$\overbrace{~~~~~~~~~~~~~~~~~~}^{n+1}$}
\end{picture}}}

\newcommand{\marctwojplusoneplusnminusonetwojplusonebox}
{\stln \lower2.6ex \hbox{
\begin{picture}(12.6,4.1)
\multiput(.3,1.3)(1,0){2}{\fr}
\put(2.3,.3){\framebox(3,2){$\cdots$}}
\multiput(5.3,1.3)(1,0){3}{\fr}
\put(8.3,1.3){\framebox(3,1){$\cdots$}}
\put(11.3,1.3){\fr}
\multiput(.3,.3)(1,0){2}{\fr}
\put(5.3,.3){\fr}
\put(.3,2.4){$\overbrace{~~~~~~~~~~~~~~~~~~}^{2j+1}$}
\put(6.3,2.4){$\overbrace{~~~~~~~~~~~~~~~~~~}^{n-1}$}
\end{picture}}}

\newcommand{\marctwojplusnminusonetwojbox}
{\stln \lower2.6ex \hbox{
\begin{picture}(12.6,4.1)
\multiput(.3,1.3)(1,0){2}{\fr}
\put(2.3,.3){\framebox(3,2){$\cdots$}}
\multiput(5.3,1.3)(1,0){3}{\fr}
\put(8.3,1.3){\framebox(3,1){$\cdots$}}
\put(11.3,1.3){\fr}
\multiput(.3,.3)(1,0){2}{\fr}
\put(5.3,.3){\fr}
\put(.3,2.4){$\overbrace{~~~~~~~~~~~~~~~~~~}^{2j}$}
\put(6.3,2.4){$\overbrace{~~~~~~~~~~~~~~~~~~}^{n-1}$}
\end{picture}}}

\newcommand{\marctwojminusoneplusnplusonetwojminusonebox}
{\stln \lower2.6ex \hbox{
\begin{picture}(12.6,4.1)
\multiput(.3,1.3)(1,0){2}{\fr}
\put(2.3,.3){\framebox(3,2){$\cdots$}}
\multiput(5.3,1.3)(1,0){3}{\fr}
\put(8.3,1.3){\framebox(3,1){$\cdots$}}
\put(11.3,1.3){\fr}
\multiput(.3,.3)(1,0){2}{\fr}
\put(5.3,.3){\fr}
\put(.3,2.4){$\overbrace{~~~~~~~~~~~~~~~~~~}^{2j-1}$}
\put(6.3,2.4){$\overbrace{~~~~~~~~~~~~~~~~~~}^{n+1}$}
\end{picture}}}

\newcommand{\marctwojplusnplustwotwojbox}
{\stln \lower2.6ex \hbox{
\begin{picture}(12.6,4.1)
\multiput(.3,1.3)(1,0){2}{\fr}
\put(2.3,.3){\framebox(3,2){$\cdots$}}
\multiput(5.3,1.3)(1,0){3}{\fr}
\put(8.3,1.3){\framebox(3,1){$\cdots$}}
\put(11.3,1.3){\fr}
\multiput(.3,.3)(1,0){2}{\fr}
\put(5.3,.3){\fr}
\put(.3,2.4){$\overbrace{~~~~~~~~~~~~~~~~~~}^{2j}$}
\put(6.3,2.4){$\overbrace{~~~~~~~~~~~~~~~~~~}^{n+2}$}
\end{picture}}}

\newcommand{\marctwojplusoneplusntwojplusonebox}
{\stln \lower2.6ex \hbox{
\begin{picture}(12.6,4.1)
\multiput(.3,1.3)(1,0){2}{\fr}
\put(2.3,.3){\framebox(3,2){$\cdots$}}
\multiput(5.3,1.3)(1,0){3}{\fr}
\put(8.3,1.3){\framebox(3,1){$\cdots$}}
\put(11.3,1.3){\fr}
\multiput(.3,.3)(1,0){2}{\fr}
\put(5.3,.3){\fr}
\put(.3,2.4){$\overbrace{~~~~~~~~~~~~~~~~~~}^{2j+1}$}
\put(6.3,2.4){$\overbrace{~~~~~~~~~~~~~~~~~~}^{n}$}
\end{picture}}}

\newcommand{\marctwojplustwoplusnminustwotwojplustwobox}
{\stln \lower2.6ex \hbox{
\begin{picture}(12.6,4.1)
\multiput(.3,1.3)(1,0){2}{\fr}
\put(2.3,.3){\framebox(3,2){$\cdots$}}
\multiput(5.3,1.3)(1,0){3}{\fr}
\put(8.3,1.3){\framebox(3,1){$\cdots$}}
\put(11.3,1.3){\fr}
\multiput(.3,.3)(1,0){2}{\fr}
\put(5.3,.3){\fr}
\put(.3,2.4){$\overbrace{~~~~~~~~~~~~~~~~~~}^{2j+2}$}
\put(6.3,2.4){$\overbrace{~~~~~~~~~~~~~~~~~~}^{n-2}$}
\end{picture}}}

\newcommand{\marctwojplusoneplusnminustwotwojplusonebox}
{\stln \lower2.6ex \hbox{
\begin{picture}(12.6,4.1)
\multiput(.3,1.3)(1,0){2}{\fr}
\put(2.3,.3){\framebox(3,2){$\cdots$}}
\multiput(5.3,1.3)(1,0){3}{\fr}
\put(8.3,1.3){\framebox(3,1){$\cdots$}}
\put(11.3,1.3){\fr}
\multiput(.3,.3)(1,0){2}{\fr}
\put(5.3,.3){\fr}
\put(.3,2.4){$\overbrace{~~~~~~~~~~~~~~~~~~}^{2j+1}$}
\put(6.3,2.4){$\overbrace{~~~~~~~~~~~~~~~~~~}^{n-2}$}
\end{picture}}}

\newcommand{\marctwojplusnminustwotwojbox}
{\stln \lower2.6ex \hbox{
\begin{picture}(12.6,4.1)
\multiput(.3,1.3)(1,0){2}{\fr}
\put(2.3,.3){\framebox(3,2){$\cdots$}}
\multiput(5.3,1.3)(1,0){3}{\fr}
\put(8.3,1.3){\framebox(3,1){$\cdots$}}
\put(11.3,1.3){\fr}
\multiput(.3,.3)(1,0){2}{\fr}
\put(5.3,.3){\fr}
\put(.3,2.4){$\overbrace{~~~~~~~~~~~~~~~~~~}^{2j}$}
\put(6.3,2.4){$\overbrace{~~~~~~~~~~~~~~~~~~}^{n-2}$}
\end{picture}}}

\newcommand{\marctwojminusoneplusnplustwotwojminusonebox}
{\stln \lower2.6ex \hbox{
\begin{picture}(12.6,4.1)
\multiput(.3,1.3)(1,0){2}{\fr}
\put(2.3,.3){\framebox(3,2){$\cdots$}}
\multiput(5.3,1.3)(1,0){3}{\fr}
\put(8.3,1.3){\framebox(3,1){$\cdots$}}
\put(11.3,1.3){\fr}
\multiput(.3,.3)(1,0){2}{\fr}
\put(5.3,.3){\fr}
\put(.3,2.4){$\overbrace{~~~~~~~~~~~~~~~~~~}^{2j-1}$}
\put(6.3,2.4){$\overbrace{~~~~~~~~~~~~~~~~~~}^{n+2}$}
\end{picture}}}

\newcommand{\marctwojminusoneplusnplustwojminusonebox}
{\stln \lower2.6ex \hbox{
\begin{picture}(12.6,4.1)
\multiput(.3,1.3)(1,0){2}{\fr}
\put(2.3,.3){\framebox(3,2){$\cdots$}}
\multiput(5.3,1.3)(1,0){3}{\fr}
\put(8.3,1.3){\framebox(3,1){$\cdots$}}
\put(11.3,1.3){\fr}
\multiput(.3,.3)(1,0){2}{\fr}
\put(5.3,.3){\fr}
\put(.3,2.4){$\overbrace{~~~~~~~~~~~~~~~~~~}^{2j-1}$}
\put(6.3,2.4){$\overbrace{~~~~~~~~~~~~~~~~~~}^{n}$}
\end{picture}}}

\newcommand{\marctwojminustwoplusnplustwotwojminustwobox}
{\stln \lower2.6ex \hbox{
\begin{picture}(12.6,4.1)
\multiput(.3,1.3)(1,0){2}{\fr}
\put(2.3,.3){\framebox(3,2){$\cdots$}}
\multiput(5.3,1.3)(1,0){3}{\fr}
\put(8.3,1.3){\framebox(3,1){$\cdots$}}
\put(11.3,1.3){\fr}
\multiput(.3,.3)(1,0){2}{\fr}
\put(5.3,.3){\fr}
\put(.3,2.4){$\overbrace{~~~~~~~~~~~~~~~~~~}^{2j-2}$}
\put(6.3,2.4){$\overbrace{~~~~~~~~~~~~~~~~~~}^{n+2}$}
\end{picture}}}

\newcommand{\marctwojplusoneplusnplusonetwojplusonebox}
{\stln \lower2.6ex \hbox{
\begin{picture}(12.6,4.1)
\multiput(.3,1.3)(1,0){2}{\fr}
\put(2.3,.3){\framebox(3,2){$\cdots$}}
\multiput(5.3,1.3)(1,0){3}{\fr}
\put(8.3,1.3){\framebox(3,1){$\cdots$}}
\put(11.3,1.3){\fr}
\multiput(.3,.3)(1,0){2}{\fr}
\put(5.3,.3){\fr}
\put(.3,2.4){$\overbrace{~~~~~~~~~~~~~~~~~~}^{2j+1}$}
\put(6.3,2.4){$\overbrace{~~~~~~~~~~~~~~~~~~}^{n+1}$}
\end{picture}}}

\newcommand{\marctwojplustwoplusnminusonetwojplustwobox}
{\stln \lower2.6ex \hbox{
\begin{picture}(12.6,4.1)
\multiput(.3,1.3)(1,0){2}{\fr}
\put(2.3,.3){\framebox(3,2){$\cdots$}}
\multiput(5.3,1.3)(1,0){3}{\fr}
\put(8.3,1.3){\framebox(3,1){$\cdots$}}
\put(11.3,1.3){\fr}
\multiput(.3,.3)(1,0){2}{\fr}
\put(5.3,.3){\fr}
\put(.3,2.4){$\overbrace{~~~~~~~~~~~~~~~~~~}^{2j+2}$}
\put(6.3,2.4){$\overbrace{~~~~~~~~~~~~~~~~~~}^{n-1}$}
\end{picture}}}

\newcommand{\marctwojplustwoplusnminusthreetwojplustwobox}
{\stln \lower2.6ex \hbox{
\begin{picture}(12.6,4.1)
\multiput(.3,1.3)(1,0){2}{\fr}
\put(2.3,.3){\framebox(3,2){$\cdots$}}
\multiput(5.3,1.3)(1,0){3}{\fr}
\put(8.3,1.3){\framebox(3,1){$\cdots$}}
\put(11.3,1.3){\fr}
\multiput(.3,.3)(1,0){2}{\fr}
\put(5.3,.3){\fr}
\put(.3,2.4){$\overbrace{~~~~~~~~~~~~~~~~~~}^{2j+2}$}
\put(6.3,2.4){$\overbrace{~~~~~~~~~~~~~~~~~~}^{n-3}$}
\end{picture}}}

\newcommand{\marctwojplusoneplusnminusthreetwojplusonebox}
{\stln \lower2.6ex \hbox{
\begin{picture}(12.6,4.1)
\multiput(.3,1.3)(1,0){2}{\fr}
\put(2.3,.3){\framebox(3,2){$\cdots$}}
\multiput(5.3,1.3)(1,0){3}{\fr}
\put(8.3,1.3){\framebox(3,1){$\cdots$}}
\put(11.3,1.3){\fr}
\multiput(.3,.3)(1,0){2}{\fr}
\put(5.3,.3){\fr}
\put(.3,2.4){$\overbrace{~~~~~~~~~~~~~~~~~~}^{2j+1}$}
\put(6.3,2.4){$\overbrace{~~~~~~~~~~~~~~~~~~}^{n-3}$}
\end{picture}}}

\newcommand{\marctwojminusoneplusnminusonetwojminusonebox}
{\stln \lower2.6ex \hbox{
\begin{picture}(12.6,4.1)
\multiput(.3,1.3)(1,0){2}{\fr}
\put(2.3,.3){\framebox(3,2){$\cdots$}}
\multiput(5.3,1.3)(1,0){3}{\fr}
\put(8.3,1.3){\framebox(3,1){$\cdots$}}
\put(11.3,1.3){\fr}
\multiput(.3,.3)(1,0){2}{\fr}
\put(5.3,.3){\fr}
\put(.3,2.4){$\overbrace{~~~~~~~~~~~~~~~~~~}^{2j-1}$}
\put(6.3,2.4){$\overbrace{~~~~~~~~~~~~~~~~~~}^{n-1}$}
\end{picture}}}

\newcommand{\marctwojminusoneplusnplusthreetwojminusonebox}
{\stln \lower2.6ex \hbox{
\begin{picture}(12.6,4.1)
\multiput(.3,1.3)(1,0){2}{\fr}
\put(2.3,.3){\framebox(3,2){$\cdots$}}
\multiput(5.3,1.3)(1,0){3}{\fr}
\put(8.3,1.3){\framebox(3,1){$\cdots$}}
\put(11.3,1.3){\fr}
\multiput(.3,.3)(1,0){2}{\fr}
\put(5.3,.3){\fr}
\put(.3,2.4){$\overbrace{~~~~~~~~~~~~~~~~~~}^{2j-1}$}
\put(6.3,2.4){$\overbrace{~~~~~~~~~~~~~~~~~~}^{n+3}$}
\end{picture}}}

\newcommand{\marctwojminustwoplusnplusonetwojminustwobox}
{\stln \lower2.6ex \hbox{
\begin{picture}(12.6,4.1)
\multiput(.3,1.3)(1,0){2}{\fr}
\put(2.3,.3){\framebox(3,2){$\cdots$}}
\multiput(5.3,1.3)(1,0){3}{\fr}
\put(8.3,1.3){\framebox(3,1){$\cdots$}}
\put(11.3,1.3){\fr}
\multiput(.3,.3)(1,0){2}{\fr}
\put(5.3,.3){\fr}
\put(.3,2.4){$\overbrace{~~~~~~~~~~~~~~~~~~}^{2j-2}$}
\put(6.3,2.4){$\overbrace{~~~~~~~~~~~~~~~~~~}^{n+1}$}
\end{picture}}}

\newcommand{\marctwojminustwoplusnplusthreetwojminustwobox}
{\stln \lower2.6ex \hbox{
\begin{picture}(12.6,4.1)
\multiput(.3,1.3)(1,0){2}{\fr}
\put(2.3,.3){\framebox(3,2){$\cdots$}}
\multiput(5.3,1.3)(1,0){3}{\fr}
\put(8.3,1.3){\framebox(3,1){$\cdots$}}
\put(11.3,1.3){\fr}
\multiput(.3,.3)(1,0){2}{\fr}
\put(5.3,.3){\fr}
\put(.3,2.4){$\overbrace{~~~~~~~~~~~~~~~~~~}^{2j-2}$}
\put(6.3,2.4){$\overbrace{~~~~~~~~~~~~~~~~~~}^{n+3}$}
\end{picture}}}

\newcommand{\marctwojplustwoplusnminusfourtwojplustwobox}
{\stln \lower2.6ex \hbox{
\begin{picture}(12.6,4.1)
\multiput(.3,1.3)(1,0){2}{\fr}
\put(2.3,.3){\framebox(3,2){$\cdots$}}
\multiput(5.3,1.3)(1,0){3}{\fr}
\put(8.3,1.3){\framebox(3,1){$\cdots$}}
\put(11.3,1.3){\fr}
\multiput(.3,.3)(1,0){2}{\fr}
\put(5.3,.3){\fr}
\put(.3,2.4){$\overbrace{~~~~~~~~~~~~~~~~~~}^{2j+2}$}
\put(6.3,2.4){$\overbrace{~~~~~~~~~~~~~~~~~~}^{n-4}$}
\end{picture}}}

\newcommand{\marctwojminustwoplusntwojminustwobox}
{\stln \lower2.6ex \hbox{
\begin{picture}(12.6,4.1)
\multiput(.3,1.3)(1,0){2}{\fr}
\put(2.3,.3){\framebox(3,2){$\cdots$}}
\multiput(5.3,1.3)(1,0){3}{\fr}
\put(8.3,1.3){\framebox(3,1){$\cdots$}}
\put(11.3,1.3){\fr}
\multiput(.3,.3)(1,0){2}{\fr}
\put(5.3,.3){\fr}
\put(.3,2.4){$\overbrace{~~~~~~~~~~~~~~~~~~}^{2j-2}$}
\put(6.3,2.4){$\overbrace{~~~~~~~~~~~~~~~~~~}^{n}$}
\end{picture}}}

\newcommand{\marctwojminustwoplusnplusfourtwojminustwobox}
{\stln \lower2.6ex \hbox{
\begin{picture}(12.6,4.1)
\multiput(.3,1.3)(1,0){2}{\fr}
\put(2.3,.3){\framebox(3,2){$\cdots$}}
\multiput(5.3,1.3)(1,0){3}{\fr}
\put(8.3,1.3){\framebox(3,1){$\cdots$}}
\put(11.3,1.3){\fr}
\multiput(.3,.3)(1,0){2}{\fr}
\put(5.3,.3){\fr}
\put(.3,2.4){$\overbrace{~~~~~~~~~~~~~~~~~~}^{2j-2}$}
\put(6.3,2.4){$\overbrace{~~~~~~~~~~~~~~~~~~}^{n+4}$}
\end{picture}}}

\newcommand{\marctwojplustwoplusntwojplustwobox}
{\stln \lower2.6ex \hbox{
\begin{picture}(12.6,4.1)
\multiput(.3,1.3)(1,0){2}{\fr}
\put(2.3,.3){\framebox(3,2){$\cdots$}}
\multiput(5.3,1.3)(1,0){3}{\fr}
\put(8.3,1.3){\framebox(3,1){$\cdots$}}
\put(11.3,1.3){\fr}
\multiput(.3,.3)(1,0){2}{\fr}
\put(5.3,.3){\fr}
\put(.3,2.4){$\overbrace{~~~~~~~~~~~~~~~~~~}^{2j+2}$}
\put(6.3,2.4){$\overbrace{~~~~~~~~~~~~~~~~~~}^{n}$}
\end{picture}}}

\newcommand{\marctwojminusoneplusntwojminusonebox}
{\stln \lower2.6ex \hbox{
\begin{picture}(12.6,4.1)
\multiput(.3,1.3)(1,0){2}{\fr}
\put(2.3,.3){\framebox(3,2){$\cdots$}}
\multiput(5.3,1.3)(1,0){3}{\fr}
\put(8.3,1.3){\framebox(3,1){$\cdots$}}
\put(11.3,1.3){\fr}
\multiput(.3,.3)(1,0){2}{\fr}
\put(5.3,.3){\fr}
\put(.3,2.4){$\overbrace{~~~~~~~~~~~~~~~~~~}^{2j-1}$}
\put(6.3,2.4){$\overbrace{~~~~~~~~~~~~~~~~~~}^{n}$}
\end{picture}}}
\newcommand{\OS}{$OSp(8^*|4)$}

\newcommand{\eq}{\begin{equation}}
\newcommand{\en}{\end{equation}}
\newcommand{\eqn}{\begin{eqnarray}}
\newcommand{\enn}{\end{eqnarray}}
\newcommand{\nn}{\nonumber }
\newcommand{\beq}{\begin{equation}}
\newcommand{\eeq}{\end{equation}}
\begin{document}
\begin{titlepage}
\begin{flushright}
  PSU-TH-205\\
CERN-TH/99-347\\
hep-th/9910110\\
\end{flushright}
\begin{center}
{\bf UNITARY SUPERMULTIPLETS OF $OSp(8^*|4)$ AND THE $AdS_{7}/CFT_{6}$
DUALITY }\footnote{Work supported in part by the
National Science Foundation under Grant Number PHY-9802510} \\
\vspace{1cm}
{\bf Murat G\"{u}naydin}\footnote{
e-mail: murat@phys.psu.edu }\\
\vspace{.4cm} CERN , Theory Division \\
1211 Geneva 23,
Switzerland\\
and \\
Physics Department \\
Penn State University \\
University Park, PA 16802 \\
and \\
\vspace{.4cm}
{\bf Seiji Takemae} \footnote{
e-mail: takemae@phys.psu.edu.}
  \\ \vspace{.4cm} Physics Department \\ Penn State University\\
University Park, PA 16802 \\ \vspace{.5cm} {\bf Abstract}
\end{center}

We study the unitary supermultiplets of the ${\cal{N}}=4$ $d=7$ anti-de
Sitter ($AdS_7$) superalgebra \OS, with the even subalgebra
 $SO(6,2) \times USp(4)$, which is the symmetry superalgebra of
M-theory on $AdS_7 \times S^4$. We give a complete classification of the
positive energy doubleton and massless supermultiplets of \OS . The
ultra-short doubleton supermultiplets do not have a Poincar\'{e} limit in
$AdS_7$ and correspond to superconformal field theories on the
boundary of $AdS_7$ which can be identified with $d=6$ Minkowski space.
We show that the six dimensional Poincare mass operator $m^2=P_{\mu}P^{\mu}$
vanishes identically for the doubleton
representations. By going from the compact $U(4)$ basis of
$SO^*(8)=SO(6,2)$ to the noncompact basis $SU^*(4)\times \mathcal{D}$ ($d=6$ Lorentz group times dilatations)  one can associate the positive
(conformal) energy representations of $SO^*(8)$ with conformal fields
transforming covariantly under the Lorentz group in $d=6$. The oscillator
method used for the construction of the unitary supermultiplets of \OS ~can be given a dynamical realization in terms of chiral super-twistor
fields.

\vspace{3cm}
\begin{flushleft}
CERN-TH/99-347 \\
November, 1999 \\
\end{flushleft}

\end{titlepage}

\renewcommand{\theequation}{\arabic{section} - \arabic{equation}}
\section{Introduction} \setcounter{equation}{0} Recently,  a
great deal of work has been done on AdS/CFT (anti-de Sitter/conformal
field theory) dualities in various dimensions. The current interest in
AdS/CFT dualities was sparked by the conjecture of Maldacena \cite{mald}
relating the large $N$ limits of certain conformal field theories in $d$
dimensions to M-theory/string theory compactified to $d+1$-dimensional AdS
spacetimes. Maldacena's conjecture was based on earlier work on properties
of the physics of $N$ $Dp$-branes in the near horizon limit \cite{ads} and
the much earlier work on 10-$d$ IIB supergravity compactified on $AdS_5
\times S^5$ and 11-$d$ supergravity compactified on $AdS_7 \times S^4$ and
$AdS_4 \times S^7$ \cite{dfhn, mgnm, krv, ptn, mgnw, gnw, ss}. Maldacena's
conjecture was formulated in a more precise manner in \cite{pol,witt}. The
relation between Maldacena's conjecture and the dynamics of the singleton
and doubleton fields that live on the boundary of AdS spacetimes was
reviewed in \cite{sfcf,mgdm} and its relation to the spectra of maximal
supergravities in eleven and ten dimensions in \cite{mgdm,dupo}. For a
recent review of AdS/CFT dualities and an extensive list of references on
the subject see \cite{adscft}.

The best studied example of this AdS/CFT duality is the duality between
the large $N$ limit of the conformally invariant ${\cal{N}}=4$ $SU(N)$
super Yang-Mills theory in $d=4$ and type IIB superstring theory on $AdS_5
\times S^5$. The ${\cal {N}} =4$ Yang-Mills supermultiplet is simply the
unique CPT self-conjugate doubleton supermultiplet of the symmetry
superalgebra $SU(2,2|4)$ of type IIB superstring over $AdS_5 \times S^5$
\cite{mgnm,gmz1}.
     Doubleton supermultiplets do not have a smooth Poincar\'{e} limit in
$d=5$ and correspond to superconformal field theories on the
boundary of $AdS_{5}$, which can be identified with the four
dimensional Minkowski space.  The $SU(2,2|4)$ algebra acts as the
$\mathcal{N}$=4 extended superconformal algebra on the boundary of
$AdS_{5}$.  A complete list of doubleton and massless (in $AdS_5$
sense) supermultiplets
 of $SU(2,2|4)$ was given in
\cite{gmz1,gmz2}.  The CPT self-conjugate massless supermultiplet of
$SU(2,2|4)$ is the massless graviton supermultiplet in $d=5$.  This
massless multiplet sits at the bottom of an infinite tower of massive
short supermultiplets of $SU(2,2|4)$ corresponding to the Kaluza-Klein
spectrum of IIB supergravity over $S^{5}$ \cite{mgnm}.
 The oscillator construction
yields not only the short multiplets but the massless and massive
intermediate and long multiplets as well \cite{mgnm,gmz2,cgkrz} .

    M-theory compactified on the four sphere $S^4$ is similarly believed to
be dual to (2,0) superconformal field theory in six dimensions
 (in a certain limit). The  symmetry superalgebra of
 M-theory compactified to $AdS_{7}$
over $S^{4}$ is \OS ~\cite{ptn,gnw}.
 The general method for the oscillator construction of
unitary supermultiplets of \OS ~was given in \cite{gnw} with emphasis on short
supermultiplets that appear in the Kaluza-Klein compactification
 of 11 dimensional
supergravity theory. The entire Kaluza-Klein spectrum of the eleven dimensional
supergravity over $AdS_7 \times S^4$ can be obtained by a simple tensoring
procedure from  the ``CPT self-conjugate'' doubleton supermultiplet \cite{gnw}.
This ``CPT self-conjugate'' doubleton supermultiplet is simply the
$(2,0)$ superconformal multiplet of the dual field theory in six dimensions.
The $AdS_7/CFT_6$ duality has been studied from various points of view more
recently \cite{recent}.

 In this paper, we shall give a complete study of
 the doubleton
and massless (in $AdS_7$ sense) supermultiplets of \OS ~which we believe are
relevant to the understanding of the full spectrum of M-theory
 over $AdS_7 \times S^4$.   Our  results  extend
 those  of  \cite{gmz1,gmz2} on $AdS_5/CFT_4$ dualities to
$AdS_{7}/CFT_{6}$ dualities and their
 relation to the unitary supermultiplets of
\OS. In particular, in section 2, we review the oscillator
construction of the positive energy UIRs of $SO(6,2)$ and show how to go
from the compact ($SU(4) \times U(1)$) basis to the
noncompact $SU^*(4) \times \cal{D}$ basis.  Considered as the $d=6$
conformal group, the $SU^*(4) \times \cal{D}$ subgroup of $SO(6,2)$ can be
identified with $d=6$ Lorentz transformations times dilatations.  We show
that doubleton representations of $SO(6,2)$ are all massless in $d=6$.
Furthermore we show that the UIR's of the lowest weight type constructed
by the oscillator method are equivalent to unitary representations of
the conformal group induced by a finite dimensional representation
of the Lorentz group $SU^*(4)$ with a definite conformal dimension.
Remarkably,  the compact $SU(4)$ labelling and the non-compact
$SU^*(4)$ labelling of the UIRs coincide. Furthermore, the conformal
dimension, $l$, which is given by the eigenvalues of the dilatation
operator, is simply $-E$, where $E$ is the (conformal) energy. Thus, the
positive energy UIRs with energy $E$ can be identified with $d=6$
conformal fields with conformal dimension $l = -E$.  In section 3, we
review the superoscillator construction of the superalgebra \OS.  In
section 4, we give a complete list of the doubleton supermultiplets of \OS ~which correspond to massless conformal supermultiplets in $d=6$.  In
section 5, we give a complete list of massless supermultiplets of
 \OS ~considered as
the seven dimensional AdS supergroup.  The short
massless supermultiplets have spin range 2 in the $d=4$ sense, the long
massless supermultiplets have spin range 4 and the intermediate massless
supermultiplets have spin ranges between 2 and 4.  We conclude with a
discussion of our results as they relate to AdS/CFT dualities.

\section{Compact ($SU(4)\times U(1)$) versus non-compact
$(SU^*(4) \times {\mathcal{D}})$ bases for the positive
 energy unitary representations of the  group $SO^*(8)=SO(6,2)$ }
\setcounter{equation}{0}

The seven dimensional AdS group $SO^*(8)=SO(6,2)$ (with the covering group
$Spin^*(8)= Spin(6,2)$) is isomorphic to the six dimensional conformal
group (its covering group).\footnote{Below we shall sometimes denote the
covering group by the same symbol as the group itself. Unless otherwise
specified we shall however always be dealing with the covering groups in
question.} As the conformal group in d=6, $SO^*(8)$ is generated
by the Lorentz group ($SO(5,1)$ with the covering group $SU^*(4)$)
generators $M_{\mu\nu}$, the six translation generators $P_{\mu}$, the
dilatation generator $D$ and the generators of special conformal
transformations $K_{\mu}$ ($\mu, \nu, \dots = 0, 1, 2, 3,4,5$). The
commutation relations of these generators are
\eqn
[M_{\mu\nu},M_{\rho\sigma}] & =
&i (\eta_{\nu\rho}M_{\mu\sigma}- \eta_{\mu\rho}M_{\nu\sigma}
-\eta_{\nu\sigma}M_{\mu\rho}+ \eta_{\mu\sigma}M_{\nu\rho})\cr
[ P_{\mu}, M_{\rho\sigma} ] & = & i
(\eta_{\mu\rho}P_{\sigma}-\eta_{\mu\sigma}
P_{\rho})\cr
[K_{\mu},M_{\rho\sigma}]& = &i
(\eta_{\mu\rho}K_{\sigma}-\eta_{\mu\sigma} K_{\rho})\cr
[D,M_{\mu\nu}]& = & [P_{\mu},P_{\nu}] = [K_{\mu},K_{\nu}]=0\cr
[P_{\mu},D] & = &iP_{\mu}; \quad [K_{\mu},D]=-iK_{\mu}\cr
[P_{\mu},K_{\nu}]& = &2i(\eta_{\mu\nu}D-M_{\mu\nu})
\enn
with $\eta_{\mu\nu}=\textrm{diag}(+,-,-,-,-,-)$. The rotation subgroup
$SO(5)$ ( or its covering group $USp(4)$) is generated by $M_{ij}$ with
$i,j=1,2,..,5$.

Defining
\eq
M_{\mu 6} = {1 \over 2} (P_{\mu} - K_{\mu}), \quad
M_{\mu 7} = {1 \over 2} (P_{\mu} + K_{\mu}), \quad M_{67} = -D,
\en
the commutation relations of the $d=6$ conformal algebra
can be written as
\eq
[M_{ab}, M_{cd}] = i(\eta_{bc}M_{ad} - \eta_{ac}M_{bd}
-\eta_{bd}M_{ac} + \eta_{ad}M_{bc}),\label{SO42}
\en
where $ - \eta_{66}=\eta_{77}=1$ and $a,b,..=0,1,...,7$.

Considered as the isometry group of seven dimensional anti-de Sitter
space, the generators of $SO^*(8)$ have a different physical
interpretation. In particular, the rotation group becomes $SO(6)$, with
the covering group $SU(4)$, generated by $M_{mn}$ ($m, n, \dots =
1,2,..,6$).  The generator $E\equiv M_{07}$ becomes the AdS energy
generating translations along the timelike Killing vector field of
$AdS_{7}$, and the non-compact generators $M_{0m}$, $M_{m7}$ correspond to
``boosts'' and spacelike ``translations'' in $AdS_{7}$.

There are two different subgroups of the $d=6$ conformal group $SO(6,2)$
which play an important role in the classification of its physically
relevant representations. First is the maximal compact subgroup
$SU(4)\times U(1)_{E}$ generated by $M_{mn}$ and $E\equiv M_{07}$ with the
$U(1)_{E}$ generator $E=\frac{1}{2}(P_{0}+K_{0})$ being simply the
conformal Hamiltonian.  Denoting the Lie algebra of $SU(4)\times U(1)_{E}$
as $L^{0}$, the conformal algebra $g=SO(6,2)$ has a three graded
decomposition with respect to $L^0$:
\eq
g = L^{+} \oplus L^{0} \oplus L^{-},
\en
where
\eqn
[L^{0},L^{\pm}] &\subseteq & L^{\pm} \cr
[L^{+},L^{-}] &\subseteq&L^{0} \cr
[L^{+},L^{+}] &=& 0=[L^{-},L^{-}]\cr
[E,L^{\pm}]&=&\pm L^{\pm}, \quad [E,L^{0}]=0.
\enn
The general construction of the positive energy unitary representations of
$SO^*(8)$ uses the realization of its generators as bilinears of an
arbitrary number $P$ (generations or colors)  of pairs of bosonic annihilation
(${\bf a}_i, {\bf b}_i$) and creation operators (${\bf a}^i, {\bf b}^i$)
transforming in the fundamental representation of $SU(4)$ and its
conjugate, respectively: \cite{gnw, mgcs,ibmg}
\eq
\begin{array}{c}
M^{i}_{j} = \bf{a}^{i}\cdot \bf{a}_{j} + \bf{b}_{j}\cdot \bf{b}^{i}, \\
A_{ij} = \bf{a}_{i}\cdot \bf{b}_{j} - \bf{a}_{j}\cdot \bf{b}_{i}, \\
A^{ij} = \bf{a}^{i}\cdot \bf{b}^{j} - \bf{a}^{j}\cdot \bf{b}^{i} \\
\end{array}
\en
where ${\bf a}_i \cdot {\bf b}_j = \sum_{K=1}^{P} a_i(K) b_j(K)$ etc.. The
bosonic annihilation and creation operators $a^{i}(K)=
a_{i}(K)^{\dagger}$, $b^{i}(K)= b_{i}(K)^{\dagger}$ satisfy the
commutation relations:
\eq
[a_i(K), a^j(L)] = \delta_{i}^{j} \delta_{KL}, \quad
[b_i(K), b^j(L)] = \delta_{i}^{j} \delta_{KL}
\en
with $i,j=1,2,3,4$ and $K,L= 1, \dots, P$. $M^i_j$ are the generators of
the maximal compact subgroup $U(4)$. Hermitian linear combinations of the
$A_{ij}$ and $A^{ij}$ are the non-compact generators of $SO(6,2)$.
\cite{mgcs,gnw}

Physically relevant representations of the conformal group are unitary
irreducible representations (UIRs) of the lowest weight type in which the
spectrum of the conformal Hamiltonian (resp. the AdS energy), $E$, is
bounded from below. The natural basis for constructing them is the compact
basis in which the lowest weight (positive energy) property becomes
manifest.  The lowest weight UIRs of $SO(6,2)$ can then be constructed in
a simple way by using the oscillator realization of the generators given
above. Each lowest weight UIR is uniquely determined by the quantum
numbers of a lowest weight vector, $|\Omega\rangle$, provided that
$|\Omega\rangle$ transforms irreducibly under $SU(4)\times U(1)_{E}$ and
is also annihilated by all the elements of $L^{-}$\cite{gnw}.

On the other hand, in $d=6$ conformal field theory, one would like to work
with fields that transform covariantly under the Lorentz group, $SU^*(4)$,
and the dilatations \footnote{ Note that the conformal group SO(6,2) has a three
graded structure with respect to the subgroup
 $SU^*(4)\times \mathcal{D}$ just as it
has with respect to its maximal compact subgroup $SU(4)\times U(1)$ \cite{mgm}}.
 In $d=4$, the standard way to construct these fields
is via the method of induced representations \cite{gmz2, macksalam}.  In
this method, a representation (typically finite dimensional) of the
stability group, $H$, of $x_{\mu}=0$ (where $x_{\mu}$ is the coordinate four
vector) induces a representation of the
conformal group $SO(4,2)$. When $SO(6,2)$ acts in the standard way on the
(conformal compactification of) $6d$ Minkowski spacetime, the stability
group, $H$, of the coordinate six-vector $x^{\mu}=0$ becomes the semi-direct
 product $(SU^*(4)\times
{\mathcal{D}})\odot {\mathcal{K}}_6$. Furthermore, ${\mathcal{K}}_6$
represents the Abelian subgroup generated by the special conformal
generators, $K_{\mu}$. The Lie algebra of $H$ consists of the generators,
$M_{\mu\nu}$, of the Lorentz group, $SU^*(4)$, the dilatation operator,
$D$, and the generators of the special conformal transformations,
$K_{\mu}$.  In our case the d=6 conformal group, $G=SO(6,2)$, will be
realized on fields that live on the coset space, $G/H$.  These fields are
labelled by their transformation properties under the Lorentz group,
$SU^*(4)$, their conformal dimension, $l$, and certain
matrices, $\kappa_{\mu}$.  In particular, these matrices, $\kappa_{\mu}$,
are related to their behavior under special conformal transformations,
$K_{\mu}$, as in $d=4$ \cite{gmz2, macksalam}.

To establish a dictionary between the compact and non-compact viewpoints, we shall now reformulate the construction of positive energy representations of
$SO(6,2)$ \cite{gnw} in terms of twistorial operators.  The twistorial
operators we introduce will involve eight dimensional bosonic spinors.  This is analogous to the twistorial construction in the case of the $d=4$ conformal group \cite{gmz2}.
A dynamical  realization of the twistorial construction of the
representations of $SU(2,2|4)$ was given in \cite{cgkrz,ckr}.
  To this end, let $\Gamma_{\mu}$ be
the $6d$ gamma matrices ($\{ \Gamma_{\mu},\Gamma_{\nu}\}=2\eta_{\mu\nu}$)
with $\Gamma_{7}=-i \Gamma_{0}\Gamma_{1}\Gamma_{2}\cdots\Gamma_{5}$. Then
the matrices,
\eqn
\Sigma(M_{\mu\nu}) &:=
&\frac{i}{4}\left[\Gamma_{\mu},\Gamma_{\nu}\right],\nn\cr
\Sigma(M_{\mu 6}) &:= & \frac{i}{2}\Gamma_{\mu}\Gamma_{7},\nn\cr
\Sigma(M_{\mu 7}) &:= & \frac{1}{2}\Gamma_{\mu},\nn\cr
\Sigma(M_{67}) &:= & \frac{1}{2}\Gamma_{7},
\enn
generate an eight dimensional (non-unitary) spinor representation of the
conformal algebra, $SO(6,2)$.  We shall refer to this representation as
the left-handed spinor representation. The right handed spinor
representation is obtained by defining $\Gamma_{7}$ with the opposite sign
(i.e. $\Gamma_{7}=+i \Gamma_{0}\Gamma_{1}\Gamma_{2}\cdots\Gamma_{5}$).
Unless otherwise specified we shall work with the left-handed spinor
representation below. For our gamma matrices we choose
\eqn
\Gamma_{0} &=& \sigma_{3} \otimes I_{2} \otimes I_{2}   \nn     \\
\Gamma_{1} &=& i\sigma_{1} \otimes \sigma_{2} \otimes I_{2} \nn  \\
\Gamma_{2} &=& i\sigma_{1} \otimes \sigma_{1} \otimes \sigma_{2} \nn \\
\Gamma_{3} &=& i\sigma_{1} \otimes \sigma_{3} \otimes \sigma_{2} \nn \\
\Gamma_{4} &=& i\sigma_{2} \otimes I_{2} \otimes \sigma_{2} \nn \\
\Gamma_{5} &=& i\sigma_{2} \otimes \sigma_{2} \otimes \sigma_{1} \nn \\
\Gamma_{7} &=& -i\sigma_{2} \otimes \sigma_{2} \otimes \sigma_{3}
\enn
where $\sigma_1,\sigma_2$ and $\sigma_3$ are the Pauli matrices.

We now regroup these bosonic oscillators into left-handed spinorial operators,
$\Psi(K)$, which we define as
\eq
\Psi(K) := \left(\begin{array}{c}
a_{1}(K)\cr
a_{2}(K)\cr
a_3(K) \cr
a_4(K) \cr
b^{1}(K)\cr
b^{2}(K) \cr
b^3 (K) \cr
b^4 (K),
\end{array}\right)
\en
so that
\eqn
\bar{\Psi}(K) & \equiv & {\Psi}^{\dagger}(K)\Gamma^{0} \nonumber \\
        & = & \left(a^{1}(K),a^{2}(K),a^3(K),a^4(K),
                -b_{1}(K), -b_{2}(K), -b_3 (K), -b_4 (K) \right).
\enn
Consider now the bilinears of these twistorial operators involving the $8 \times 8$ matrices
$\Sigma(M_{ab})$ :
\eq
\bar{\Psi}\Sigma(M_{ab})\Psi :=
\sum_{K=1}^{P} \bar{\Psi}(K)\Sigma(M_{ab})\Psi(K).
\en
One finds that they satisfy the commutation relations of $SO(6,2)$
\eq
\left[ \bar{\Psi}\Sigma(M_{ab})\Psi,\bar{\Psi}\Sigma(M_{cd})\Psi\right]=
\bar{\Psi}\left[ \Sigma(M_{ab}),\Sigma(M_{cd})\right]\Psi.
\en
and yield an infinite dimensional unitary
representation in the Fock space of the oscillators $a^{i}(K)$ and
$b^{i}(K)$ .

  We have the following relations between the generators realized as bilinears
of twistorial operators  and the bilnears given in terms of $U(4)$ covariant
oscillators:
\eqn
\bar{\psi}\Gamma_{0}\psi &=& M^{1}_{1} + M^{2}_{2} + M^{3}_{3} +
M^{4}_{4} \nn   \\
\bar{\psi}\Gamma_{01}\psi &=& A^{13} + A^{24} + A_{31} + A_{42}
\nn \\
\bar{\psi}\Gamma_{1}\psi &=& A^{13} + A^{24} + A_{13} + A_{24}  \nn \\
\bar{\psi}\Gamma_{02}\psi &=& A^{14} + A^{32} + A_{41} + A_{23} \nn \\
\bar{\psi}\Gamma_{2}\psi &=& A^{14} + A^{32} + A_{14} + A_{32}  \nn \\
\bar{\psi}\Gamma_{03}\psi &=& A^{12} + A^{43} + A_{21} + A_{34} \nn \\
\bar{\psi}\Gamma_{3}\psi &=& A^{12} + A^{43} + A_{12} + A_{43}  \nn \\
\bar{\psi}\Gamma_{04}\psi &=& i\, \left( A^{21} + A^{43} + A_{21} + A_{43}
\right) \nn \\
\bar{\psi}\Gamma_{4}\psi &=& i\, \left( A^{21} + A^{43} + A_{12} + A_{34}
\right) \nn \\
\bar{\psi}\Gamma_{05}\psi &=& i \left( A^{41} + A^{32} + A_{41}
+ A_{32} \right)        \nn \\
\bar{\psi}\Gamma_{5}\psi &=& i\, \left( A^{41} + A^{32} + A_{14} + A_{23}
\right) \nn \\
\bar{\psi}\Gamma_{07}\psi &=& i \left( A^{13} + A^{42} + A_{13}
+ A_{42} \right)        \nn \\
\bar{\psi}\Gamma_{7}\psi &=& i\, \left( A^{13} + A^{42} + A_{31} + A_{24}
\right).
\enn
As stated above, the positive energy UIRs of $SO(6,2)$  are easily obtained by
constructing an irreducible representation $|\Omega\rangle$ (lowest weight
vector) of $SU(4)\times U(1)_E$ in the Fock space of the oscillators that
is annihilated by all the operators $A_{ij}$ of $L^{-}$ subspace:
\eq
A_{ij}|\Omega\rangle=0.
\en
Acting repeatedly with the di-creation operators $A^{ij}$ of $L^{+}$ on
$|\Omega\rangle$, one generates the basis of a positive energy UIR of the
group $SO^*(8)$.  To give  concrete examples of lowest weight
vectors, consider the case $P=1$ (doubletons). In this case the possible
lowest weight vectors are of the form:
\eqn
|0\rangle & &   \nn     \\
a^{i_1}|0\rangle &=& |\onebox\rangle    \nn     \\
a^{i_1}a^{i_2}|0\rangle &=& |\twobox\rangle     \nn     \\
      ~ & \vdots & ~            \nn \\
a^{i_1}a^{i_2}\cdots a^{i_n}|0\rangle &=& |\marcgenrowbox\rangle.
\enn
or of the form
\eq
a^{(i} b^{j)} |0\rangle = |\twobox \rangle
\en
This shows that  the possible lowest weight vectors $|\Omega\rangle$ of
the doubleton UIR's of $SO^*(8)$ transform in the symmetric tensor representations of
$SU(4)$.

\indent The positive energy UIRs of $SO^*(8)$ can be
identified with conformal fields in $d=6$ with positive conformal energy
and a definite conformal dimension transforming covariantly under the six
dimensional Lorentz group. To establish this connection we need to find a
mapping from the $SU^*(4)$- and $D$-covariant basis to the compact
$U(4)$ basis.  For this we define the following set of generators $J_{mn}$
(m,n, $\ldots$ = 1, $\ldots$, 6)
of another compact $SU(4)$ basis in terms of the $SU^*(4)$ generators,
$M_{\mu\nu} = \{M_{ij}, M_{i0}\}$ (i,j, $\ldots$ = 1, $\ldots$, 5),
\eqn
J_{ij} :=M_{ij}, \nonumber  \\
J_{i6}:=i M_{i0}= -J_{6i}.
\enn
The two $SU(4)$s generated by $J_{mn}$ and $M_{mn}$ have as a common
subgroup the rotation group $USp(4)$ in $d=6$.  Consider now the operator
\eq
U:=e^{\bar{\Psi} \Sigma (M_{06}+iM_{67})\Psi}.
\en
It has the following important property
\begin{eqnarray}
J_{mn}U & = & U(M_{mn}+L^{-}),  \nonumber\\
D U    & = & U(-iE+L^{-}),      \nonumber\\
K_{\mu}U & = & U(L^{-}),\label{UL}
\end{eqnarray}
where $L^{-}$ stands for certain linear combinations of di-annihilation
operators $A_{ij}$. Specifically,  for $J_{mn}U = U(M_{mn} +
L^-)$ we find,
\eqn
\bar{\psi} \Sigma (J_{16}) \psi U &=& U \left(-\bar{\psi} \Sigma(M_{16})
\psi + \frac{1}{2} \left(A_{31} + A_{42}\right) \right) \nn     \\
\bar{\psi} \Sigma (J_{26}) \psi U &=& U \left(-\bar{\psi} \Sigma(M_{26})
\psi + \frac{1}{2} \left(A_{41} + A_{23}\right) \right) \nn     \\
\bar{\psi} \Sigma (J_{36}) \psi U &=& U \left(-\bar{\psi} \Sigma(M_{36})
\psi + \frac{1}{2} \left(A_{21} + A_{34}\right) \right) \nn     \\
\bar{\psi} \Sigma (J_{46}) \psi U &=& U \left(-\bar{\psi} \Sigma(M_{46})
\psi + \frac{i}{2} \left(A_{21} + A_{43}\right) \right) \nn     \\
\bar{\psi} \Sigma (J_{56}) \psi U &=& U \left(-\bar{\psi} \Sigma(M_{56})
\psi + \frac{i}{2} \left(A_{41} + A_{32}\right) \right) \nn     \\
\bar{\psi} \Sigma (J_{ij}) \psi U &=& U \left( \bar{\psi} \Sigma(M_{ij}) \psi \right). \enn

 For $DU=U(-iE +L^-)$,
 we find, \eqn \bar{\psi} \Sigma \left(D\right) \psi U &=& U \left(-iE + \frac{-i}{2} \left(A_{31} +
A_{24}\right) \right). \enn For $K_{\mu}U = U(L^-)$, we find, \eqn \bar{\psi} \Sigma (K_{0}) \psi U &=& U \left(
\frac{1}{2} \left(A_{13} + A_{42}\right) \right) \nn     \\
\bar{\psi} \Sigma (K_{1}) \psi U &=& U \left(
\frac{1}{2} \left(A_{13} + A_{24}\right) \right) \nn     \\
\bar{\psi} \Sigma (K_{2}) \psi U &=& U \left(
\frac{1}{2} \left(A_{14} + A_{32}\right) \right) \nn     \\
\bar{\psi} \Sigma (K_{3}) \psi U &=& U \left(
\frac{1}{2} \left(A_{12} + A_{43}\right) \right) \nn     \\
\bar{\psi} \Sigma (K_{4}) \psi U &=& U \left(
\frac{i}{2} \left(A_{12} + A_{34}\right) \right) \nn     \\
\bar{\psi} \Sigma (K_{5}) \psi U &=& U \left(
\frac{i}{2} \left(A_{14} + A_{23}\right) \right).
\enn
Acting with $U$ on a lowest weight vector $|\Omega \rangle$ corresponds to
a (complex) rotation in the corresponding representation space of
$SO^*(8)$:
\eq
U|\Omega \rangle = e^{\bar{\Psi} \Sigma (M_{06}+iM_{67})
\Psi} |\Omega \rangle.
\en
Since $L^{-}|\Omega\rangle=0$, it then follows from (\ref{UL}) that
$\Phi(0):= U|\Omega\rangle$ is an irreducible representation of the little
group $H$ with conformal dimension \footnote{In our conventions, $l$ has dimension of
length (or inverse mass).} $l=-E$ and trivially represented
special conformal transformations $K_{\mu}$ (i.e. $\kappa_{\mu}=0$).
Acting with $e^{-i x^{\mu}P_{\mu}}$ on $\Phi(0)$ then translates the field
in Minkowski space:
\eq \Phi(x^{\mu})=e^{-i x^{\mu}P_{\mu}}\cdot \Phi(0) =
e^{-i x^{\mu}P_{\mu}}U|\Omega\rangle
\en
and generates a (group theoretically equivalent) induced representation of
$SO^*(8)$. We should note that the state $\Phi(x^{\mu})$ can be thought
of as a coherent state labelled by $\Omega$ and the coordinate $x^{\mu}$.

Since the bosonic oscillators, in terms of which we realized
the generators, transform in the spinor representation of $SO^*(8)$, the
oscillator construction  can be given a dynamical realization as was
done for $SU(2,2)$ \cite{cgkrz,ckr}.

 We recall that the doubleton representations of $SO^*(8)$
correspond to taking a single pair ($P=1$) of bosonic oscillators and
they do not have a smooth Poincar\'{e} limit in
$d=7$.  Let us now  show that they correspond to massless fields
in $d=6$ with mass defined in the usual Poincar\'{e} sense,
\eq
m^{2}=P_{\mu}P^{\mu}.
\en
The translation generators $P_{\mu}$ have the following realization in
terms of the oscillators:
\eqn
P_0 &=& \bar{\psi} \left( \Sigma\left(M_{06}\right) +
                        \Sigma\left(M_{07}\right) \right) \psi \nn \\
    &=& \frac{1}{2} \bar{\psi} \Gamma_{0} \left( 1 + i\Gamma_{7} \right)
                                                \psi    \nn \\
P_1 &=& \bar{\psi} \left( \Sigma\left(M_{16}\right) +
                        \Sigma\left(M_{17}\right) \right) \psi \nn \\
&=& \frac{1}{2} \bar{\psi} \Gamma_{1} \left( 1 + i\Gamma_{7} \right)
                                                \psi    \nn \\
P_2 &=& \bar{\psi} \left( \Sigma\left(M_{26}\right) +
                        \Sigma\left(M_{27}\right) \right) \psi \nn \\
&=& \frac{1}{2} \bar{\psi}  \Gamma_{2} \left( 1 + i\Gamma_{7} \right)
                                                \psi    \nn \\
P_3 &=& \bar{\psi} \left( \Sigma\left(M_{36}\right) +
                        \Sigma\left(M_{37}\right) \right) \psi \nn \\
&=&\frac{1}{2} \bar{\psi}  \Gamma_{3} \left( 1 + i\Gamma_{7} \right)
                                                \psi    \nn \\
P_4 &=& \bar{\psi} \left( \Sigma\left(M_{46}\right) +
                        \Sigma\left(M_{47}\right) \right) \psi \nn \\
&=&\frac{1}{2} \bar{\psi}  \Gamma_{4} \left( 1 + i\Gamma_{7} \right)
                                                \psi    \nn \\
P_5 &=& \bar{\psi} \left( \Sigma\left(M_{56}\right) +
                        \Sigma\left(M_{57}\right) \right) \psi \nn \\
&=&\frac{1}{2} \bar{\psi}  \Gamma_{5} \left( 1 + i\Gamma_{7} \right)
                                                \psi.
\enn
Substituting in the above expressions for $P_{\mu}$ one finds after some
lengthy calculation the mass operator $m^2$ vanishes identically for
$P=1$. Thus all the doubleton irreps of $SO^*(8)$ are massless in $d=6$.
For $P\neq 1$ the mass operator is non-vanishing and the lowest weight
vectors of $SO^*(8)$ are in general not eigenstates of the mass operator.
Hence they correspond to massive conformal fields in $d=6$. We should
stress that this is in complete parallel to the situation in $d=4$ where
the doubleton representations of $SO(4,2)$ are all massless
\cite{binegar,gmz2}.

\section{Oscillator Construction of the UIRs of \OS}
\indent    In this section, we give the commutation relations of \OS ~in a  non-compact as well as a compact basis.  Bearing in mind that it was shown above how to go from the compact basis to a non-compact basis for the UIR's of $SO(6,2)$, we then present a realization of the \OS ~generators in the
compact basis in terms of superoscillators. For further details, we refer the reader to \cite{gnw, mgrs}.

\subsection{The superalgebra \OS}

\indent  The symmetry group of M-theory on $AdS_7 \times S^4$ is the
supergroup \OS ~with the even subgroup $SO^{*}(8) \times USp(4)$, where
$USp(4)$ is isomorphic to $SO(5)$, the isometry group of the four sphere.
\OS ~can be interpreted as the $\mathcal{N}$=4 extended $AdS$ superalgebra in
$d=7$ or as the $(2,0)$ extended conformal superalgebra in $d=6$.
Because of triality property of the representations of $SO(8)$ and hence
of the finite dimensional representations of $SO(6,2)$ there exist three
different forms of the superalgebra \OS, according to whether the supersymmetry generators transform in the vector, left-handed or right-handed spinor representation of $SO(6,2)$. The anti-symmetric tensor of any one of these three
representations  transform like the adjoint representation of $SO(6,2)$.
For M-theory on $AdS_7\times S^4$ the relevant form is the one  in which
the supersymmetry generators transform in the left-handed spinor representation
of $SO(6,2)$, such that it decomposes as $(4+\bar{4})$ under the compact
subgroup $SU(4)$. Denoting the supersymmetry generators of \OS ~as
$Q_{\hat{\alpha}I}$ we can write their anticommutators as \cite{ckp}
\eqn
\left\{ Q_{\hat\alpha I}, Q_{\hat\beta J} \right\} = -\frac{1}{2} \left(
\Omega_{IJ}
M_{\hat\alpha \hat\beta} + C_{\hat\alpha \hat\beta} U_{IJ} \right).
\enn
where $I,J,..=1,..,4$ and $\hat\alpha, \hat\beta =1,...8$. $U_{IJ}=U_{JI}$
are the $USp(4)$ generators and $\Omega_{IJ}=-\Omega_{JI}$ is the symplectic
invariant tensor. The tensor $C_{\hat\alpha \hat\beta}$ is the charge conjugation matrix in $(6,2)$ dimensions and is symmetric \cite{ckp}. Furthermore, $M_{\hat\alpha \hat\beta} $ are related to the $SO(6,2)$
generators $M_{ab}$ given in the previous section as follows:
\eq
M_{\hat\alpha \hat\beta} = \frac{1}{8} \left( \Gamma^{ab}
 \right)_{\hat\alpha \hat\beta}
 M_{ab}
\en
The generators  of $USp(4)$ satisfy:
\eqn
\left[ U_{IJ}, U_{KL} \right] = \Omega_{I(K}U_{L)J} + \Omega_{J(K}U_{L)I}
\enn
The commutation  relations of $SO(6,2)$ and $USp(4)$ with the supersymmetry
generators are then of the form \cite{ckp}
\eqn
\left[ M_{ab}, Q_{\hat\alpha I} \right] =  \left( \Sigma(M^{ab})
\right)_{\hat\alpha}^{\hat\beta} Q_{\hat\beta I}, \nn \\
\left[ U_{IJ}, Q_{K \hat\alpha} \right] = -\Omega_{K(I} Q_{J)\hat\alpha}.
\enn
\indent  The superalgebra \OS ~has a three graded decomposition with
respect to its compact subsuperalgebra $U(4|2)$
\eq
g = L^{+} \oplus L^{0} \oplus L^{-},
\en
where
\eqn
{[}L^{0},L^{\pm}{]} & \subseteq & L^{\pm}  \nn \\
{[}L^{+},L^{-}{]} & \subseteq & L^{0} \nn \\
{[}L^{+},L^{+}{]} & = & 0 = {[}L^{-},L^{-}{]}.
\enn
Here $L^{0}$ represents the generators of $U(4|2)$.
\newline
\indent  Generalizing the (purely bosonic) oscillator construction
for $SO^{*}(8)$ in section 2, the Lie superalgebra \OS ~can be realized
in terms of bilinear combinations of bosonic and fermionic
annihilation and creation operators $\xi_{A}(K)$ ($\xi^{A}(K) =
\xi^{\dagger}_{A}(K)$) and $\eta_{M}(K)$ ($\eta^{M}(K) =
\eta^{\dagger}_{M}(K)$)
which transform covariantly and contravariantly, respectively, under the $U(4|2)$ subsupergroup of \OS
\eqn
\xi_{A}(K) = \left( \begin{array}{c}
                          a_{i}(K) \\
                                      \alpha_{\mu}(K)
                                                  \end{array}     \right), &
\xi^{A}(K) = \left( \begin{array}{c}
                          a^{i}(K) \\
                                                  \alpha^{\mu}(K)
                                                  \end{array}     \right),  \nn \\
\eta_{B}(L) = \left( \begin{array}{c}
                          b_{j}(L) \\
                                                  \beta_{\nu}(L)
                                                  \end{array}     \right), &
\eta^{B}(L) = \left( \begin{array}{c}
                          b^{j}(L) \\
                                                  \beta^{\nu}(L)
                                                  \end{array}     \right).
\enn
with i,j = 1,2,3,4; $\mu, \nu$=1,2 and
\eqn
\left\{ \xi_{A}(K), \xi^{B}(L) \right] & = & \delta^{B}_{A} \delta_{KL} \nn \\
\left\{ \eta_{A}(K), \eta^{B}(L) \right] & = & \delta^{B}_{A} \delta_{KL}
K,L = 1,\ldots,P \nn \\
\left\{ \xi_{A}(K), \eta^{B}(L) \right] & = & \left\{ \xi_{A}(K), \xi_{B}(L)
\right] = \left\{ \eta_{A}(K), \eta_{B}(L) \right] = 0,
\enn
where $\{~,~{]}$ represents an anti-commutator among any two fermionic
oscillators and a commutator otherwise.  Moreover, annihilation and
creation operators are labelled by lower and upper indices,
respectively.  The generators of \OS ~are given in terms of the above
superoscillators schematically as
\eqn
A_{AB} & = & \mbox{\boldmath $\xi$}_{A} \cdot \mbox{\boldmath $\eta$}_{B}
 - \mbox{\boldmath $\eta$}_{A} \cdot
\mbox{\boldmath $\xi$}_{B} = A_{ij} \oplus A_{\mu\nu} \oplus Q_{i\mu}
\nn \\
A^{AB} & = & A_{AB}^{\dagger} = \mbox{\boldmath $\eta$}^{B} \cdot \mbox{\boldmath $\xi$}^{A} -
\mbox{\boldmath $\xi$}^{B} \cdot\mbox{\boldmath $ \eta$}^{A} = A^{ij} \oplus A^{\mu\nu}
\oplus Q^{i\mu} \nn \\
M^{A}_{B} & = & \mbox{\boldmath $\xi$}^{A} \cdot \mbox{\boldmath $\xi$}_{B} +
(-1)^{(degA)(degB)}\mbox{\boldmath $\eta$}_{B} \cdot \mbox{\boldmath $\eta$}^{A} =
M^{i}_{j} \oplus M^{\mu}_{\nu} \oplus Q^{i}_{\mu}
\oplus Q^{\mu}_{i}
\enn
where the boldfaced $\mathbf{\xi}$'s and $\mathbf{\eta}$'s
 indicate that we are taking
an arbitrary number $P$ of ``generations'' of superoscillators and
the dot represents the summation over the internal index
$K=1,\ldots,P$ (i.e.
 $\mbox{\boldmath $\xi$}_{A}\cdot \mbox{\boldmath $\eta$}_{M} \equiv
\Sigma^{P}_{K=1}\xi_{A}(K)\eta_{M}(K)$).
\newline
\indent  The even subgroup $SO^{*}(8) \times USp(4)$ is generated by
the di-bosonic and di-fermionic generators.  In particular, one
recovers the $SO^{*}(8)$ generators of section 2 in terms of the bosonic
oscillators in their $SU(4) \times U(1)_{E}$ basis:
\eqn
A_{ij} & = & \mathbf{a_{i}\cdot b_{j}} - \mathbf{a_{j}\cdot b_{i}}, \nn \\
A^{ij} & = & \mathbf{a^{i}\cdot b^{j}} - \mathbf{a^{j}\cdot b^{i}}, \nn \\
M^{i}_{j} &=& \mathbf{a^{i}\cdot a_{j}} + \mathbf{b_{j}\cdot b^{i}}, \nn \\
E & = & Q_{B} = \frac{1}{2} M^{i}_{i} = \frac{1}{2} (N_{B} + 4P)
\enn
satisfying
\eqn
[ A_{ij}, A^{kl} ] & = & \delta^{k}_{i} M^{l}_{j} + \delta^{l}_{j} M^{k}_{i}
                  -\delta^{k}_{j} M^{l}_{i} - \delta^{l}_{i} M^{k}_{j}.
\enn
Here, $N_{B} = \bf{a^{i}\cdot a_{i}} +  \bf{b^{i}\cdot b_{i}}$
is the bosonic number operator and $E$ (which
corresponds to the $AdS$ energy) is the generator of $U(1)_{E}$.

\indent  Similarly, the $USp(4)$ generators in their $SU(2) \times
U(1)$ basis are expressed in terms of the fermionic oscillators
$\alpha$ and $\beta$:
\eqn
A_{\mu\nu} & = & \mbox{\boldmath $\xi$}_{\mu} \cdot \mbox{\boldmath $\eta$}_{\nu}
 - \mbox{\boldmath $\eta$}_{\mu} \cdot
\mbox{\boldmath $\xi$}_{\nu}   \nn \\
A^{\mu\nu} & = & \mbox{\boldmath $\eta$}^{\mu} \cdot \mbox{\boldmath $\xi$}^{\nu}
 - \mbox{\boldmath $\xi$}^{\mu} \cdot
\mbox{\boldmath $\eta$}^{\nu}  \nn \\
M^{\mu}_{\nu} & = & \mbox{\boldmath $\xi$}^{\mu} \cdot \mbox{\boldmath $\xi$}_{\nu} -
               \mbox{\boldmath $\eta$}_{\nu} \cdot \mbox{\boldmath $\eta$}^{\mu} \nn
                                                                     \\
Q_{F} & = & \frac{1}{2}(\mbox{\boldmath $\alpha$}^{\mu}\cdot
\mbox{\boldmath $ \alpha$}_{\mu} -
\mbox{\boldmath $\beta$}_{\mu} \cdot \mbox{\boldmath $\beta$}^{\mu})
 = \frac{1}{2}(N_{F} - 2P)
\label{USp4}
\enn
with the closure relation
\eq
[ A_{\mu\nu}, A^{\rho\sigma} ] =
-M^{\rho}_{\mu}\delta^{\sigma}_{\nu} -
M^{\sigma}_{\nu}\delta^{\rho}_{\mu} -
M^{\sigma}_{\mu}\delta^{\rho}_{\nu} -
M^{\rho}_{\nu}\delta^{\sigma}_{\mu}.
\en
Here $N_{F} = \mbox{\boldmath $\alpha$}^{\mu}\cdot \mbox{\boldmath $\alpha$}_{\mu} +
\mbox{\boldmath $\beta$}^{\mu} \cdot \mbox{\boldmath $\beta$}_{\mu}$ is the fermionic number operator
and $Q_{F}$ is the generator of $U(1)$.
\indent  Analogously, the odd generators are given by products of
bosonic and fermionic oscillators ($Q^{i\mu} = {\mathbf a}^{i}\cdot
\mbox{\boldmath $\beta$}^{\mu} -
{\mathbf b}^{i} \cdot \mbox{\boldmath $\alpha$}^{\mu}$) and satisfy the following closure relations
\eqn
\{Q^{i\mu}, Q^{j\nu} \} & = & 0 \nn \\
\{Q_{i\mu}, Q^{j\nu} \} & = & \delta^{\nu}_{\mu} M^{j}_{i} -
\delta^{j}_{i} M^{\nu}_{\mu}.
\enn
\subsection{Unitary Supermultiplets of \OS}
\indent  To construct a basis for a lowest weight UIR of \OS, one
starts from a set of states, collectively denoted by
$|\Omega\rangle$, in the Fock space of the oscillators a, b,
$\alpha$, $\beta$ that transforms irreducibly under $U(4|2)$ and that
is annihilated by all the generators $A_{AB}
\equiv (A_{ij} \oplus A_{\mu\nu} \oplus Q_{i\mu})$ of $L^{-}$
\eq
A_{AB}|\Omega\rangle = 0. \label{lwv}
\en
By acting on $|\Omega\rangle$ repeatedly with $L^{+}$, one then
generates an infinite set of states that form a UIR of \OS
\eq
|\Omega\rangle, L^{+1}|\Omega\rangle, L^{+1}L^{+1}|\Omega\rangle,
\ldots
\en
The irreducibility of the resulting representation of \OS ~follows
from the irreducibility of $|\Omega\rangle$ under $U(4|2)$.  Because
of the property (\ref{lwv}), $|\Omega\rangle$ as a whole will be
referred to as the ``lowest weight vector (lwv)'' of the corresponding
UIR of \OS.
\newline
\indent  In the restriction to the subspace involving purely bosonic
oscillators, the above construction reduces to the subalgebra
$SO^{*}(8)$ and its positive energy UIRs as described in section 2.
Similarly, when restricted to the subspace involving purely
fermionic oscillators, one gets the compact internal symmetry group
$USp(4)$ (\ref{USp4}), and the above construction yields the
representations of $USp(4)$ in its $SU(2)$ basis.
\newline
\indent  Accordingly, a lowest weight UIR of \OS ~decomposes into a
direct sum of finitely many positive energy UIRs of $SO^{*}(8)$
transforming in certain representations of the internal symmetry
group $USp(4)$.  Thus, each positive energy UIR of \OS ~corresponds to
a supermultiplet of fields living in $AdS_{7}$ or on its boundary.
Interpreted as a UIR of the $\mathcal{N}$=2 conformal superalgebra in
d=6, each lowest weight UIR of \OS ~corresponds to a supermultiplet of
massless or massive fields.

\indent
We now briefly describe the form in which our results are presented.
Consider an \OS ~lwv ($|\Omega\rangle$) which transforms in an irrep
of $U(4|2)$.  $|\Omega\rangle$ can be decomposed into its
irreducible components under the even subgroup $U(4)\times U(2)$.
This is most simply done by decomposing the supertableaux of $|\Omega\rangle$
into the Young tableaux of $U(4)\times U(2)$ using the results of
\cite{bbars}. These $U(4)\times U(2)$ irreps  correspond to lwv's of
UIRs of $SO^{*}(8) \times USp(4)$.  In the tables given in subsequent
sections, such lwv's
will be marked by an asterisk adjacent to their corresponding
$U(4)\times U(2)$ Young tableaux.
The additional lwv's of $SO^{*}(8) \times USp(4)$ are obtained by
acting on the lwv $|\Omega\rangle$ with the supersymmetry generators
$Q^{i\mu}$ repeatedly.  These lwv's are listed in the tables without
an asterisk. Furthermore, the tables below have been written in
``ascending order''.  That is,
immediately following each lwv with an asterisk are its ``descendant''
lwv's i.e, those  lwv's obtained by acting on it with
supersymmetry generators.  Furthermore the
 $SU(4)_{D}$ and $SU^*(4)_{D}$ transformation properties of the lwv's
of $SO^*(8)$ are indicated by their  Dynkin labels .  As explained above,
 the $SU(4)_{D}$
labels coincide with the $SU^*(4)_{D}$ labels.

\section{The Doubleton Supermultiplets of \OS}

\indent
By choosing one pair of super oscillators ($\xi$ and $\eta$) in the
oscillator realization of \OS ~(i.e. for $P$ = 1), one obtains the
so-called doubleton supermultiplets.  These supermultiplets
contain only doubleton representations of $SO^{*}(8)$, i.e. they correspond to
multiplets of fields living on the boundary of $AdS_{7}$ without a 7d
Poincar\'{e} limit.  Equivalently, they can be characterized as
multiplets of massless fields in 6d Minkowski space that form a UIR of
the $\mathcal{N} = 2$ superconformal algebra \OS.
\newline

\indent
The doubleton supermultiplet of \OS, which is defined
by the lowest weight vector, $|\Omega\rangle = |0\rangle$,
is the (2,0) conformal supermultiplet and is the analog of the
${\cal{N}}=4$ super Yang-Mills multiplet in $d=6$ \cite{gnw}.  The content of
the (2,0) supermultiplet is given in Table 1.
\newline

\vspace{.2cm}
%\begin{table}[ht]
\begin{center}
\begin{tabular}{|c|c|c|c|c|}
\hline
$SU(4) \times SU(2)$ lwv & $Q_{B}$=E & $Q_{F}$ & ${SU(4)}_{D}$ or
${SU^{*}(4)}_{D}$ & $SO(5)$
\\ \hline

*$|0>$  & 2     & -1    & (0,0,0)       & 5     \\ \hline

$|\onebox, \onebox>$    & $\frac{5}{2}$ & $\frac{-1}{2}$ & (1,0,0)
& 4 \\ \hline

$|\twobox, \oneonebox>$ & 3     & 0     & (2,0,0)       & 1     \\ \hline

\end{tabular}
\end{center}
Table 1. The doubleton supermultiplet corresponding to the lwv
$|\Omega\rangle = |0\rangle$.  The first column indicates the lwv's
of $SO^{*}(8) \times USp(4)$ with $U(4) \times U(2)$ Young tableaux.
The second column shows the $AdS$ energies $Q_{B} =
\frac{1}{2}(N_{B} + 4P)$, the third column lists the fermion $U(1)$
charge, the fourth column shows the $SU(4)$ Dynkin labels, and the
fifth column shows the representation of SO(5) induced by the lwv.
%\end{table}
\vspace{.2cm}

In Table 2 we give the doubleton supermultiplet defined by the lowest
weight vector  $|\Omega\rangle =
|\sonebox\rangle$.

\vspace{.2cm}
%\begin{table}[ht]
\begin{center}
\begin{tabular}{|c|c|c|c|c|}
\hline
$SU(4) \times SU(2)$ lwv & $Q_{B}$=E & $Q_{F}$ & ${SU(4)}_{D}$ or
${SU^{*}(4)}_{D}$ & $SO(5)$
\\ \hline
*$|\onebox, 1>$ & $\frac{5}{2}$ & -1    & (1,0,0)
& 5     \\ \hline

$|\twobox, \onebox>$    & 3     & $\frac{-1}{2}$        & (2,0,0)
& 4     \\ \hline

$|\threebox, \oneonebox>$       & $\frac{7}{2}$ & 0     & (3,0,0)
& 1     \\ \hline

*$|1, \onebox>$ & 2     & $\frac{-1}{2}$        & (0,0,0)
& 4     \\ \hline

$|\onebox, \oneonebox>$ & $\frac{5}{2}$ & 0     & (1,0,0)
& 1     \\ \hline
\end{tabular}
\end{center}
Table 2. Doubleton Supermultiplet defined by lwv $|\Omega\rangle =
|\sonebox\rangle$.

Table 3 gives the general doubleton supermultiplet whose lowest
weight vector is \linebreak $|\Omega\rangle = |\smarctwojbox\rangle$ for $ j>\frac{1}{2}$.
\vspace{.7cm}

\vspace{.2cm}
%\begin{table}[ht]
\begin{center}
\begin{tabular}{|c|c|c|c|c|}
\hline
$SU(4) \times SU(2)$ lwv & $Q_{B}$=E & $Q_{F}$ & ${SU(4)}_{D}$ or
${SU^{*}(4)}_{D}$ & $SO(5)$
\\ \hline
*$|\marctwojbox, 1>$    & j+2   & -1    & (2j,0,0)      & 5     \\ \hline

$|\marctwojplusonebox, \onebox>$        & j + $\frac{5}{2}$     &
$\frac{-1}{2}$  & (2j+1,0,0)    &4      \\ \hline

$|\marctwojplustwobox, \oneonebox>$     & j+3   & 0     & (2j+2,0,0)
& 1     \\ \hline

*$|\marctwojminusonebox, \onebox>$      & j+$\frac{3}{2}$       &
$\frac{-1}{2}$  & (2j-1,0,0)    & 4     \\ \hline

$|\marctwojbox, \oneonebox>$    & j+2   & 0     & (2j,0,0)      & 1
\\ \hline

*$|\marctwojminustwobox, \oneonebox>$   & j+1   & 0     & (2j-2,0,0)
& 1     \\ \hline
\end{tabular}
\end{center}
Table 3. General Doubleton Supermultiplet defined by the lwv
$|\Omega\rangle = |\smarctwojbox\rangle$ for $j >\frac{1}{2}$ where j is
an integer or half-odd integer.
\vspace{.7cm}

\section{The Massless Supermultiplets}
The doubleton supermultiplets described in the last section are
fundamental in the sense that all other lowest weight UIRs of \OS
occur in the tensor product of two or more doubleton supermultiplets.
Instead of trying to identify these irreducible submultiplets in the
(in general reducible, but not fully reducible) tensor products, one
simply increases the number $P$ of oscillator generations so that the
tensoring becomes implicit while the irreducibility stays manifest.
\newline
\indent
The simplest class, corresponding to $P=2$, contains the
supermultiplets that are known to be  ``massless'' in the
7d $AdS$ sense.  We will therefore label all
supermultiplets that are obtained by taking $P=2$ in the oscillator
construction despite some problems with the invariant definition of the
notion of ``mass'' in $AdS$ spacetimes \cite{gmz1}.
\newline
\indent
We will now give a complete list of the allowed \OS lowest weight
vectors $|\Omega\rangle$ for $P$=2.
\newline
\indent
The condition $L^{-}|0\rangle$ leaves the following possibilities:

\begin{itemize}
\item   $|\Omega\rangle$ = $|0\rangle$.  This lwv gives rise
$\mathcal{N}$=4 graviton supermultiplet in $AdS_{7}$ and occurs in
the tensor product of two CPT self-conjugate doubleton supermultiplets.

\item   $\xi^{A_{1}}(1)\xi^{A_{2}}(1)|0\rangle$ = $|\stwobox\rangle$.
Alternatively, one could use the states
$\xi^{A_{1}}(2)\xi^{A_{2}}(2)|0\rangle$,\\
$\eta^{B_{1}}(1)\eta^{B_{2}}(1)|0\rangle$,
$\eta^{B_{1}}(2)\eta^{B_{2}}(2)|0\rangle$,
$\xi^{(A_{1}}(1)\eta^{B_{1})}(2)|0\rangle$, or
$\xi^{(A_{1}}(2)\eta^{B_{1})}(1)|0\rangle$.

\item   $|\Omega\rangle$ =
$\xi^{A_{1}}(1)\xi^{A_{2}}(1) \cdots \xi^{A_{2j}}(1)|0\rangle$ =
$|\smarctwojbox\rangle$.  Equivalent lwv's are \\
$\xi^{A_{1}}(2)\xi^{A_{2}}(2) \cdots \xi^{A_{2j}}(2)|0\rangle$,
$\eta^{B_{1}}(1)\eta^{B_{2}}(1)\ldots \eta^{B_{2j}}(1)|0\rangle$,
$\eta^{B_{1}}(2)\eta^{B_{2}}(2)\ldots \eta^{B_{2j}}(2)|0\rangle$,\\
$\xi^{(A_{1}}(1)\xi^{A_{2}}(1)\cdots\xi^{A_{r}}(1)\eta^{B_{r+1}}(2)
\eta^{B_{r+2}}(2)\cdots \eta^{B_{2j})}(2)|0\rangle$,\\ or
$\xi^{(A_{1}}(2)\xi^{A_{2}}(2)\cdots\xi^{A_{r}}(2)\eta^{B_{r+1}}(1)
\eta^{B_{r+2}}(1)\cdots \eta^{B_{2j})}(1)|0\rangle$. \\
Increasing j leads to multiplets with higher and higher spins and
$AdS$ energies.
\end{itemize}

In addition to these purely (super)symmetrized lwv's, one can also
anti-(super)symmetrize pairs of superoscillators, since $P$=2.
The condition $L^{-}|0\rangle$ leaves the following possibilities:

\begin{itemize}
\item  $|\Omega\rangle$ = $\xi^{[A_{1}}(1)\xi^{B_{1}]}(2)\ldots
\xi^{[A_{2j}}(1)\xi^{B_{2j}]}(2)\xi^{A_{2j+1}}(1)\ldots
\xi^{A_{2j+n}}(1)|0\rangle$\\
= $|\smarctwojplusntwojbox\rangle$.\\
Equivalent lwv's are
$\eta^{[A_{1}}(1)\eta^{B_{1}]}(2) \ldots
\eta^{[A_{2j}}(1)\eta^{B_{2j}]}(2) \eta^{A_{2j+1}}(1) \ldots
\eta^{A_{2j+n}}(1) |0\rangle$, \\
$\xi^{[A_{1}}(1)\eta^{B_{1}]}(2)\cdots
\xi^{[A_{2j}}(1)\eta^{B_{2j}]}(2)\xi^{A_{2j+1}}(1) \cdots
\xi^{A_{2j+n}}(1)|0\rangle$, \\
$\xi^{[A_{1}}(1)\eta^{B_{1}]}(2)\cdots
\xi^{[A_{2j}}(1)\eta^{B_{2j}]}(2) \xi^{B_{2j+1}}(2) \cdots
\xi^{B_{2j+n}}(2)|0\rangle$.
\end{itemize}

In the following tables we list the  massless supermultiplets  defined
by the above lowest weight vectors\footnote{I am deeply grateful to Sudarshan Fernando for having brought to my attention states which were missing from my initial calculation.  Those missing states have been added to the following tables.}.
\vspace{.2cm}
%\begin{table}[ht]
\begin{center}
\begin{tabular}{|c|c|c|c|c|}
\hline
$SU(4) \times SU(2)$ lwv & $Q_{B}$=E & $Q_{F}$ & ${SU(4)}_{D}$ or
${SU^{*}(4)}_{D}$ & $SO(5)$
\\ \hline
*$|1,1>$        & 4     & -2    & (0,0,0)       & 14    \\ \hline

$|\onebox, \onebox>$    & $\frac{9}{2}$ & $\frac{-3}{2}$        & (1,0,0)       &
16 \\ \hline

$|\oneonebox, \twobox>$ & 5     & -1    & (0,1,0)       & 10    \\ \hline

$|\twobox, \oneonebox>$ & 5     & -1    & (2,0,0)       & 5     \\ \hline

$|\twoonebox, \twoonebox>$ & $\frac{11}{2}$     & $\frac{-1}{2}$        &
(1,1,0) & 4     \\ \hline

$|\twotwobox, \twotwobox>$      & 6     & 0     & (0,2,0)       & 1
\\ \hline
\end{tabular}
\end{center}
Table 4. Massless Graviton Supermultiplet defined by lwv
$|\Omega> = |0>$.
\vspace{.7cm}

\vspace{.2cm}
%\begin{table}[ht]
\begin{center}
\begin{tabular}{|c|c|c|c|c|}
\hline
SU(4)$\otimes$SU(2) lwv   & $Q_{B}$       & $Q_{F}$       & $SU(4)_{D}$ or
$SU^*(4)_{D}$    & SO(5)
\\ \hline
*$|\onebox, 1>$ & $\frac{9}{2}$ & -2    & (1,0,0)       & 14    \\ \hline

$|\twobox, \onebox>$    & 5     & $\frac{-3}{2}$        & (2,0,0)
& 16    \\ \hline

$|\oneonebox, \onebox>$ & 5     & $\frac{-3}{2}$        & (0,1,0)
& 16    \\ \hline

$|\twoonebox, \twobox>$ & $\frac{11}{2}$        & -1    & (1,1,0)
& 10    \\ \hline

$|\twoonebox, \oneonebox>$      & $\frac{11}{2}$        & -1
& (1,1,0)       & 5     \\ \hline

$|\threebox, \oneonebox>$       & $\frac{11}{2}$        & -1
& (3,0,0)       & 5     \\ \hline

$|\threeonebox, \twoonebox>$    & 6     & $\frac{-1}{2}$
& (2,1,0)       & 4     \\ \hline

$|\twotwobox, \twoonebox>$      & 6     & $\frac{-1}{2}$
& (0,2,0)       & 4     \\ \hline

$|\threetwobox, \twotwobox>$    & $\frac{13}{2}$        & 0
& (1,2,0)       & 1     \\ \hline

*$|1, \onebox>$ & 4     & $\frac{-3}{2}$        & (0,0,0)
& 16    \\ \hline

$|\onebox, \oneonebox>$ & $\frac{9}{2}$ & -1    & (1,0,0)
& 5     \\ \hline

$|\onebox, \twobox>$    & $\frac{9}{2}$ & -1    & (1,0,0)
& 10    \\ \hline

$|\twobox, \twoonebox>$ & 5     & $\frac{-1}{2}$        & (2,0,0)
& 4     \\ \hline

$|\oneonebox, \twoonebox>$      & 5     & $\frac{-1}{2}$
& (0,1,0)       & 4     \\ \hline

$|\twoonebox, \twotwobox>$      & $\frac{11}{2}$        & 0
& (1,1,0)       & 1     \\ \hline
\end{tabular}
\end{center}
Table 5. Massless Supermultiplet defined by the lwv $|\Omega>=
|\sonebox>$.
\vspace{.7cm}

\vspace{.2cm}
%\begin{table}[ht]
\begin{center}
\begin{tabular}{|c|c|c|c|c|}
\hline
SU(4)$\otimes$SU(2) lwv & $Q_{B}$       & $Q_{F}$       & $SU(4)_{D}$ or
$SU^*(4)_{D}$ & SO(5)
\\ \hline
*$|\oneonebox, 1>$     & 5     & -2    & (0,1,0)        & 14    \\ \hline

$|\twoonebox, \onebox>$       & $\frac{11}{2}$        & $\frac{-3}{2}$
& (1,1,0)       & 16    \\ \hline

$|\twotwobox, \twobox>$       & 6     & -1    & (0,2,0) & 10    \\ \hline

$|\threeonebox, \oneonebox>$  & 6     & -1    & (2,1,0) &5      \\ \hline

$|\threetwobox, \twoonebox>$  & $\frac{13}{2}$        & $\frac{-1}{2}$
& (1,2,0)       & 4     \\ \hline

$|\threethreebox, \twotwobox>$        & 7     & 0     & (0,3,0) & 1     \\
\hline

*$|\onebox, \onebox>$ & $\frac{9}{2}$ & $\frac{-3}{2}$        & (1,0,0) &
16 \\ \hline

$|\twobox, \twobox>$ & 5     & -1    & (2,0,0)  & 10    \\ \hline

$|\twobox, \oneonebox>$      & 5     & -1    & (2,0,0)  & 5     \\
\hline

$|\oneonebox, \oneonebox>$   & 5     & -1    & (0,1,0)  & 5     \\ \hline

$|\oneonebox, \twobox>$ & 5 & -1 & (0,1,0) & 10 \\ \hline

$|\threebox, \twoonebox>$ & $\frac{11}{2}$   & $\frac{-1}{2}$        &
(3,0,0) & 4       \\ \hline

$|\twoonebox, \twoonebox>$   & $\frac{11}{2}$        &
$\frac{-1}{2}$ & (1,1,0)       & 4     \\ \hline

$|\twotwobox, \twotwobox>$ & 6 & 0 & (0,2,0) & 1 \\ \hline

$|\threeonebox, \twotwobox>$    & 6 & 0     & (2,1,0)   & 1     \\
\hline

*$|1, \twobox>$      & 4     & -1    & (0,0,0)  & 10    \\ \hline

$|\onebox, \twoonebox>$     & $\frac{9}{2}$ & $\frac{-1}{2}$        &
(1,0,0) & 4       \\ \hline

$|\oneonebox, \twotwobox>$  & 5     & 0     & (0,1,0)   & 1     \\ \hline

\end{tabular}
\end{center}
Table 6. Massless Supermultiplet defined by lwv $|\Omega>=
\normalsize{|\soneonebox>}$.
\vspace{.7cm}

\vspace{.2cm}
%\begin{table}[ht]
\begin{center}
\begin{tabular}{|c|c|c|c|c|}
\hline
SU(4)$\otimes$SU(2) lwv & $Q_{B}$       & $Q_{F}$       & $SU(4)_{D}$ or
$SU^*(4)_{D}$    & $SO(5)$
\\ \hline
*$|\twobox, 1>$        & 5     & -2    & (2,0,0)        & 14    \\ \hline

$|\threebox, \onebox>$        & $\frac{11}{2}$        & $\frac{-3}{2}$
& (3,0,0)       & 16    \\ \hline

$|\twoonebox, \onebox>$       & $\frac{11}{2}$        & $\frac{-3}{2}$
& (1,1,0)       & 16    \\ \hline

$|\fourbox, \oneonebox>$      & 6     & -1    & (4,0,0) & 5     \\ \hline

$|\threeonebox, \oneonebox>$  & 6     & -1    & (2,1,0) & 5     \\ \hline

$|\twotwobox, \oneonebox>$    & 6     & -1    & (0,2,0) & 5     \\ \hline

$|\threeonebox, \twobox>$     & 6     & -1    & (2,1,0) & 10    \\ \hline

$|\fouronebox, \twoonebox>$   & $\frac{13}{2}$        & $\frac{-1}{2}$
& (3,1,0)       & 4     \\ \hline

$|\threetwobox, \twoonebox>$  & $\frac{13}{2}$        & $\frac{-1}{2}$
& (1,2,0)       & 4     \\ \hline

$|\fourtwobox, \twotwobox>$  & 7     & 0     & (2,2,0)  & 1     \\ \hline

*$|\onebox, \onebox>$ & $\frac{9}{2}$ & $\frac{-3}{2}$        & (1,0,0) &
16 \\ \hline

$|\oneonebox, \oneonebox>$   & 5     & -1    & (0,1,0)  & 5     \\ \hline

$|\oneonebox, \twobox>$      & 5     & -1    & (0,1,0)  & 10    \\ \hline

$|\twobox, \oneonebox>$      & 5     & -1    & (2,0,0)  & 5     \\ \hline

$|\twobox, \twobox>$ & 5     & -1    & (2,0,0)  & 10    \\ \hline

$|\threebox, \twoonebox>$    & $\frac{11}{2}$        & $\frac{-1}{2}$
& (3,0,0)       & 4     \\ \hline

$|\twoonebox, \twoonebox>$   & $\frac{11}{2}$        & $\frac{-1}{2}$
& (1,1,0)       & 4     \\ \hline

$|\threeonebox, \twotwobox>$ & 6     & 0     & (2,1,0)  & 1     \\ \hline

$|\twotwobox, \twotwobox>$   & 6     & 0     & (0,2,0)  & 1     \\ \hline

*$|1, \oneonebox>$   & 4     & -1    & (0,0,0)  & 5     \\ \hline

$|\onebox, \twoonebox>$     & $\frac{9}{2}$ & $\frac{-1}{2}$        &
(1,0,0) & 4       \\ \hline

$|\twobox, \twotwobox>$             & 5     & 0     & (2,0,0)   & 1
\\ \hline
\end{tabular}
\end{center}
Table 7. Massless Supermultiplet Defined by the lwv
$|\Omega>=\normalsize{|\stwobox>}$.
\vspace{.7cm}

\vspace{.2cm}
%\begin{table}[ht]
\small{
\begin{center}
\begin{tabular}{|c|c|c|c|c|}
\hline
SU(4)$\otimes$SU(2) lwv & $Q_{B}$       & $Q_{F}$       & $SU(4)_{D}$ or
$SU^*(4)_{D}$    & $SO(5)$
\\ \hline
*$|\threebox, 1>$      & $\frac{11}{2}$        & -2    & (3,0,0)
& 14    \\ \hline

$|\fourbox, \onebox>$ & 6     & $\frac{-3}{2}$        & (4,0,0) & 16    \\
\hline

$|\threeonebox, \onebox>$     & 6     & $\frac{-3}{2}$        & (2,1,0) &
16 \\   \hline

$|\fivebox, \oneonebox>$      & $\frac{13}{2}$        & -1    & (5,0,0) &
 5 \\ \hline

$|\fouronebox, \oneonebox>$   & $\frac{13}{2}$        & -1    & (3,1,0) &
 5      \\ \hline

$|\threetwobox, \oneonebox>$  & $\frac{13}{2}$        & -1    & (1,2,0) &
 5      \\ \hline

$|\fouronebox, \twobox>$      & $\frac{13}{2}$        & -1      & (3,1,0)
& 10    \\ \hline

$|\fiveonebox, \twoonebox>$   & 7     & $\frac{-1}{2}$        & (4,1,0)
& 4     \\ \hline

$|\fourtwobox, \twoonebox>$   & 7     & $\frac{-1}{2}$        & (2,2,0)
& 4     \\ \hline

$|\fivetwobox, \twotwobox>$  & $\frac{15}{2}$        & 0     & (3,2,0)
& 1     \\ \hline

*$|\twobox, \onebox>$ & 5     & $\frac{-3}{2}$        & (2,0,0) & 16    \\
\hline

$|\threebox, \twobox>$       & $\frac{11}{2}$        & -1    & (3,0,0)
& 10    \\ \hline

$|\twoonebox, \twobox>$      & $\frac{11}{2}$        & -1    & (1,1,0)
& 10    \\ \hline

$|\threebox, \oneonebox>$    & $\frac{11}{2}$        & -1    & (3,0,0)
& 5     \\ \hline

$|\twoonebox, \oneonebox>$   & $\frac{11}{2}$        & -1    & (1,1,0)  & 5
\\ \hline

$|\fourbox, \twoonebox>$     & 6     & $\frac{-1}{2}$        & (4,0,0)  & 4
\\ \hline

$|\threeonebox, \twoonebox>$ & 6     & $\frac{-1}{2}$        & (2,1,0)  & 4
\\ \hline

$|\twotwobox, \twoonebox>$   & 6     & $\frac{-1}{2}$        & (0,2,0)  & 4
\\ \hline

$|\fouronebox, \twotwobox>$  & $\frac{13}{2}$        & 0     & (3,1,0)  & 1
\\ \hline

$|\threetwobox, \twotwobox>$        & $\frac{13}{2}$        & 0     &
(1,2,0) & 1       \\ \hline

*$|\onebox, \oneonebox>$     & $\frac{9}{2}$ & -1    & (1,0,0)  & 5     \\
\hline
$|\twobox, \twoonebox>$     & 5     & $\frac{-1}{2}$        & (2,0,0)   & 4
\\ \hline

$|\oneonebox, \twoonebox>$  & 5     & $\frac{-1}{2}$        & (0,1,0)   & 4
\\ \hline

$|\threebox, \twotwobox>$   & $\frac{11}{2}$        & 0     & (3,0,0)   & 1
\\ \hline

$|\twoonebox, \twotwobox>$  & $\frac{11}{2}$        & 0     & (1,1,0)   & 1
\\ \hline
\end{tabular}
\end{center}}
Table 8. Massless Supermultiplet Defined by the lwv
$|\Omega>=\normalsize{|\sthreebox>}$.
\vspace{.7cm}

\vspace{.2cm}
%\begin{table}[ht]
\begin{center}
\begin{tabular}{|c|c|c|c|c|}
\hline
SU(4)$\otimes$SU(2) lwv & $Q_{B}$       & $Q_{F}$       & $SU(4)_{D}$ or
$SU^*(4)_{D}$    & $SO(5)$
\\ \hline
*$|\twoonebox, 1>$     & $\frac{11}{2}$        & -2    & (1,1,0)        & 14
\\ \hline

$|\threeonebox, \onebox>$     & 6     & $\frac{-3}{2}$        & (2,1,0) &
16 \\ \hline

$|\twotwobox, \onebox>$       & 6     & $\frac{-3}{2}$        & (0,2,0) &
16 \\ \hline

$|\fouronebox, \oneonebox>$   & $\frac{13}{2}$        & -1    & (3,1,0) &
5 \\ \hline

$|\threetwobox, \oneonebox>$  & $\frac{13}{2}$        & -1    & (1,2,0) &
5 \\ \hline

$|\threetwobox, \twobox>$     & $\frac{13}{2}$        & -1    & (1,2,0) &
10 \\ \hline

$|\fourtwobox, \twoonebox>$   & 7     & $\frac{-1}{2}$        & (2,2,0) &
4 \\ \hline

$|\threethreebox, \twoonebox>$        & 7     & $\frac{-1}{2}$        &
(0,3,0) & 4       \\ \hline

$|\fourthreebox, \twotwobox>$ & $\frac{15}{2}$        & 0     & (1,3,0) &
1 \\ \hline

*$|\twobox, \onebox>$ & 5     & $\frac{-3}{2}$        & (2,0,0) & 16
\\ \hline

$|\twoonebox, \twobox>$ & $\frac{11}{2}$ & -1 & (1,1,0) & 10 \\ \hline

$|\threebox, \twobox>$       & $\frac{11}{2}$        & -1    & (3,0,0)  &
10 \\ \hline

$|\threebox, \oneonebox>$    & $\frac{11}{2}$        & -1    & (3,0,0)  &
5 \\ \hline

$|\twoonebox, \oneonebox>$   & $\frac{11}{2}$        & -1    & (1,1,0)  &
5 \\ \hline

$|\fourbox, \twoonebox>$     & 6     & $\frac{-1}{2}$        & (4,0,0)  &
4 \\ \hline

$|\threeonebox, \twoonebox>$ & 6     & $\frac{-1}{2}$        & (2,1,0)  &
4 \\ \hline

$|\twotwobox, \twoonebox>$ & 6 & $\frac{-1}{2}$ & (0,2,0) & 4 \\ \hline

$|\fouronebox, \twotwobox>$  & $\frac{13}{2}$        & 0     & (3,1,0)  &
1 \\ \hline

$|\threetwobox, \twotwobox>$ & $\frac{13}{2}$ & 0 & (1,2,0) & 1 \\ \hline

*$|\onebox, \oneonebox>$     & $\frac{9}{2}$ & -1    & (1,0,0)  & 5     \\
\hline

$|\twobox, \twoonebox>$     & 5     & $\frac{-1}{2}$        & (2,0,0)
& 4     \\ \hline

$|\oneonebox, \twoonebox>$  & 5     & $\frac{-1}{2}$        & (0,1,0)   & 4
\\ \hline

$|\threebox, \twotwobox>$   & $\frac{11}{2}$        & 0     & (3,0,0)   & 1
\\ \hline

$|\twoonebox, \twotwobox>$ & $\frac{11}{2}$ & 0 & (1,1,0) & 1 \\ \hline
\end{tabular}
\end{center}
Table 9. Massless Supermultiplet Defined by the lwv $|\Omega>=
\normalsize{|\stwoonebox>}$(cont'd).
\vspace{.7cm}

\vspace{.2cm}
%\begin{table}[ht]
\begin{center}
\begin{tabular}{|c|c|c|c|c|}
\hline
SU(4)$\otimes$SU(2) lwv & $Q_{B}$       & $Q_{F}$       & $SU(4)_{D}$ or
$SU^*(4)_{D}$    & $SO(5)$
\\ \hline
*$|1, \twoonebox>$  & 4     & $\frac{-1}{2}$        & (0,0,0)   & 4     \\
\hline

$|\onebox, \twotwobox>$    & $\frac{9}{2}$ & 0     & (1,0,0)    & 1     \\
\hline

*$|\oneonebox, \onebox>$   & 5     & $\frac{-3}{2}$        & (0,1,0)
& 16    \\ \hline

*$|\onebox, \twobox>$     & $\frac{9}{2}$ & -1    & (1,0,0)     & 10    \\
\hline
\end{tabular}
\end{center}
Table 9. Massless Supermultiplet defined by the lwv $|\Omega>=
\normalsize{|\stwoonebox>}$.
\vspace{.7cm}

\vspace{.2cm}
%\begin{table}[ht]
\begin{center}
\begin{tabular}{|c|c|c|c|}
\hline
SU(4)$\otimes$SU(2) lwv & $Q_{B}$       & $SU(4)_{D}$ or $SU^*(4)_{D}$  &
$SO(5)$ \\ \hline
*$|\threeonebox, 1>$   & 6     & (2,1,0)        & 14    \\ \hline

$|\fouronebox, \onebox>$      & $\frac{13}{2}$        & (3,1,0) & 16
\\ \hline

$|\threetwobox, \onebox>$     & $\frac{13}{2}$        & (1,2,0) & 16
\\ \hline

$|\fiveonebox, \oneonebox>$   & 7     & (4,1,0) & 5     \\ \hline

$|\fourtwobox, \oneonebox>$   & 7     & (2,2,0) & 5     \\ \hline

$|\threethreebox, \oneonebox>$        & 7     & (0,3,0) & 5     \\ \hline

$|\fourtwobox, \twobox>$      & 7     & (2,2,0) & 10    \\ \hline

$|\fivetwobox, \twoonebox>$   & $\frac{15}{2}$        & (3,2,0) & 4
\\ \hline

$|\fourthreebox, \twoonebox>$ & $\frac{15}{2}$        & (1,3,0) & 4
\\ \hline

$|\fivethreebox, \twotwobox>$  & 8  & (2,3,0)        & 1\\  \hline

*$|\twoonebox, \onebox>$    & $\frac{11}{2}$        & (1,1,0)   & 16    \\
\hline

$|\twotwobox, \twobox>$ & 6 & (0,2,0) & 10 \\ \hline

$|\threeonebox, \oneonebox>$       & 6     & (2,1,0)    & 5     \\ \hline

$|\twotwobox, \oneonebox>$ & 6     & (0,2,0)    & 5     \\ \hline

$|\threetwobox, \twoonebox>$       & $\frac{13}{2}$        & (1,2,0)    & 4
\\ \hline

$|\fouronebox, \twoonebox>$ & $\frac{13}{2}$ & (3,1,0) & 4 \\ \hline
\end{tabular}
\end{center}
Table 10. Massless Supermultiplet defined by the lwv $|\Omega>=
\normalsize{|\sthreeonebox>}$(cont'd).
\vspace{.7cm}

\vspace{.2cm}
%\begin{table}[ht]
\begin{center}
\begin{tabular}{|c|c|c|c|}
\hline
SU(4)$\otimes$SU(2) lwv & $Q_{B}$       & $SU(4)_{D}$ or $SU^*(4)_{D}$  &
$SO(5)$
\\ \hline
$|\threethreebox, \twotwobox>$     & 7     & (0,3,0)    & 1     \\ \hline

*$|\threebox, \onebox>$    & $\frac{11}{2}$        & (3,0,0)    & 16    \\
\hline

$|\threeonebox, \twobox>$ & 6 & (2,1,0) & 10 \\ \hline

$|\fourbox, \twobox>$             & 6     & (4,0,0)     & 10    \\ \hline

$|\fourbox, \oneonebox>$  & 6     & (4,0,0)     & 5     \\ \hline

$|\fivebox, \twoonebox>$  & $\frac{13}{2}$        & (5,0,0)     & 4     \\
\hline

$|\fiveonebox, \twotwobox>$       & 7     & (4,1,0)     & 1     \\ \hline

$|\fourtwobox, \twotwobox>$ & 7 & (2,2,0) & 1 \\ \hline

*$|\oneonebox, \oneonebox>$       & 5     & (0,1,0)     & 5     \\ \hline

$|\twoonebox, \twoonebox>$ & $\frac{11}{2}$ & (1,1,0) & 4 \\ \hline

$|\threeonebox, \twotwobox>$ & 6 & (2,1,0) & 1 \\ \hline

*$|\twobox, \twobox>$     & 5     & (2,0,0)      & 10    \\ \hline

$|\threebox, \twoonebox>$        & $\frac{11}{2}$        & (3,0,0)      & 4
\\ \hline

*$|\twobox, \oneonebox>$           & 5     & (2,0,0)    & 5     \\ \hline

$|\twotwobox, \twotwobox>$ & 6 & (0,2,0) & 1 \\ \hline

$|\fourbox, \twotwobox>$ & 6 & (4,0,0) & 1 \\ \hline

*$|\onebox, \twoonebox>$   & $\frac{9}{2}$ & (1,0,0)     & 4     \\ \hline

$|\twobox, \twotwobox>$   & 5     & (2,0,0)     & 1     \\ \hline

$|\oneonebox, \twotwobox>$        & 5     & (0,1,0)     & 1     \\ \hline
\end{tabular}
\end{center}
Table 10. Massless Supermultiplet defined by the lwv $|\Omega> =
\normalsize{|\sthreeonebox>}$.
\vspace{.7cm}

\vspace{.2cm}
%\begin{table}[ht]
\begin{center}
\begin{tabular}{|c|c|c|c|}
\hline
SU(4)$\otimes$SU(2) lwv & $Q_{B}$       & $SU(4)_{D}$ or $SU^*(4)_{D}$  &
$SO(5)$
\\ \hline
*$|\fouronebox, 1>$    & $\frac{13}{2}$        & (3,1,0)        & 14    \\
\hline

$|\fiveonebox, \onebox>$      & 7     & (4,1,0) & 16    \\ \hline

$|\fourtwobox, \onebox>$      & 7     & (2,2,0) & 16    \\ \hline

$|\sixonebox, \oneonebox>$    & $\frac{15}{2}$        & (5,1,0) &
5 \\ \hline

$|\fivetwobox, \oneonebox>$   & $\frac{15}{2}$        & (3,2,0) & 5
\\ \hline

$|\fourthreebox, \oneonebox>$ & $\frac{15}{2}$        & (1,3,0) & 5
\\ \hline

$|\fivetwobox, \twobox>$      & $\frac{15}{2}$        & (3,2,0) & 10
\\ \hline

$|\sixtwobox, \twoonebox>$    & 8     & (4,2,0) & 4     \\ \hline

$|\fivethreebox, \twoonebox>$ & 8     & (2,3,0) & 4     \\ \hline

$|\sixthreebox, \twotwobox>$ & $\frac{17}{2}$        & (3,3,0)  & 1     \\
\hline

*$|\threeonebox, \onebox>$  & 6     & (2,1,0)   & 16    \\ \hline

$|\fouronebox, \twobox>$ & $\frac{13}{2}$ & (3,1,0) & 10 \\ \hline

$|\fouronebox, \oneonebox>$        & $\frac{13}{2}$        & (3,1,0)    & 5
\\ \hline

$|\threetwobox, \twobox>$  & $\frac{13}{2}$        & (1,2,0)    & 10    \\
\hline

$|\threetwobox, \oneonebox>$       & $\frac{13}{2}$        & (1,2,0)    & 5
\\ \hline

$|\fivetwobox, \twotwobox>$ & $\frac{15}{2}$ & (3,2,0) & 1 \\ \hline

$|\fiveonebox, \twoonebox>$ & 7 & (4,1,0) & 4 \\ \hline

$|\threethreebox, \twoonebox>$     & 7     & (0,3,0)    & 4     \\ \hline

$|\fourtwobox, \twoonebox>$        & 7     & (2,2,0)    & 4     \\ \hline

$|\fourthreebox, \twotwobox>$      & $\frac{15}{2}$        & (1,3,0)    & 1
\\ \hline

*$|\twoonebox, \oneonebox>$        & $\frac{11}{2}$        & (1,1,0)    & 5
\\ \hline
\end{tabular}
\end{center}
Table 11. Massless Supermultiplet Defined by the lwv $|\Omega> =
\normalsize{|\sfouronebox>}$ (cont'd).
\vspace{.7cm}

\vspace{.2cm}
%\begin{table}[ht]
\begin{center}
\begin{tabular}{|c|c|c|c|}
\hline
SU(4)$\otimes$SU(2) lwv & $Q_{B}$       & $SU(4)_{D}$ or $SU^*(4)_{D}$  &
$SO(5)$
\\ \hline
$|\threeonebox, \twoonebox>$ & 6 & (2,1,0) & 4 \\ \hline

$|\twotwobox, \twoonebox>$        & 6     & (0,2,0)     & 4     \\ \hline

$|\fouronebox, \twotwobox>$ & $\frac{13}{2}$ & (3,1,0) & 1 \\ \hline

*$|\threebox, \twobox>$   & $\frac{11}{2}$        & (3,0,0)     & 10    \\
\hline

$|\fourbox, \twoonebox>$ & 6     & (4,0,0)      & 4     \\ \hline

*$|\twobox, \twoonebox>$   & 5     & (2,0,0)    & 4     \\ \hline

$|\threebox, \twotwobox>$ & $\frac{11}{2}$        & (3,0,0)     & 1     \\
\hline

$|\twoonebox, \twotwobox>$        & $\frac{11}{2}$        & (1,1,0)     & 1
\\ \hline

*$|\fourbox, \onebox>$      & 6     & (4,0,0)   & 16    \\ \hline

$|\fivebox, \twobox>$      & $\frac{13}{2}$        & (5,0,0)    & 10    \\
\hline

$|\fivebox, \oneonebox>$   & $\frac{13}{2}$        & (5,0,0)    & 5     \\
\hline

$|\sixbox, \twoonebox>$    & 7     & (6,0,0)    & 4     \\ \hline

$|\sixonebox, \twotwobox>$ & $\frac{15}{2}$        & (5,1,0)    & 1     \\
\hline

*$|\threebox, \oneonebox>$ & $\frac{11}{2}$        & (3,0,0)    & 5     \\
\hline

$|\fivebox, \twotwobox>$ & $\frac{13}{2}$ & (5,0,0) & 1 \\ \hline

$|\threetwobox, \twotwobox>$ & $\frac{13}{2}$ & (1,2,0) & 1 \\ \hline
\end{tabular}
\end{center}
Table 11. Massless Supermultiplet Defined by the lwv $|\Omega> =
\normalsize{|\sfouronebox>}$.
\vspace{.7cm}

\vspace{.2cm}
%\begin{table}[ht]
\begin{center}
\begin{tabular}{|c|c|c|c|c|}
\hline
SU(4)$\otimes$SU(2) lwv & $Q_{B}$       & $Q_{F}$       & $SU(4)_{D}$ or
$SU^*(4)_{D}$ & $SO(5)$
\\ \hline
*$|\twotwobox, 1>$     & 6     & -2    & (0,2,0)        & 14
\\ \hline

$|\threetwobox, \onebox>$     & $\frac{13}{2}$        & $\frac{-3}{2}$
& (1,2,0)       & 16    \\ \hline

$|\fourtwobox, \oneonebox>$   & 7     & -1    & (2,2,0) & 5     \\ \hline

$|\threethreebox, \twobox>$   & 7     & -1    & (0,3,0) & 10    \\ \hline

$|\fourthreebox, \twoonebox>$ & $\frac{15}{2}$        & $\frac{-1}{2}$
& (1,3,0)       & 4     \\ \hline

$|\fourfourbox, \twotwobox>$  & 8     & 0     & (0,4,0) & 1     \\ \hline

*$|\twoonebox, \onebox>$      & $\frac{11}{2}$        & $\frac{-3}{2}$
& (1,1,0)       & 16    \\ \hline

$|\twotwobox, \twobox>$ & 6 & -1 & (0,2,0) & 10 \\ \hline

$|\threeonebox, \oneonebox>$ & 6     & -1    & (2,1,0)  & 5     \\ \hline

$|\twotwobox, \oneonebox>$   & 6     & -1    & (0,2,0)  & 5     \\ \hline

$|\threeonebox, \twobox>$    & 6     & -1    & (2,1,0)  & 10    \\ \hline

$|\threetwobox, \twoonebox>$ & $\frac{13}{2}$        & $\frac{-1}{2}$
& (1,2,0)       & 4     \\ \hline

$|\fouronebox, \twoonebox>$  & $\frac{13}{2}$        & $\frac{-1}{2}$
& (3,1,0)       & 4     \\ \hline

$|\fourtwobox, \twotwobox>$ & 7 & 0 & (2,2,0) & 1 \\ \hline

$|\threethreebox, \twotwobox>$ & 7 & 0  & (0,3,0) & 1 \\ \hline

*$|\twobox, \oneonebox>$     & 5     & -1    & (2,0,0)  & 5     \\ \hline

$|\threebox, \twoonebox>$   & $\frac{11}{2}$        & $\frac{-1}{2}$
& (3,0,0)       & 4     \\ \hline

$|\twoonebox, \twoonebox>$ & $\frac{11}{2}$ & $\frac{-1}{2}$ & (1,1,0) & 4 \\ \hline

$|\fourbox, \twotwobox>$    & 6     & 0     & (4,0,0)   & 1     \\ \hline

$|\twotwobox, \twotwobox>$ & 6 & 0 & (0,2,0) & 1 \\ \hline

$|\threeonebox, \twotwobox>$    & 6 & 0 & (2,1,0)   & 1     \\
\hline

*$|\oneonebox, \twobox>$            & 5     & -1    & (0,1,0)   & 10    \\
\hline
\end{tabular}
\end{center}
Table 12. Massless Supermultiplet defined by the lwv $|\Omega>=
\normalsize{|\stwotwobox>}$ (cont'd).
\vspace{.7cm}

\vspace{.2cm}
%\begin{table}[ht]
\begin{center}
\begin{tabular}{|c|c|c|c|c|}
\hline
SU(4)$\otimes$SU(2) lwv & $Q_{B}$       & $Q_{F}$       & $SU(4)_{D}$ or
$SU^*(4)_{D}$ & $SO(5)$
\\ \hline
*$|\onebox, \twoonebox>$   & $\frac{9}{2}$ & $\frac{-1}{2}$        &
(1,0,0) & 4       \\ \hline

$|\oneonebox, \twotwobox>$ & 5 & 0 & (0,1,0) & 1 \\ \hline

$|\twobox, \twotwobox>$        & 5     & 0     & (2,0,0)
& 1     \\ \hline

*$|1, \twotwobox>$        & 4     & 0     & (0,0,0)     & 1     \\ \hline
\end{tabular}
\end{center}
Table 12. Massless Supermultiplet defined by the lwv $|\Omega>=
\normalsize{|\stwotwobox>}$.
\vspace{.7cm}

\vspace{.2cm}
%\begin{table}[ht]
\begin{center}
\begin{tabular}{|c|c|c|c|c|}
\hline
SU(4)$\otimes$SU(2) lwv & $Q_{B}$       & $Q_{F}$       & $SU(4)_{D}$ or
$SU^*(4)_{D}$    & $SO(5)$
\\ \hline
*$|\marctwojbox, 1>$   & j+4   & -2    & (2j,0,0)       & 14    \\ \hline

$|\marctwojplusonebox, \onebox>$      & j+$\frac{9}{2}$       &
$\frac{-3}{2}$  & (2j+1,0,0)    & 16    \\ \hline

$|\marctwojonebox, \onebox>$  & j+$\frac{9}{2}$       & $\frac{-3}{2}$
& (2j-1,1,0)    & 16    \\ \hline

$|\marctwojplustwobox, \oneonebox>$   & j+5   & -1    & (2j+2,0,0)      & 5
\\ \hline

$|\marctwojplusoneonebox, \oneonebox>$        & j+5   & -1    &
(2j,1,0)        & 5     \\ \hline

$|\marctwojtwobox, \oneonebox>$       & j+5   & -1    & (2j-2,2,0)      & 5
\\ \hline

$|\marctwojplusoneonebox, \twobox>$   & j+5   & -1    & (2j,1,0)        & 10
\\ \hline

$|\marctwojplustwoonebox, \twoonebox>$        & j+$\frac{11}{2}$      &
$\frac{-1}{2}$  & (2j+1,1,0)    & 4     \\ \hline

$|\marctwojplusonetwobox, \twoonebox>$        & j+$\frac{11}{2}$      &
$\frac{-1}{2}$  & (2j-1,2,0)    & 4     \\ \hline

$|\marctwojplustwotwobox, \twotwobox>$       & j+6   & 0     & (2j,2,0) &
1       \\ \hline

*$|\marctwojminusonebox, \onebox>$    & j+$\frac{7}{2}$       &
$\frac{-3}{2}$  & (2j-1,0,0)    & 16    \\ \hline

$|\marctwojbox, \twobox>$    & j+4   & -1    & (2j,0,0) & 10    \\ \hline

$|\marctwojminusoneonebox, \twobox>$ & j+4   & -1    & (2j-2,1,0)       & 10
\\ \hline

$|\marctwojbox, \oneonebox>$ & j+4   & -1    & (2j,0,0) & 5     \\ \hline

$|\marctwojminusoneonebox, \oneonebox>$      & j+4   & -1    &
(2j-2,1,0)      & 5
\\ \hline
\end{tabular}
\end{center}
Table 13. Massless Supermultiplet Defined by lwv
$|\Omega>=\normalsize{|\smarctwojbox>}$ (cont'd).
\vspace{.7cm}

\vspace{.2cm}
%\begin{table}[ht]
\begin{center}
\begin{tabular}{|c|c|c|c|c|}
\hline
SU(4)$\otimes$SU(2) lwv   & $Q_{B}$       & $Q_{F}$       & $SU(4)_{D}$ or
$SU^*(4)_{D}$    & $SO(5)$
\\ \hline
$|\marctwojplusonebox, \twoonebox>$  & j+$\frac{9}{2}$       &
$\frac{-1}{2}$  & (2j+1,0,0)    & 4     \\ \hline

$|\marctwojonebox, \twoonebox>$      & j+$\frac{9}{2}$       &
$\frac{-1}{2}$  & (2j-1,1,0)    & 4     \\ \hline

$|\marctwojminusonetwobox, \twoonebox>$      & j+$\frac{9}{2}$       &
$\frac{-1}{2}$  & (2j-3,2,0)    & 4     \\ \hline

$|\marctwojplusoneonebox, \twotwobox>$       & j+5   & 0     & (2j,1,0) &
1       \\ \hline

$|\marctwojtwobox, \twotwobox>$     & j+5   & 0     & (2j-2,2,0)        & 1
\\ \hline

*$|\marctwojminustwobox, \oneonebox>$        & j+3   & -1    &
(2j-2,0,0)      & 5     \\ \hline

$|\marctwojminusonebox, \twoonebox>$        & j+$\frac{7}{2}$       &
$\frac{-1}{2}$  & (2j-1,0,0)    & 4     \\ \hline

$|\marctwojminustwoonebox, \twoonebox>$     & j+$\frac{7}{2}$       &
$\frac{-1}{2}$  & (2j-3,1,0)    & 4     \\ \hline

$|\marctwojbox, \twotwobox>$        & j+4   & 0     & (2j,0,0)  & 1     \\
\hline

$|\marctwojminusoneonebox, \twotwobox>$     & j+4       &
0       & (2j-2,1,0)    & 1     \\ \hline

$|\marctwojminustwotwobox, \twotwobox>$     & j+4   & 0     &
(2j-4,2,0)      & 1     \\ \hline
\end{tabular}
\end{center}
Table 13. General Doubleton Supermultiplet defined by the lwv
$|\Omega>=\normalsize{|\smarctwojbox>}$ .
\vspace{.7cm}

\vspace{.2cm}
%\begin{table}[ht]
\begin{center}
\begin{tabular}{|c|c|c|c|c|}
\hline
SU(4)$\otimes$SU(2) lwv & $Q_{B}$       & $Q_{F}$       & $SU(4)_{D}$ or
$SU^*(4)_{D}$    & $SO(5)$
\\ \hline
*$|\smarctwojtwojbox, 1>$      & 2j+4  & -2    & (0,2j,0)       & 14    \\
\hline

$|\smarctwojplusonetwojbox, \onebox>$ & 2j+$\frac{9}{2}$      &
$\frac{-3}{2}$  & (1,2j,0)      & 16    \\ \hline

$|\marctwojplustwotwojbox, \oneonebox>$       & 2j+5  & -1    &
(2,2j,0)        & 5     \\ \hline

$|\marctwojplusonetwojplusonebox, \twobox>$   & 2j+5  & -1    &
(0,2j+1,0)      & 10 \\ \hline

$|\marctwojplustwotwojplustwobox, \twotwobox>$        & 2j+6  & 0     &
(0,2j+2,0)      & 1 \\ \hline

$|\marctwojplustwotwojplusonebox, \twoonebox>$        &
2j+$\frac{11}{2}$       &       $\frac{-1}{2}$  & (1,2j+1,0)    & 4     \\
\hline

*$|\marctwojminusonetwojminusonebox, \twobox>$      & 2j+3  & -1    &
(0,2j-1,0)      & 10 \\ \hline

$|\marctwojtwojminusonebox, \twoonebox>$   & 2j+$\frac{7}{2}$      &
$\frac{-1}{2}$  & (1,2j-1,0)    & 4     \\ \hline

$|\marctwojtwojbox, \twotwobox>$   & 2j+4  & 0     & (0,2j,0)   & 1     \\
\hline

*$|\marctwojtwojminusonebox, \onebox>$     & 2j+$\frac{7}{2}$      &
$\frac{-3}{2}$  & (1,2j-1,0)    & 16    \\ \hline

$|\marctwojtwojbox, \twobox>$      & 2j+4  & -1    &
(0,2j,0)      & 10 \\ \hline

$|\marctwojplusonetwojminusonebox, \twobox>$ & 2j+4 & -1 & (2,2j-1,0) & 10 \\ \hline

$|\marctwojplusonetwojminusonebox, \oneonebox>$   & 2j+4  & -1    &
(2,2j-1,0)      & 5       \\ \hline
\end{tabular}
\end{center}
Table 14. Massless Supermultiplet Defined by the lwv $|\Omega>=
\normalsize{|\smarctwojtwojbox>}$ (cont'd).
\vspace{.7cm}

\vspace{.2cm}
%\begin{table}[ht]
\begin{center}
\begin{tabular}{|c|c|c|c|c|}
\hline
SU(4)$\otimes$SU(2) lwv & $Q_{B}$       & $Q_{F}$       & $SU(4)_{D}$ or
$SU^*(4)_{D}$    & $SO(5)$
\\ \hline
$|\marctwojtwojbox, \oneonebox>$  & 2j+4  & -1    & (0,2j,0)    & 5     \\
\hline

$|\marctwojplustwotwojminusonebox, \twoonebox>$   &
2j+$\frac{9}{2}$        &       $\frac{-1}{2}$  & (3,2j-1,0)    & 4     \\
\hline

$|\marctwojplusonetwojbox, \twoonebox>$   & 2j+$\frac{9}{2}$      &
$\frac{-1}{2}$  & (1,2j,0)      & 4     \\ \hline

$|\marctwojplustwotwojbox, \twotwobox>$   & 2j+5  & 0     & (2,2j,0)    & 1
\\ \hline

$|\marctwojplusonetwojplusonebox, \twotwobox>$ & 2j+5 & 0 & (0,2j+1,0) & 1 \\ \hline

*$|\marctwojminustwotwojminustwobox, \twotwobox>$ & 2j+2  & 0     &
(0,2j-2,0)      & 1       \\ \hline

*$|\marctwojminusonetwojminustwobox, \twoonebox>$  &
2j+$\frac{5}{2}$        & $\frac{-1}{2}$        & (1,2j-2,0)
& 4     \\ \hline

$|\marctwojminusonetwojminusonebox, \twotwobox>$ & 2j+3 & 0 & (0,2j-1,0) & 4 \\ \hline

$|\marctwojtwojminustwobox, \twotwobox>$  & 2j+3  & 0     & (2,2j-2,0)  &
1       \\ \hline

*$|\marctwojtwojminustwobox, \oneonebox>$   & 2j+3  & -1    &
(2,2j-2,0)      & 5     \\ \hline

$|\marctwojplusonetwojminustwobox, \twoonebox>$    &
2j+$\frac{7}{2}$        &       $\frac{-1}{2}$  & (3,2j-2,0)    & 4     \\
\hline

$|\marctwojplustwotwojminustwobox, \twotwobox>$    & 2j+4  & 0     &
(4,2j-2,0)      & 1       \\ \hline

$|\marctwojplusonetwojminusonebox, \twotwobox>$ & 2j+4 & 0 & (2,2j-1,0) & 1 \\ \hline
\end{tabular}
\end{center}
Table 14. Massless Supermultiplet Defined by the lwv $|\Omega>=
\normalsize{|\smarctwojtwojbox>}$.
\vspace{.7cm}

\vspace{.2cm}
%\begin{table}[ht]
\begin{center}
\begin{tabular}{|c|c|c|c|}
\hline
SU(4)$\otimes$SU(2) lwv & $Q_{B}$       & $SU(4)_{D}$ or $SU^*(4)_{D}$  &
$SO(5)$
\\ \hline
*$|\marctwojplusonetwojbox, 1>$        & 2j+$\frac{9}{2}$      &
(1,2j,0)        & 14    \\ \hline

$|\marctwojplustwotwojbox, \onebox>$  & 2j+5  & (2,2j,0)        & 16    \\
\hline

$|\marctwojplusonetwojplusonebox, \onebox>$   & 2j+5  & (0,2j+1,0)      & 16
\\ \hline

$|\marctwojplustwotwojplusonebox, \twobox>$   & 2j+$\frac{11}{2}$     &
(1,2j+1,0)      & 10      \\ \hline

$|\marctwojplusthreetwojbox, \oneonebox>$     & 2j+$\frac{11}{2}$     &
(3,2j,0)        & 5       \\ \hline

$|\marctwojplustwotwojplusonebox, \oneonebox>$        &
2j+$\frac{11}{2}$       & (1,2j+1,0)    & 5       \\ \hline

$|\marctwojplusthreetwojplusonebox, \twoonebox>$      & 2j+6  &
(2,2j+1,0)      & 4     \\ \hline

$|\marctwojplustwotwojplustwobox, \twoonebox>$        & 2j+6  &
(0,2j+2,0)      & 4     \\ \hline

$|\marctwojplusthreetwojplustwobox, \twotwobox>$      &
2j+$\frac{13}{2}$       & (1,2j+2,0)    & 1     \\ \hline

*$|\marctwojtwojbox, \onebox>$      & 2j+4  & (0,2j,0)  & 16    \\ \hline

$|\marctwojplusonetwojbox, \twobox>$ & 2j+$\frac{9}{2}$ & (1,2j,0) & 10 \\ \hline

$|\marctwojplusonetwojbox, \oneonebox>$    & 2j+$\frac{9}{2}$      &
(1,2j,0)        & 5       \\ \hline

$|\marctwojplusonetwojplusonebox, \twoonebox>$     & 2j+5  & (0,2j+1,0) &
4       \\ \hline

$|\marctwojplustwotwojbox, \twoonebox>$ & 2j+5 & (2,2j,0) & 4 \\ \hline
\end{tabular}
\end{center}
Table 15. Massless Supermultiplet Defined by the lwv $|\Omega> =
\normalsize{|\smarctwojplusonetwojbox>}$ where j is an          \\
integer or half-odd integer such that $j >\frac{1}{2}$ (cont'd).
\vspace{.7cm}

\vspace{.2cm}
%\begin{table}[ht]
\begin{center}
\begin{tabular}{|c|c|c|c|}
\hline
SU(4)$\otimes$SU(2) lwv & $Q_{B}$       & $SU(4)_{D}$ or $SU^*(4)_{D}$  &
$SO(5)$
\\ \hline
$|\marctwojplustwotwojplusonebox, \twotwobox>$ & 2j+$\frac{11}{2}$ & (1,2j+1,0) & 1 \\ \hline

*$|\marctwojtwojminusonebox, \twobox>$     & 2j+$\frac{7}{2}$      &
(1,2j-1,0)      & 10 \\ \hline

$|\marctwojplusonetwojminusonebox, \twoonebox>$   & 2j+4  & (2,2j-1,0)  &
4 \\ \hline

$|\marctwojtwojbox, \twoonebox>$ & 2j+4 & (0,2j,0) & 4 \\ \hline

$|\marctwojplusonetwojbox, \twotwobox>$   & 2j+$\frac{9}{2}$      &
(1,2j,0)        & 1       \\ \hline

*$|\marctwojminusonetwojminusonebox, \twoonebox>$ & 2j+3  & (0,2j-1,0)  &
4       \\ \hline

*$|\marctwojplusonetwojminusonebox, \onebox>$      & 2j+4  & (2,2j-1,0)   & 16    \\ \hline

$|\marctwojplustwotwojminusonebox, \twobox>$      &
2j+$\frac{9}{2}$        & (3,2j-1,0)    & 10      \\ \hline

$|\marctwojplustwotwojminusonebox, \oneonebox>$   &
2j+$\frac{9}{2}$        & (3,2j-1,0)    & 5       \\ \hline

$|\marctwojplusthreetwojminusonebox, \twoonebox>$ & 2j+$\frac{9}{2}$  & (4,2j-1,0)  & 4       \\ \hline

$|\marctwojplusthreetwojbox, \twotwobox>$ & 2j+$\frac{11}{2}$     &
(3,2j,0)        & 1       \\ \hline
\end{tabular}
\end{center}
Table 15. Massless Supermultiplet Defined by the lwv $|\Omega> =
\normalsize{|\smarctwojplusonetwojbox>}$ (cont'd).
\vspace{.7cm}

\vspace{.2cm}
%\begin{table}[ht]
\begin{center}
\begin{tabular}{|c|c|c|c|}
\hline
SU(4)$\otimes$SU(2) lwv & $Q_{B}$       & $SU(4)_{D}$ or $SU^*(4)_{D}$  &
$SO(5)$
\\ \hline
*$|\marctwojtwojminusonebox, \oneonebox>$   & 2j+$\frac{7}{2}$      &
(1,2j-1,0)      & 5       \\ \hline

*$|\marctwojminusonetwojminustwobox, \twotwobox>$  &
2j+$\frac{5}{2}$        & (1,2j-2,0)    & 1     \\ \hline

*$|\marctwojtwojminustwobox, \twoonebox>$ & 2j+3  & (2,2j-2,0)  & 4     \\
\hline

$|\marctwojtwojminusonebox, \twotwobox>$ & 2j+$\frac{7}{2}$ & (1,2j-1,0) & 1 \\ \hline

$|\marctwojplusonetwojminustwobox, \twotwobox>$  &
2j+$\frac{7}{2}$        & (3,2j-2,0)    & 1       \\ \hline

*$|\marctwojplusonetwojminustwobox, \oneonebox>$ &
2j+$\frac{7}{2}$        & (3,2j-2,0)    & 5       \\ \hline

$|\marctwojplustwotwojminustwobox, \twoonebox>$ & 2j+4  & (4,2j-2,0)    & 4
\\ \hline

$|\marctwojplusthreetwojminustwobox, \twotwobox>$       &
2j+$\frac{9}{2}$        & (5,2j-2,0)    & 1     \\ \hline

$|\marctwojplustwotwojminusonebox, \twotwobox>$ & 2j+$\frac{9}{2}$ & (3,2j-1,0) & 1 \\ \hline
\end{tabular}
\end{center}
Table 15. Massless Supermultiplet Defined by the lwv $|\Omega> =
\normalsize{|\smarctwojplusonetwojbox>}$.
\vspace{.7cm}

\vspace{.2cm}
%\begin{table}[ht]
\begin{center}
\begin{tabular}{|c|c|c|c|}
\hline
SU(4)$\otimes$SU(2) lwv & $Q_{B}$       & $SU(4)_{D}$ or $SU^*(4)_{D}$  &
$SO(5)$
\\ \hline
*$|\marctwojplustwotwojbox, 1>$        & 2j+5  & (2,2j,0)       & 14    \\
\hline

$|\marctwojplusthreetwojbox, \onebox>$        & 2j+$\frac{11}{2}$     &
(3,2j,0)        & 16 \\ \hline

$|\marctwojplustwotwojplusonebox, \onebox>$   & 2j+$\frac{11}{2}$     &
(1,2j+1,0)      & 16      \\ \hline

$|\marctwojplusfourtwojbox, \oneonebox>$      & 2j+6  & (4,2j,0)        & 5
\\ \hline

$|\marctwojplusthreetwojplusonebox, \oneonebox>$      & 2j+6  &
(2,2j+1,0)      & 5     \\ \hline

$|\marctwojplustwotwojplustwobox, \oneonebox>$        & 2j+6  &
(0,2j+2,0)      & 5     \\ \hline

$|\marctwojplusthreetwojplusonebox, \twobox>$ & 2j+6 & (2,2j+1,0)
& 10 \\ \hline

$|\marctwojplusfourtwojplusonebox, \twoonebox>$  &
2j+$\frac{13}{2}$ & (3,2j+1,0)    & 4      \\ \hline

$|\marctwojplusthreetwojplustwobox, \twoonebox>$      &
2j+$\frac{13}{2}$       & (1,2j+2,0)    & 4     \\ \hline

$|\marctwojplusfourtwojplustwobox, \twotwobox>$      & 2j+7  &
(2,2j+2,0)      & 1    \\ \hline

*$|\marctwojplusonetwojbox, \onebox>$       & 2j+$\frac{9}{2}$      &
(1,2j,0)        & 16 \\ \hline

$|\marctwojplustwotwojbox, \twobox>$ & 2j+5 & (2,2j,0) & 10 \\ \hline

$|\marctwojplustwotwojbox, \oneonebox>$    & 2j+5  & (2,2j,0)   & 5     \\
\hline

$|\marctwojplusonetwojplusonebox, \twobox>$        & 2j+5  & (0,2j+1,0) &
10      \\ \hline
\end{tabular}
\end{center}
Table 16. Massless Supermultiplet Defined by the lwv $|\Omega>=
\normalsize{|\smarctwojplustwotwojbox>}$ where $j$  is an  \\
integer or half-odd integer such that $ j>\frac{1}{2}$ (cont'd).
\vspace{.7cm}

\vspace{.2cm}
%\begin{table}[ht]
\begin{center}
\begin{tabular}{|c|c|c|c|}
\hline
SU(4)$\otimes$SU(2) lwv & $Q_{B}$       & $SU(4)_{D}$ or $SU^*(4)_{D}$  &
$SO(5)$
\\ \hline
$|\marctwojplusonetwojplusonebox, \oneonebox>$     & 2j+5  & (0,2j+1,0) &
5 \\ \hline

$|\marctwojplusthreetwojbox, \twoonebox>$ & 2j+$\frac{11}{2}$ & (3,2j,0) & 4 \\ \hline

$|\marctwojplustwotwojplusonebox, \twoonebox>$     &
2j+$\frac{11}{2}$       & (1,2j+1,0)    & 4       \\ \hline

$|\marctwojplustwotwojplustwobox, \twotwobox>$     & 2j+6  & (0,2j+2,0) &
1     \\ \hline

$|\marctwojplusthreetwojplusonebox, \twotwobox>$ & 2j+6 & (2,2j+1,0) & 1 \\ \hline

*$|\marctwojtwojbox, \oneonebox>$  & 2j+4  & (0,2j,0)   & 5     \\ \hline

*$|\marctwojplusonetwojminusonebox, \twobox>$     & 2j+4  & (2,2j-1,0)  &
10      \\ \hline

$|\marctwojplustwotwojbox, \twotwobox>$ & 2j+5 & (2,2j,0) & 1
\\ \hline

$|\marctwojplustwotwojminusonebox, \twoonebox>$  &
2j+$\frac{9}{2}$        & (3,2j-1,0)    & 4       \\ \hline

$|\marctwojplusonetwojbox, \twoonebox>$  & 2j+$\frac{9}{2}$      &
(1,2j,0)        & 4       \\ \hline

*$|\marctwojtwojminusonebox, \twoonebox>$  & 2j+$\frac{7}{2}$      &
(1,2j-1,0)      & 4       \\ \hline

$|\marctwojplusonetwojminusonebox, \twotwobox>$ & 2j+4 & (2,2j-1,0) & 1 \\ \hline

$|\marctwojtwojbox, \twotwobox>$  & 2j+4  & (0,2j,0)    & 1   \\ \hline

*$|\marctwojplustwotwojminusonebox, \onebox>$       &
2j+$\frac{9}{2}$        & (3,2j-1,0)    & 16      \\ \hline
\end{tabular}
\end{center}
Table 16. Massless Supermultiplet Defined by the lwv $|\Omega> =
\normalsize{|\smarctwojplustwotwojbox>}$(cont'd).
\vspace{.7cm}

\vspace{.2cm}
%\begin{table}[ht]
\begin{center}
\begin{tabular}{|c|c|c|c|}
\hline
SU(4)$\otimes$SU(2) lwv & $Q_{B}$       & $SU(4)_{D}$ or $SU^*(4)_{D}$  &
$SO(5)$
\\ \hline
$|\marctwojplusthreetwojminusonebox, \twobox>$     & 2j+5  & (4,2j-1,0) &
10 \\ \hline

$|\marctwojplusthreetwojminusonebox, \oneonebox>$  & 2j+5  & (4,2j-1,0) &
5 \\ \hline

$|\marctwojplusfourtwojminusonebox, \twoonebox>$   &
2j+$\frac{11}{2}$       & (5,2j-1,0)    & 4     \\ \hline

$|\marctwojplusfourtwojbox, \twotwobox>$   & 2j+6  & (4,2j,0)   & 1    \\
\hline

*$|\marctwojplusonetwojminusonebox, \oneonebox>$   & 2j+4  & (2,2j-1,0) &
5       \\ \hline

$|\marctwojplustwotwojbox, \twotwobox>$ & 2j+5 & (2,2j,0) & 1 \\ \hline

$|\marctwojplusthreetwojminusonebox, \twotwobox>$ & 2j+5 & (4,2j-1,0) & 1 \\ \hline

$|\marctwojplusonetwojplusonebox, \twotwobox>$ & 2j+5 & (0,2j+1,0) & 1 \\ \hline

*$|\marctwojtwojminustwobox, \twotwobox>$ & 2j+3 & (2,2j-2,0) & 1 \\ \hline

*$|\marctwojplusonetwojminustwobox, \twoonebox>$ &
2j+$\frac{7}{2}$        & (3,2j-2,0)    & 4     \\ \hline

$|\marctwojplustwotwojminustwobox, \twotwobox>$ & 2j+4  & (4,2j-2,0)    & 1
\\ \hline

*$|\marctwojplustwotwojminustwobox, \oneonebox>$   & 2j+4  & (4,2j-2,0) &
5       \\ \hline

$|\marctwojplusthreetwojminustwobox, \twoonebox>$ &
2j+$\frac{9}{2}$ & (5,2j-2,0)   & 4    \\ \hline

$|\marctwojplusfourtwojminustwobox, \twotwobox>$  & 2j+5  & (6,2j-2,0)  &
1     \\ \hline
\end{tabular}
\end{center}
Table 16. Massless Supermultiplet Defined by the lwv $|\Omega> =
\normalsize{|\smarctwojplustwotwojbox>}$.
\vspace{.7cm}

\vspace{.2cm}
%\begin{table}[ht]
\begin{center}
\begin{tabular}{|c|c|c|c|}
\hline
SU(4)$\otimes$SU(2) lwv & $Q_{B}$       & $SU(4)_{D}$ or $SU^*(4)_{D}$  &
$SO(5)$
\\ \hline
*$|\marctwojplusthreetwojbox, 1>$      & 2j+$\frac{11}{2}$     &
(3,2j,0)        & 14 \\ \hline

$|\marctwojplusfourtwojbox, \onebox>$ & 2j+6  & (4,2j,0)        & 16    \\
\hline

$|\marctwojplusthreetwojplusonebox, \onebox>$ & 2j+6  & (2,2j+1,0)      & 16
\\ \hline

$|\marctwojplusfivetwojbox, \oneonebox>$      & 2j+$\frac{13}{2}$     &
(5,2j,0)      & 5       \\ \hline

$|\marctwojplusfourtwojplusonebox, \oneonebox>$       &
2j+$\frac{13}{2}$       & (3,2j+1,0)    & 5     \\ \hline

$|\marctwojplusthreetwojplustwobox, \oneonebox>$      &
2j+$\frac{13}{2}$       & (1,2j+2,0)    & 5     \\ \hline

$|\marctwojplusfourtwojplusonebox, \twobox>$  & 2j+$\frac{13}{2}$     &
(3,2j+1,0)      & 10      \\ \hline

$|\marctwojplusfivetwojplusonebox, \twoonebox>$       & 2j+7  &
(4,2j+1,0)      & 4 \\ \hline

$|\marctwojplusfourtwojplustwobox, \twoonebox>$       & 2j+7  &
(2,2j+2,0)      & 4 \\ \hline

$|\marctwojplusfivetwojplustwobox, \twotwobox>$      &
2j+$\frac{15}{2}$       & (3,2j+2,0)    & 1     \\ \hline

*$|\marctwojplustwotwojbox, \onebox>$       & 2j+5  & (2,2j,0)  & 16    \\
\hline

$|\marctwojplusthreetwojbox, \twobox>$ & 2j+$\frac{11}{2}$ & (3,2j,0) & 10 \\ \hline

$|\marctwojplusthreetwojbox, \oneonebox>$  & 2j+$\frac{11}{2}$     &
(3,2j,0)        & 5       \\ \hline

$|\marctwojplustwotwojplusonebox, \twobox>$        &
2j+$\frac{11}{2}$       & (1,2j+1,0)    & 10    \\ \hline
\end{tabular}
\end{center}
Table 17. Massless Supermultiplet Defined by the lwv $|\Omega>=
\normalsize{|\smarctwojplusthreetwojbox>}$ where $j$  is an integer or
half-odd integer such that  $j >\frac{1}{2}$ (cont'd).
\vspace{.7cm}

\vspace{.2cm}
%\begin{table}[ht]
\begin{center}
\begin{tabular}{|c|c|c|c|}
\hline
SU(4)$\otimes$SU(2) lwv & $Q_{B}$       & $SU(4)_{D}$ or $SU^*(4)_{D}$  &
$SO(5)$
\\ \hline
$|\marctwojplusfourtwojplusonebox, \twotwobox>$ & 2j+$\frac{13}{2}$ & (3,2j+1,0) & 1 \\ \hline

$|\marctwojplustwotwojplusonebox, \oneonebox>$     &
2j+$\frac{11}{2}$       & (1,2j+1,0)    & 5     \\ \hline

$|\marctwojplusfourtwojbox, \twoonebox>$ & 2j+6 & (4,2j,0) & 4 \\ \hline

$|\marctwojplustwotwojplustwobox, \twoonebox>$     & 2j+6  & (0,2j+2,0) &
4 \\ \hline

$|\marctwojplusthreetwojplusonebox, \twoonebox>$   & 2j+6  & (2,2j+1,0) &
4 \\ \hline

$|\marctwojplusthreetwojplustwobox, \twotwobox>$   &
2j+$\frac{13}{2}$       & (1,2j+2,0)    & 1     \\ \hline

*$|\marctwojplusonetwojbox, \oneonebox>$   & 2j+$\frac{9}{2}$      &
(1,2j,0)        & 5       \\ \hline

$|\marctwojplustwotwojbox, \twoonebox>$ & 2j+5 & (2,2j,0) & 4 \\ \hline

$|\marctwojplusonetwojplusonebox, \twoonebox>$    & 2j+5  & (0,2j+1,0)  &
4 \\ \hline

$|\marctwojplusthreetwojbox, \twotwobox>$ & 2j+$\frac{11}{2}$ & (3,2j,0) & 1 \\ \hline

$|\marctwojplustwotwojplusonebox, \twotwobox>$ & 2j+$\frac{11}{2}$ & (1,2j+1,0) & 1 \\ \hline

*$|\marctwojplustwotwojminusonebox, \twobox>$     & 2j+$\frac{9}{2}$
& (3,2j-1,0)    & 10    \\ \hline

$|\marctwojplusthreetwojminusonebox, \twoonebox>$        & 2j+5  &
(4,2j-1,0)      & 4       \\ \hline

*$|\marctwojplusonetwojminusonebox, \twoonebox>$   & 2j+4  & (2,2j-1,0) &
4 \\ \hline

$|\marctwojplustwotwojminusonebox, \twotwobox>$ & 2j+$\frac{9}{2}$ & (3,2j-1,0) & 1 \\ \hline
\end{tabular}
\end{center}
Table 17. Massless Supermultiplet Defined by the lwv $|\Omega> =
\normalsize{|\smarctwojplusthreetwojbox>}$(cont'd).
\vspace{.7cm}

\vspace{.2cm}
%\begin{table}[ht]
\begin{center}
\begin{tabular}{|c|c|c|c|}
\hline
SU(4)$\otimes$SU(2) lwv & $Q_{B}$       & $SU(4)_{D}$ or $SU^*(4)_{D}$  &
$SO(5)$
\\ \hline
$|\marctwojplusonetwojbox, \twotwobox>$   & 2j+$\frac{9}{2}$      &
(1,2j,0)        & 1       \\ \hline

*$|\marctwojplusthreetwojminusonebox, \onebox>$     & 2j+5  &
(4,2j-1,0)      & 16 \\ \hline

$|\marctwojplusfourtwojminusonebox, \twobox>$      &
2j+$\frac{11}{2}$       & (5,2j-1,0)    & 10    \\ \hline

$|\marctwojplusfourtwojminusonebox, \oneonebox>$   &
2j+$\frac{11}{2}$       & (5,2j-1,0)    & 5     \\ \hline

$|\marctwojplusfivetwojminusonebox, \twoonebox>$   & 2j+6  & (6,2j-1,0) &
4 \\ \hline

$|\marctwojplusfivetwojbox, \twotwobox>$  &
2j+$\frac{13}{2}$ & (5,2j,0)  & 1       \\ \hline

*$|\marctwojplustwotwojminusonebox, \oneonebox>$  & 2j+$\frac{9}{2}$
& (3,2j-1,0)    & 5     \\ \hline

$|\marctwojplusfourtwojminusonebox, \twotwobox>$ & 2j+$\frac{11}{2}$ & (5,2j-1,0) & 1 \\ \hline

*$|\marctwojplusonetwojminustwobox, \twotwobox>$ & 2j+$\frac{7}{2}$
& (3,2j-2,0)    & 1     \\ \hline

*$|\marctwojplustwotwojminustwobox, \twoonebox>$   & 2j+4  & (4,2j-2,0) &
4 \\ \hline

$|\marctwojplusthreetwojminustwobox, \twotwobox>$ & 2j+$\frac{9}{2}$
& (5,2j-2,0)    & 1     \\ \hline

*$|\marctwojplusthreetwojminustwobox, \oneonebox>$  & 2j+$\frac{9}{2}$
& (5,2j-2,0)    & 5     \\ \hline

$|\marctwojplusfourtwojminustwobox, \twoonebox>$   & 2j+5  & (6,2j-2,0) &
4 \\ \hline

$|\marctwojplusfivetwojminustwobox, \twotwobox>$   &
2j+$\frac{11}{2}$       & (7,2j-2,0)    & 1     \\ \hline
\end{tabular}
\end{center}
Table 17. Massless Supermultiplet Defined by the lwv $|\Omega> =
\normalsize{|\smarctwojplusthreetwojbox>}$.
\vspace{.7cm}

\vspace{.2cm}
%\begin{table}[ht]
\begin{center}
\begin{tabular}{|c|c|c|c|}
\hline
SU(4)$\otimes$SU(2) lwv & $Q_{B}$       & $SU(4)_{D}$ or $SU^*(4)_{D}$  &
$SO(5)$
\\ \hline
*$|\marctwojplusntwojbox, 1>$  & $\frac{1}{2}$(4j+n+8) & (n,2j,0)       & 14
\\ \hline

$|\marctwojplusnplusonetwojbox, \onebox>$     & $\frac{1}{2}$(4j+n+9) &
(n+1,2j,0)    & 16      \\ \hline

$|\marctwojplusoneplusnminusonetwojplusonebox, \onebox>$      &
$\frac{1}{2}$(4j+n+9)   & (n-1,2j+1,0)    & 16    \\ \hline

$|\marctwojplusnplustwotwojbox, \oneonebox>$  & $\frac{1}{2}$(4j+n+10)
& (n+2,2j,0)  & 5     \\ \hline

$|\marctwojplusoneplusntwojplusonebox, \oneonebox>$   &
$\frac{1}{2}$(4j+n+10)  & (n,2j+1,0)      & 5     \\ \hline

$|\marctwojplustwoplusnminustwotwojplustwobox, \oneonebox>$   &
$\frac{1}{2}$(4j+n+10)  & (n-2,2j+2,0)  & 5     \\ \hline

$|\marctwojplusoneplusntwojplusonebox, \twobox>$      &
$\frac{1}{2}$(4j+n+10)  & (n,2j+1,0)      & 10    \\ \hline

$|\marctwojplusoneplusnplusonetwojplusonebox, \twoonebox>$    &
$\frac{1}{2}$(4j+n+11)   & (n+1,2j+1,0)    & 4     \\ \hline

$|\marctwojplustwoplusnminusonetwojplustwobox, \twoonebox>$   &
$\frac{1}{2}$(4j+n+11)  & (n-1,2j+2,0)  & 4     \\ \hline

$|\marctwojplustwoplusntwojplustwobox, \twotwobox>$  &
$\frac{1}{2}$(4j+n+12)  & (n,2j+2,0)    & 1     \\ \hline

*$|\marctwojplusnminusonetwojbox, \onebox>$ & $\frac{1}{2}$(4j+n+7) &
(n-1,2j,0)    & 16      \\ \hline

$|\marctwojplusntwojbox, \twobox>$ & $\frac{1}{2}$(4j+n+8) & (n,2j,0) & 10 \\ \hline

$|\marctwojplusntwojbox, \oneonebox>$      & $\frac{1}{2}$(4j+n+8) &
(n,2j,0)      & 5       \\ \hline
\end{tabular}
\end{center}
Table 18. General Massless Supermultiplet defined by the lwv
$|\Omega>=\normalsize{|\smarctwojplusntwojbox>}$ where $j$  is an
integer or half-odd integer such that $j >\frac{1}{2}$
and n is a positive integer such that $n >3 $ (cont'd).
\vspace{.7cm}

\vspace{.2cm}
%\begin{table}[ht]
\begin{center}
\begin{tabular}{|c|c|c|c|}
\hline
SU(4)$\otimes$SU(2) lwv & $Q_{B}$       & $SU(4)_{D}$ or $SU^*(4)_{D}$  &
$SO(5)$
\\ \hline
$|\marctwojplusoneplusnminustwotwojplusonebox, \twobox>$ & $\frac{1}{2}$(4j+n+8) & (n-2, 2j+1,0) & 10 \\ \hline

$|\marctwojplusoneplusnminustwotwojplusonebox, \oneonebox>$        &
$\frac{1}{2}$(4j+n+8)   & (n-2,2j+1,0)  & 5     \\ \hline

$|\marctwojplustwoplusnminusthreetwojplustwobox, \twoonebox>$      &
$\frac{1}{2}$(4j+n+9)   & (n-3,2j+2,0)  & 4     \\ \hline

$|\marctwojplusoneplusnminusonetwojplusonebox, \twoonebox>$        &
$\frac{1}{2}$(4j+n+9)   & (n-1,2j+1,0)  & 4     \\ \hline

$|\marctwojplustwoplusnminustwotwojplustwobox, \twotwobox>$        &
$\frac{1}{2}$(4j+n+10)  & (n-2,2j+2,0)  & 1     \\ \hline

$|\marctwojplusoneplusntwojplusonebox, \twotwobox>$ & $\frac{1}{2}$(4j+n+10) & (n,2j+1,0) & 1 \\ \hline

*$|\marctwojplusnminustwotwojbox, \oneonebox>$     &
$\frac{1}{2}$(4j+n+6)   & (n-2,2j,0)    & 5     \\ \hline

$|\marctwojplusoneplusnminusthreetwojplusonebox, \twoonebox>$     &
$\frac{1}{2}$(4j+n+7)   & (n-3,2j+1,0)  & 4     \\ \hline

$|\marctwojplusoneplusnminustwotwojplusonebox, \twotwobox>$ & $\frac{1}{2}$(4j+n+8) & (n-2,2j+1,0) & 1 \\ \hline

$|\marctwojplustwoplusnminusfourtwojplustwobox, \twotwobox>$      &
$\frac{1}{2}$(4j+n+8)   & (n-4,2j+2,0)  & 1     \\ \hline

*$|\marctwojminusoneplusnplustwojminusonebox, \twobox>$   &
$\frac{1}{2}$(4j+n+6)   & (n,2j-1,0)    & 10    \\ \hline

$|\marctwojminusoneplusnplusonetwojminusonebox, \twoonebox>$     &
$\frac{1}{2}$(4j+n+7)   & (n+1,2j-1,0)  & 4     \\ \hline

$|\marctwojplusnminusonetwojbox, \twoonebox>$    &
$\frac{1}{2}$(4j+n+7)   & (n-1,2j,0)    & 4     \\ \hline
\end{tabular}
\end{center}
Table 18. General Massless Supermultiplet defined by the lwv
$|\Omega>=\normalsize{|\smarctwojplusntwojbox>}$ (cont'd).
\vspace{.7cm}

\vspace{.2cm}
%\begin{table}[ht]
\begin{center}
\begin{tabular}{|c|c|c|c|}
\hline
SU(4)$\otimes$SU(2) lwv & $Q_{B}$       & $SU(4)_{D}$ or $SU^*(4)_{D}$  &
$SO(5)$
\\ \hline
$|\marctwojplusntwojbox, \twotwobox>$    & $\frac{1}{2}$(4j+n+8) &
(n,2j,0)        & 1       \\ \hline

*$|\marctwojminusoneplusnminusonetwojminusonebox, \twoonebox>$     &
$\frac{1}{2}$(4j+n+5)   & (n-1,2j-1,0)  & 4     \\ \hline

$|\marctwojminusoneplusntwojminusonebox, \twotwobox>$ & $\frac{1}{2}$(4j+n+6) & (n, 2j-1, 0) & 1 \\ \hline

$|\marctwojplusnminustwotwojbox, \twotwobox>$     &
$\frac{1}{2}$(4j+n+6)   & (n-2,2j,0)    & 1     \\ \hline

*$|\marctwojminusoneplusnplusonetwojminusonebox, \onebox>$  &
$\frac{1}{2}$(4j+n+7)   & (n+1,2j-1,0)  & 16    \\ \hline

$|\marctwojminusoneplusnplustwotwojminusonebox, \twobox>$  &
$\frac{1}{2}$(4j+n+8)   & (n+2,2j-1,0)  & 10    \\ \hline

$|\marctwojminusoneplusnplustwotwojminusonebox, \oneonebox>$       &
$\frac{1}{2}$(4j+n+8)   & (n+2,2j-1,0)  & 5     \\ \hline

$|\marctwojminusoneplusnplusthreetwojminusonebox, \twoonebox>$     &
$\frac{1}{2}$(4j+n+9)   & (n+3,2j-1,0)  & 4     \\ \hline

$|\marctwojplusnplusonetwojbox, \twoonebox>$       &
$\frac{1}{2}$(4j+n+9)   & (n+1,2j,0)    & 4     \\ \hline

$|\marctwojplusnplustwotwojbox, \twotwobox>$       &
$\frac{1}{2}$(4j+n+10) & (n+2,2j,0)     & 1      \\ \hline

*$|\marctwojminusoneplusntwojminusonebox, \oneonebox>$ &
$\frac{1}{2}$(4j+n+6)   & (n,2j-1,0)    & 5     \\ \hline

$|\marctwojminusoneplusnplustwotwojminusonebox, \twotwobox>$ & $\frac{1}{2}$(4j+n+8) & (n+2,2j-1,0) & 1 \\ \hline

*$|\marctwojminustwoplusntwojminustwobox, \twotwobox>$    &
$\frac{1}{2}$(4j+n+4)   & (n,2j-2,0)    & 1     \\ \hline

*$|\marctwojminustwoplusnplusonetwojminustwobox, \twoonebox>$    &
$\frac{1}{2}$(4j+n+5)   & (n+1,2j-2,0)  & 4     \\ \hline
\end{tabular}
\end{center}
Table 18. General Massless Supermultiplet defined by the lwv
$|\Omega>=\normalsize{|\smarctwojplusntwojbox>}$ (cont'd).
\vspace{.7cm}

\vspace{.2cm}
%\begin{table}[ht]
\begin{center}
\begin{tabular}{|c|c|c|c|}
\hline
SU(4)$\otimes$SU(2) lwv & $Q_{B}$       & $SU(4)_{D}$ or $SU^*(4)_{D}$  &
$SO(5)$
\\ \hline
$|\marctwojminustwoplusnplustwotwojminustwobox, \twotwobox>$    &
$\frac{1}{2}$(4j+n+6)   & (n+2,2j-2,0)  & 1     \\ \hline

*$|\marctwojminustwoplusnplustwotwojminustwobox, \oneonebox>$      &
$\frac{1}{2}$(4j+n+6)   & (n+2,2j-2,0)  & 5     \\ \hline

$|\marctwojminustwoplusnplusthreetwojminustwobox, \twoonebox>$    &
$\frac{1}{2}$(4j+n+7)   & (n+3,2j-2,0)  & 4     \\ \hline

$|\marctwojminustwoplusnplusfourtwojminustwobox, \twotwobox>$     &
$\frac{1}{2}$(4j+n+8)   & (n+4,2j-2,0)  & 1     \\ \hline
\end{tabular}
\end{center}
Table 18. General Massless Supermultiplet defined by the lwv
$|\Omega>=\normalsize{|\smarctwojplusntwojbox>}$.
\vspace{.7cm}

\clearpage
\section{Discussion and Conclusions}
\normalsize{
     We saw in section [4] that there exist infinitely many doubleton
supermultiplets of \OS.  The CPT-``self-conjugate'' irreducible doubleton
supermultiplet appears as gauge modes in the Kaluza-Klein spectrum of 11-d
SUGRA over $S^4$ and decouples from the spectrum.  It is the
supermultiplet of the (2,0) superconformal theory in d=6 which is believed
to be dual to M-theory over $AdS_{7}\times S^4$.  This is in complete
parallel to the duality between the large N limit of $\mathcal{N}$=4
supersymmetric $SU(N)$ Yang-Mills theory and the IIB superstring theory
over $AdS_5 \times S^5$.  The other non-CPT self-conjugate doubleton
supermultiplets of \OS ~correspond to the doubleton supermultiplets of
$SU(2,2|4)$ with non-zero central charge \cite{gmz2}.  In the case of $SU(2,2|4)$, it
was conjectured \cite{gmz2}  that these non-CPT self-conjugate doubleton
supermultiplets are related to $(p,q)$  superstrings \cite{pqstring}.  Hence, we expect the
non-CPT self-conjugate doubleton supermultiplets of \OS ~to be related to
the sector of M-theory that is dual to $(p,q)$  superstrings.

 Since the $\Psi$ are bosonic spinors one can give a dynamical realization of  the
oscillator construction  of the UIR's of \OS ~in the language of six dimensional
 twistors as was suggested for the $d=4$  conformal group in \cite{gmz2} and
 explicitly realized in \cite{cgkrz,ckr}.
     Based on results in \cite{cgkrz,ckr}, we expect that such  a dynamical
(twistorial) realization of the oscillator construction of the UIRs of \OS
to correspond to some extreme limit of M-theory over $AdS_7 \times
S^4$.

     In section [5] we gave a complete list of massless supermultiplets of \OS ~considered
as $\mathcal{N}$=4 supersymmetry algebra in $AdS_7$.  The CPT
self-conjugate massless supermultiplet is simply the graviton
supermultiplet.  The other massless supermultiplets have spin range
between 2 and 4.

     One can easily extend the above construction to build all the positive
energy massive supermultiplets of \OS ~by taking $P>2$.  The short massive
supermultiplets of spin range 2 (vacuum supermultiplets) correspond to the
massive Kaluza-Klein spectrum of 11-d SUGRA over $S^4$ \cite{gnw}.

     We should stress that there are three different forms of the
superalgebra \OS ~related via triality. The form considered in this paper
as well as in \cite{gnw} satisfies the standard spin and statistics connection
and is the one relevant for the sector of M-theory over $AdS_7\times S^4$
containing the 11 dimensional supergravity. It is an open question whether
or not the other forms of \OS ~with exotic spin and statistics connection
are relevant to M-theory. One would have naively expected the form of
\OS ~in which the supersymmetry generators transform in the right-handed
spinor representation of $SO(6,2)$ to satisfy the usual spin and statistics
connection. However, since the right handed spinor representation decomposes
as $(1+6+\bar{1})$  under $SU(4)$ subgroup it is evident that the ``supersymmetry
generators'' in this case can not satisfy the usual spin and statistics
connection.

{\bf Acknowledgements} : We would like to thank Marco Zagermann for useful discussions .

{\it Note added:} This version of the manuscript in the arXiv hep-th incorporates the erratum sent to Nucl. Phys.
B about  missing states in the tables of some of the  supermultiplets. We are grateful to Sudarshan Fernando for
alerting us to the fact that some states were missing in the tables of the original published version.

}
\end{document}